\documentclass[authoryear,5p, times]{elsarticle}
\usepackage{graphicx}
\usepackage{multicol}
\usepackage{color}
\usepackage{tabularx}
\usepackage{pbox}
\usepackage{longtable}
\usepackage{graphics,epsfig,aas_macros,amsmath}
\newcommand{\software}[1]{\texttt{\small \MakeUppercase{#1}}}
\usepackage[authoryear]{natbib}
\newcommand{\obsid}{Obs-ID\,}
\newcommand{\obsids}{Obs-IDs\,}

\begin{document}
\begin{frontmatter}

\title{ X-ray and optical orbital modulation of EXO 0748-676 : A co-variability study using \textit{XMM-Newton}\textsuperscript{*}\corref{titleft}}
 \cortext[titleft]{based on observations obtained with \textit{XMM-Newton}, an ESA science mission with instruments and contributions 
 directly funded by ESA Member States and NASA.}
\author[Raman]{Gayathri Raman \corref{mycorrespondingauthor}}
\cortext[mycorrespondingauthor]{Corresponding author: Gayathri Raman; Email:graman@rri.res.in}
\author[Raman]{Biswajit~Paul}
\address[Raman]{Raman Research Institute, C. V. Raman Avenue, Sadashivanagar, Bangalore-560012, India }

\begin{abstract}
We present a  multi-wavelength  timing study of the eclipsing low mass X-ray binary EXO 0748-676 (UY Vol) using \textit{XMM-Newton}
 when the source was in a hard spectral state. The orbital optical and X-ray light curves show a fairly large amount of intensity 
modulation in the 7 observations taken during September-November, 2003, covering 36 complete binary orbits of EXO 0748-676. 
While assessing the non-burst variability, simultaneously in the optical and X-ray light curves, we find that they 
are not correlated at reprocessing or orbital time-scales, but are weakly correlated at a few 1000s 
of seconds time-scales. Although a large fraction of the optical emission is likely to be due to reprocessing,
the lack of significant correlation and presence of large variability in the orbital X-ray and optical light curves is 
probably due to structures and structural changes in the accretion disk that produce, and sometimes mask the reprocessed signal 
in varying amounts. These disk structures could be induced, at least partly, by irradiation. From the observed modulations seen in the 
optical light curves, there is strong evidence of accretion disk evolution at time scales of a few hours.
\end{abstract}


\end{frontmatter}

\section{Introduction}

   EXO 0748-676 is a low mass X-ray binary (LMXB) that was discovered with EXOSAT in 1985 \citep{Parmar}. 
 Soon after its discovery the optical counterpart was identified as UY Vol \citep{Wade, Pedersen}. It is a high inclination 
 ($\sim$75-82$^\circ$) eclipsing binary system with an orbital period of 3.82 hr \citep{Parmar31985circular, Crampton, Parmar}. 
  The presence of thermonuclear bursts in the X-ray light curve established the compact object in the binary as a neutron star
 \citep{Gottwald, Parmar}. Many high and low frequency variabilities have been detected in this source, including a 
 Quasi Periodic Oscillation (QPO) at 695 Hz \citep{homan_vanderklis_2000}. \citet{Galloway_2010} discovered burst oscillations at 552 Hz,
 which is likely to be the spin period of the neutron star in EXO 0748-676. This object also shows pre-eclipse intensity dips that are 
 commonly understood to be due to obscuration of  the central X-ray source by structures in the outer disk, whose azimuthal distribution 
 is variable \citep{Parmar}. Mass of the companion has been estimated to be in the range of 0.16 to 0.42 M$_{\odot}$ with the 
 upper limit corresponding to an M2V spectral type star \citep{Hynes_2009}.\\ 
 
 In addition to eclipses, dips and thermonuclear X-ray bursts, this system exhibits a lot of intensity modulation in the
 X-rays as well as in the UV/optical \citep{Crampton, Parmar, vanP}. A significant fraction of the UV/optical emission
 is understood to be reprocessed emission from the accretion disk ~\citep{vanP}.  The broad and shallow eclipses seen in the optical 
 as against the narrow and sharp X-ray eclipses indicate that the reprocessed optical photons are being emitted from an extended region, 
 namely, the accretion disk \citep{Crampton, vanP}.\\
 
 EXO 0748-676 was moderately bright and displayed a lot of short term variability since the time of its discovery in 1985
till 2008. For a span 24 years, its persistent X-ray luminosity remained at $\sim$10$^{36-37}$ergs/s \citep{degenaar_2011}, after which, 
it went into quiescence in 2008 \citep{Hynes_2009}.\\
 
Recently, a number of studies involving high inclination neutron star X-ray binaries indicate a 
correlation between the luminosity state and disk-wind outflows \citep{Ponti2014}. This is similar 
to the correlation between spectral state and radio jet outflows observed in black hole binaries \citep{Ponti2012a}. 
EXO 0748-676 and AX J1745-281 are two systems in which highly ionized Fe absorption lines (Fe XXV and Fe XXVI) 
were detected specifically during their soft spectral state \citep{Ponti2014,Ponti2015}. These absorption lines, being 
indicators of disk-winds, confirmed the fact that a correlation between luminosity state and presence of disk-wind 
indeed exists \citep{Ponti2014,Ponti2015}. There have also been instances of simultaneous detection 
of radio-jets as well as disk-wind outflow signatures from objects like GX 13+1, Sco X-1 and GX 340+0, in the hard luminosity state, 
which indicate that jets and disk-winds may not be mutually exclusive \citep{homan_2016}. \\

Soon after its discovery, it was observed that EXO 0748-676 displayed different X-ray intensity states \citep{Parmar, Gottwald}. 
Recent studies by \citet{Ponti2014}, showed that the bright state displayed less intense dipping phenomenon compared to the faint state. 
We have also explored this intensity state dependent dipping behavior of EXO 0748-676, in this current work.\\

Simultaneous X-ray and optical data have been used previously to study the co-variability in LMXBs. 
Some sources showed X-ray and optical correlation during quiescence (4U 1735-44: \citealt{Corbet}, V404 Cyg: \citealt{Hynes_2004});
some, no correlation during quiescence (Cen X-3: \citealt{Cackett}); others exhibit correlation during highest emission 
(4U 1556-605: \citealt{Motch1989}); while anti-correlated emission (4U 0614+09: \citealt{Machin}) or even bimodal behavior 
(i.e., correlation as well as anti-correlation, Sco X-1: \citealt{Ilovaisky}) is also seen. 
The lack of a consistent relation between X-ray and optical 
variations in LMXBs have usually been interpreted to be due to changes in the extent and visibility of the reprocessing structures 
that produce most of the optical variability.\\

The X-ray and optical emission for EXO 0748-676 was initially seen on an average to be correlated on short time 
scales \citep{Motch_exo}. \citet{Thomas} studied the correlated X-ray and optical flux in EXO 0748-676 to understand the reprocessed 
emission. They did not find any X-ray and optical co-variability that was indicated from earlier studies. \citet{Southwell} also 
confirmed this change in the nature of co-variability. In fact, a larger X-ray flux during the \textit{bright} state would mean an 
increased mass accretion rate that should make the outer disk bulge more prominent, thereby giving rise to the broad optical 
eclipse or the optical \textit{high} state. The lack of X-ray and optical co-variability was associated with the geometric masking 
of the reprocessed emission. Understanding how the system geometry alters the X-ray and reprocessed optical signal requires 
correlation studies, which we have carried out with very long, simultaneous X-ray and optical light curves, obtained using 
\textit{XMM-Newton}.\\

 In this paper, we probe the non-burst co-variability of the X-rays and optical emission in this quasi-persistent binary, 
 using \textit{XMM-Newton} observations, when it was in the hard state in 2003. We also present the orbit-to-orbit variations 
 seen in the X-ray and optical light curves.

\section{Observations}

EXO 0748-676 was observed during September-November, 2003 with \textit{XMM-Newton}. Details of the archival data involving 7
\obsids are shown in Table 1. We have used data from the European Photon Imaging Camera (EPIC) instrument using the PN 
CCD (0.1 - 10 keV) for all of the 7 observations. The Optical Monitor (OM, \citealt{Mason2001}) was simultaneously operated in timing mode 
using the white band (170 - 650 nm). EXO 0748-676 was observed for $\sim$600 ks spanning all the 7 \obsids
in the X-rays. The EPIC-PN and OM together thus provide a simultaneous multi-wavelength coverage to the
target in the X-ray and UV/optical bands.\\

We reduced the EPIC-PN data using the \software{SAS} version 12.0.1 software, using the latest calibration files.
The X-ray light curves were extracted in the soft (0.3 - 5 keV) and hard (5 - 10 keV) bands from a 40" radius 
centered on the position of EXO 0748-676 with 1 s binning. The optical light curve was extracted from a region of 6" around the source 
using the \textit{omichain} task which processes the data with the latest calibration files, and subsequently performs source detection 
and aperture photometry. \\

There have been previous burst studies using this dataset by \citet{Boirin}, spectroscopic studies of the dips by \citet{Diaz}, 
burst spectral studies by \citet{Cottam} and a comparison of the optical to X-ray burst emission by \citet{BP}. 
During another observation of EXO 0748-676 with \textit{XMM-Newton} in 2005 (\obsid 0212480501), the source was found to be in a 
soft spectral state \citep{Ponti2014}. We reduced the OM data for this particular dataset as well. However, because of 
the different filters used and relatively short exposures, the optical light curves from this observation were not suitable
for further analysis.

 \begin{table*}
\centering
 \resizebox{0.9\textwidth}{!}
 {\begin{tabular}{cccccccccc}
 \hline
 \obsid &  \pbox{20cm}{Date\\(dd/mm/yyyy)} & \pbox{20cm}{Start Time\\UT (hh:mm)} & t$_{exp}$ (h) &\pbox{20cm}{No. of \\X-ray\\ bursts}  & \pbox{20cm}{No. of\\Optical\\bursts} & \pbox{20cm}{No. of \\ Full orbits}  & \pbox{20cm}{Non-burst\\soft X-ray\\variability($\%$)} & \pbox{20cm}{Non-burst\\optical\\variability($\%$)} & \pbox{20cm}{Non-burst\\hard X-ray\\variability($\%$)}\\  
 \hline
0160760101 & 19/09/2003 &  13:37 &24.6 &10 & 8 & 6 &  52.19 & 13.93 & 22.68\\
0160760201 & 21/09/2003  & 13:38 &25.1 &14 & 13 & 6 & 65.25  & 16.78 & 25.19 \\
0160760301 & 23/09/2003  & 10:42 &30.0 &14 & 11 & 7 & 50.20  & 15.38 & 22.55\\
0160760401 & 25/09/2003  & 17:29 &20.4 &9 & 8 & 5 & 65.79  & 17.17 & 23.16 \\
0160760601 & 21/10/2003  & 10:02 &15.2 &8 & 6 & 3 &  55.82 & 17.07  & 22.87\\
0160760801 & 25/10/2003  & 19:19 &17.3 &9 & 6 & 3 & 73.30 & 16.11 & 23.32 \\
01607601301 & 12/11/2003 &  08:24 &25.2 &12 & 10 & 6 & 65.27  & 19.54 & 24.97\\
\hline
 \end{tabular}}
 \caption{EXO 0748-676 was observed using \textit{XMM-Newton} in 2003. This table details the \obsids, date of observation, 
 number of thermonuclear X-ray bursts, number or reprocessed optical bursts, number of complete orbital segments and non-burst 
 r.m.s variability for the soft X-rays, hard X-rays and optical.}
 
  \end{table*}

  
  
 \section{Data analysis and results}
 
 \subsection{X-ray and optical light curves}
 \subsubsection{Raw light curves}
 The X-ray light curve has been divided into two bands, soft (0.3 - 5.0 keV) and hard (5.0 - 10.0 keV). 
 A total of 76 thermonuclear X-ray bursts were reported from all the 7 observations in X-rays \citep{Boirin}, a summary of which is 
 presented in Table 1. 63 of these bursts had simultaneous optical band data as well \citep{BP}. There were gaps in the optical data
 during 13 of the X-ray bursts, because of which they were not recorded in the optical observations. \citet{BP} fitted all the burst 
 profiles with a model consisting of a constant component, a burst with a linear rise and an exponential decay. The X-ray and optical
 counts of the bursts were calculated by subtracting the pre-burst component from  the total photon counts during the bursts. 
 The bursts show an average X-ray to optical conversion factor, which is the ratio of the optical counts to the X-ray counts for each 
 of the burst, of 0.15 \citep{BP}. Simultaneous X-ray and optical light curves for the \obsid 0160760201 are shown in Figure 1 with 
 a bin size of 60 s. Optical eclipses are broad depressions as opposed to the sharp and near total X-ray eclipses, indicating that 
 a corresponding part of the optical emission originates in the inner accretion disk.
 
  \begin{figure}
 \centering
  \includegraphics[scale=0.3, angle=-90]{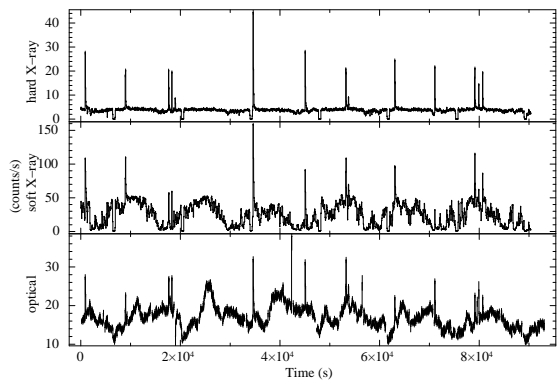}
 \caption{Simultaneous  hard X-ray (5 - 10 keV), soft X-ray (0.3 - 5.0 keV), and optical light curves with 60 s binning are shown for
 the \obsid 0160760201. The light curves show simultaneous thermonuclear bursts in all the three bands as well as significant non-burst 
 intensity modulation in the soft X-ray and optical bands. }
\end{figure}
 
 \subsubsection{Filtered non-burst light curves}

A time window file, created using the task \textit{xronwin}, was used to exclude the thermonuclear burst time intervals from 
the X-ray and optical light curves. The time windows, in some cases, contained a little more than just the X-ray burst duration 
to account for the delayed optical response and also the slow, extended decay tails. These time windows spanned a range of 
approximately 40 - 100 s depending on the duration of each of the bursts. Simultaneous hard X-ray, soft X-ray and optical 
light curves are shown in Figure 2 for all the 7 observations with a bin size of 60 s. A significant amount of variability is 
seen in the soft X-ray band compared to the hard X-ray in all observations. 
The most prominent variability observed in the light curves is the orbital modulation. The pattern of orbital modulation changed 
significantly during the observations, which we discuss below. For almost all the \obsids, the optical count rate varies between 
10 - 25 counts/s and soft X-ray count rates are nearly 2 - 3 times the optical rates. The percentage r.m.s variability with respect 
to the average values of the 60 s binned light curves in these bands for all the 7 observations are shown in Table 1. Significant 
differences in the intensity variation patterns like heavy to moderate dipping behavior, frequent short time-scale fluctuations, 
smooth orbital variability etc. are seen in all the three bands on different observation days. \\

 \subsection{Orbit-to-orbit modulation}
 
 To further study the sub-orbital short time-scale modulation and its evolution in consecutive orbits, the light curves were 
 sliced into several orbits using the X-ray orbital ephemeris provided in \citet{Wolff2009}. Plots containing soft X-ray and 
 optical light curves from different binary orbits are overlaid and are shown in Figure 3 for all the seven observations.\\
 
 The \obsid 0160760101 displays an erratic X-ray behavior with many soft X-ray dips at different phases. 
 In four of the observations (\obsids 0160760201, 0160760401, 0160760801 and 0160761301),  pre-eclipse dips are seen in soft 
 X-rays in the orbital phase range of 0.6-0.9. In some observations there are dips around phase 0.2 (\obsid 0160760601), 0.15 
 (\obsid 0160760401) and 0.45 (\obsid 0160761301). Other than the dips, most observations show large orbit-to-orbit intensity 
 variations often by a factor of a few; for example at phase $\sim$ 0.2 in \obsids 0160760201 and 0160760401 and at all phases in 
 \obsid 0160760101. In \obsid 0160760601 there are dips in phase range 0.6-0.9 but with a large difference in the dips phases 
 from orbit to orbit. The duty cycle of the soft X-ray dips is seen to be as high as 50 $\%$ in some orbits.  \\
 
 Optical light curves show dips and even peaks at different phases. \obsids 0160760201, 0160760301, 0160760401 and 0160760601
 show peaking behavior at phases 0.1 and 0.4 with slight dips around phase 0.2 in some of the orbits. 
 A single orbit each in \obsids 0160760201 and 0160760601 peaks at phase 0.2 unlike the rest of the orbits. 
 \obsids 0160760101 and 0160760301 have a single peak around 0.25, whereas \obsids 0160760801 and
 0160761301 show a peak somewhere close to phase 0.3. The soft X-ray and corresponding optical variations are broadly classified
 into five classes according to the variable dipping seen in their orbital profiles (Figure 4). This classification is based 
 on the observed soft X-ray light curves from all the 7 observations.\\

  \begin{figure*}

\begin{minipage}{1.0\linewidth}
\centering
\includegraphics[scale=0.22,angle=-90]{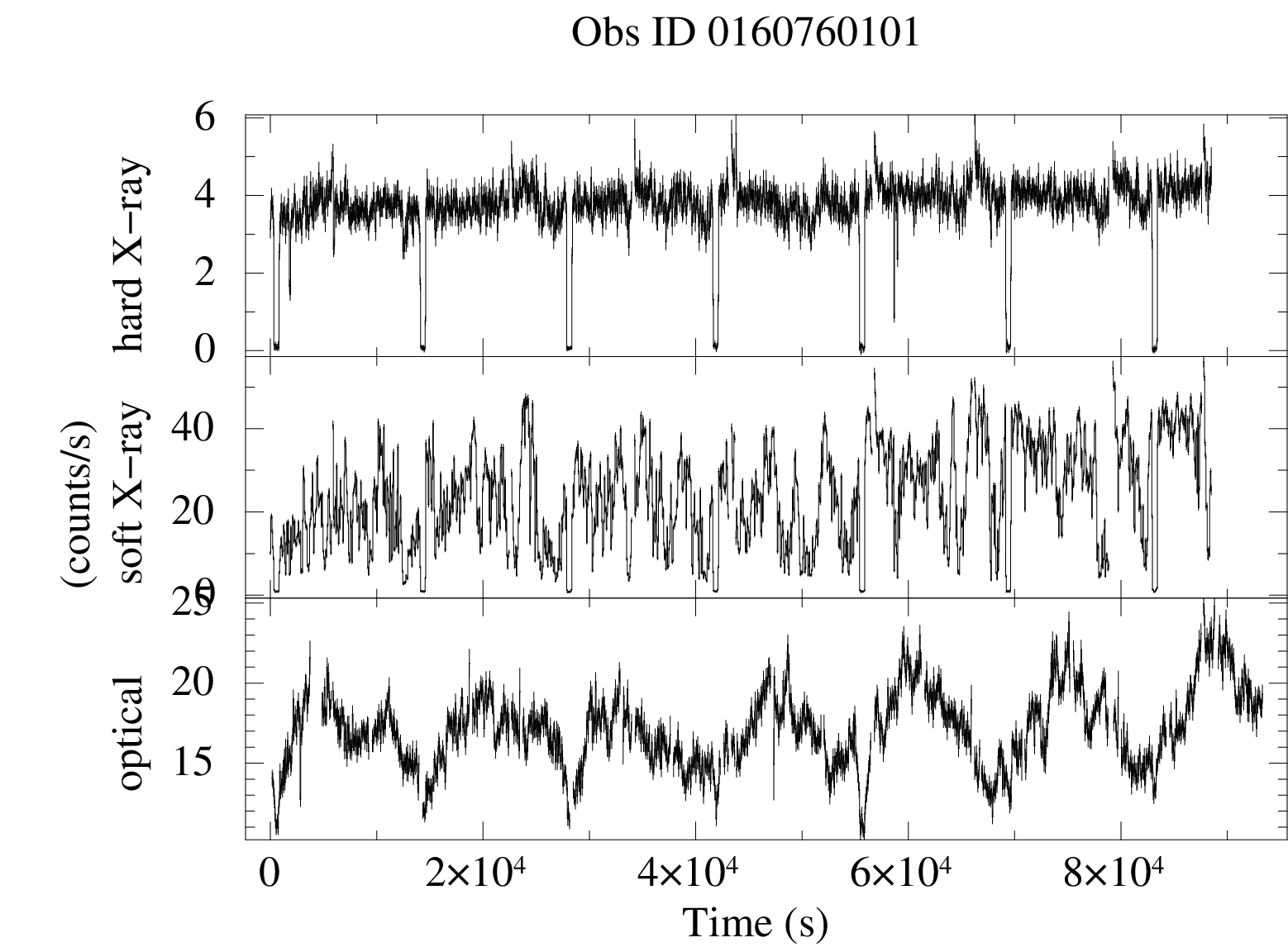}
\includegraphics[scale=0.22, angle=-90]{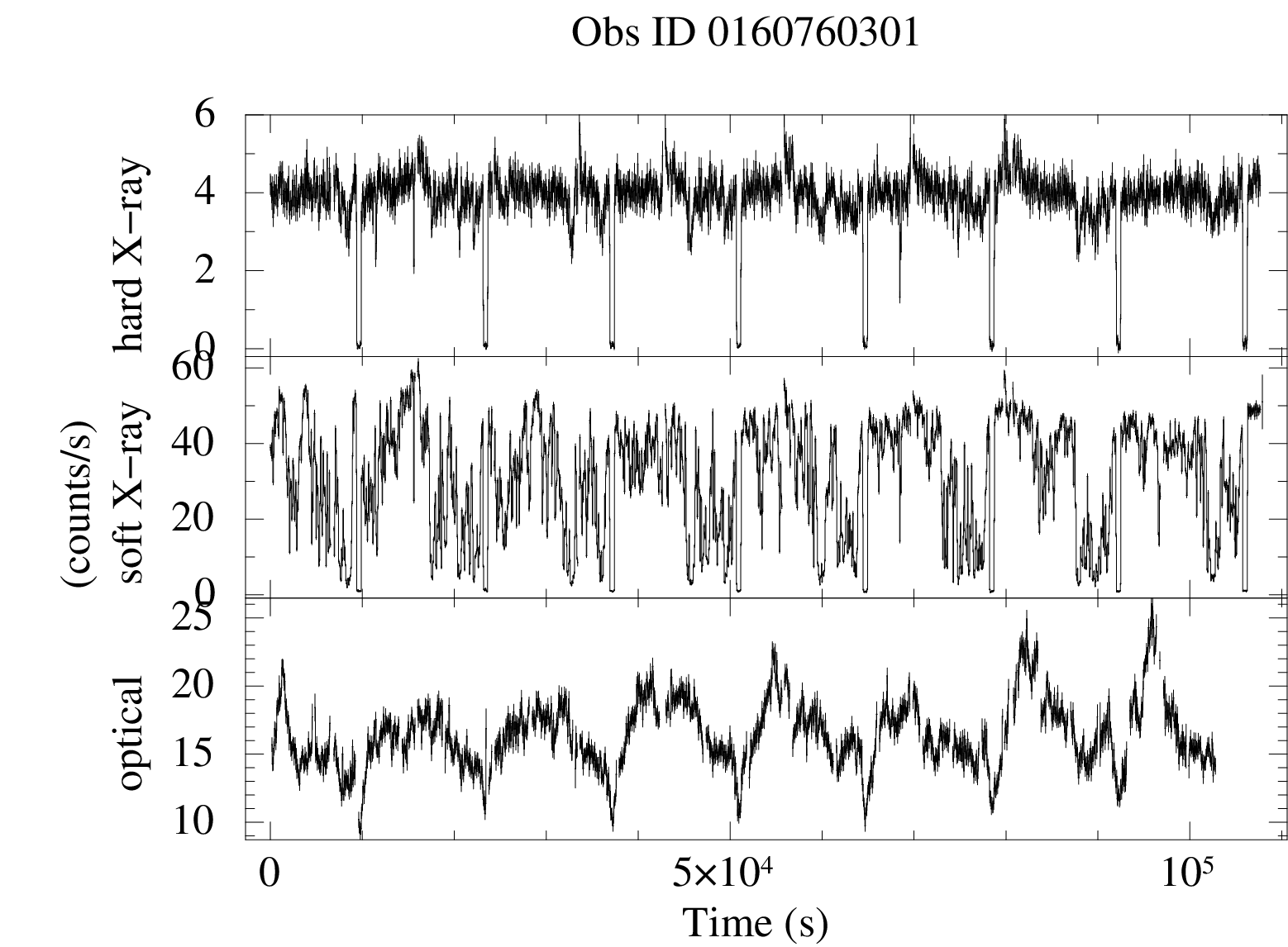}

  \end{minipage}     
   \hspace{0.8cm}    
\begin{minipage}{1.0\linewidth}   
\centering
      \includegraphics[scale=0.22,angle=-90]{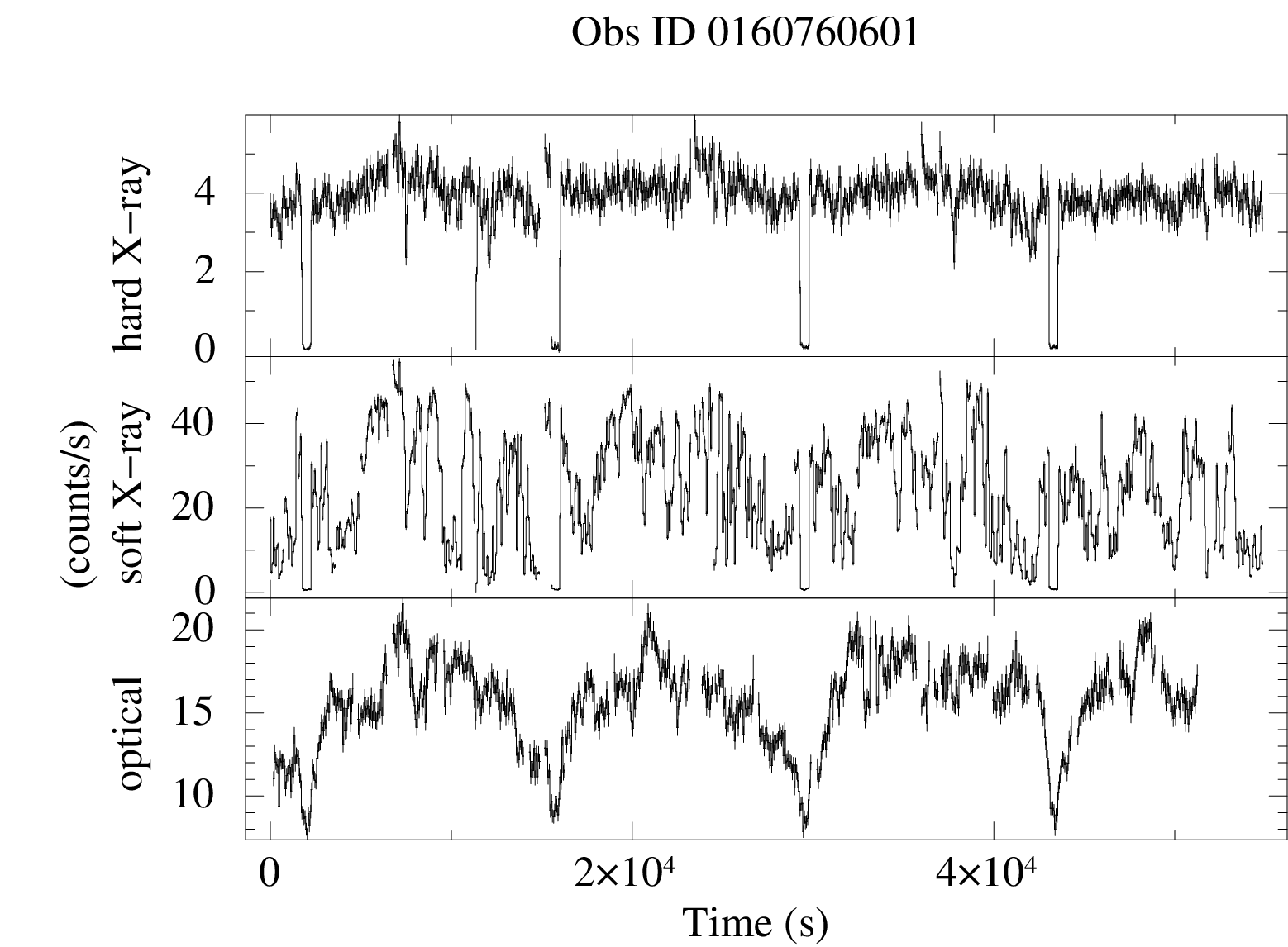}
       \includegraphics[scale=0.22,angle=-90]{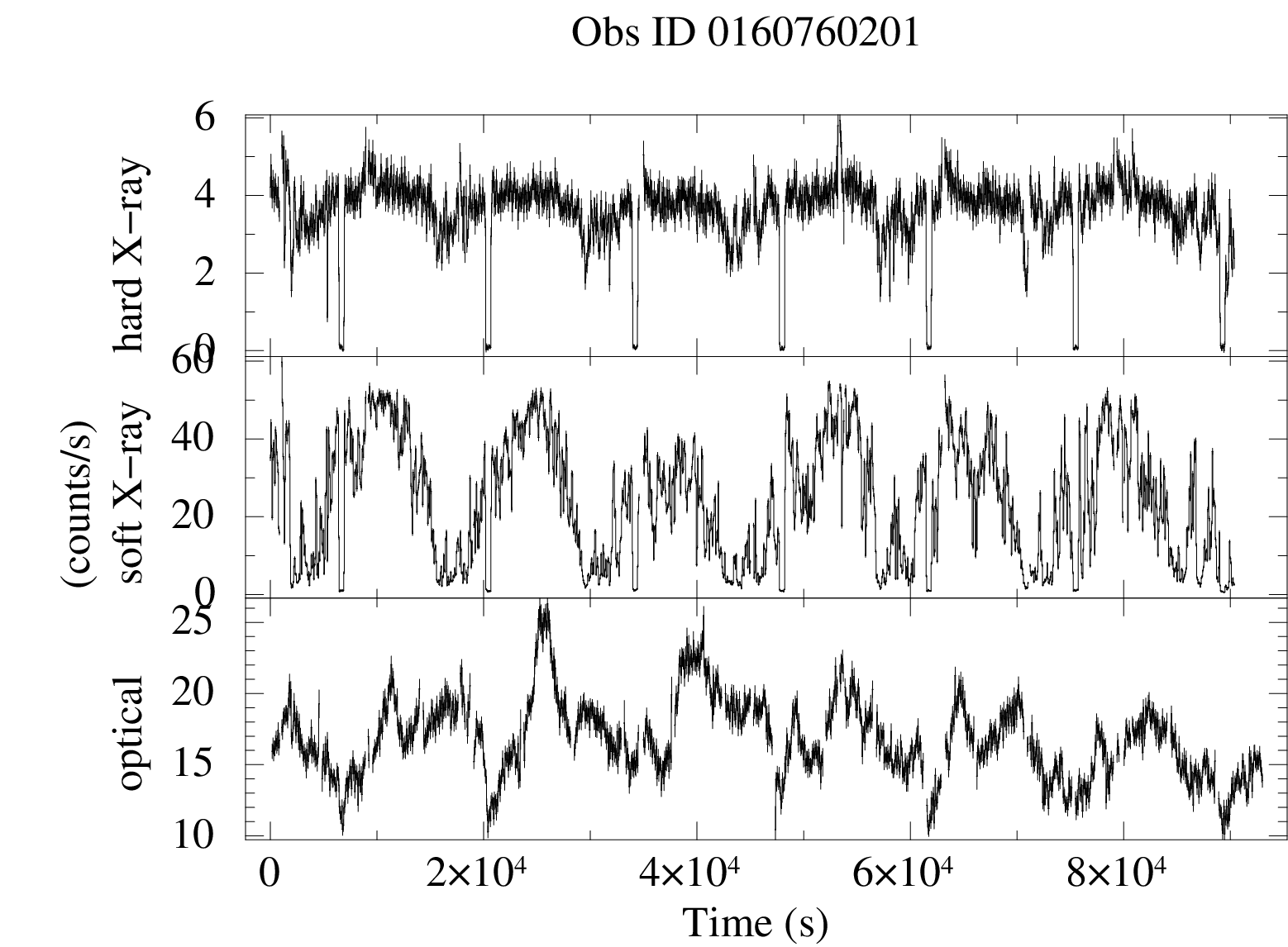}
       
  \end{minipage}     
   \hspace{0.8cm}    
\begin{minipage}{1.0\linewidth}   
\centering
       \includegraphics[scale=0.22, angle=-90]{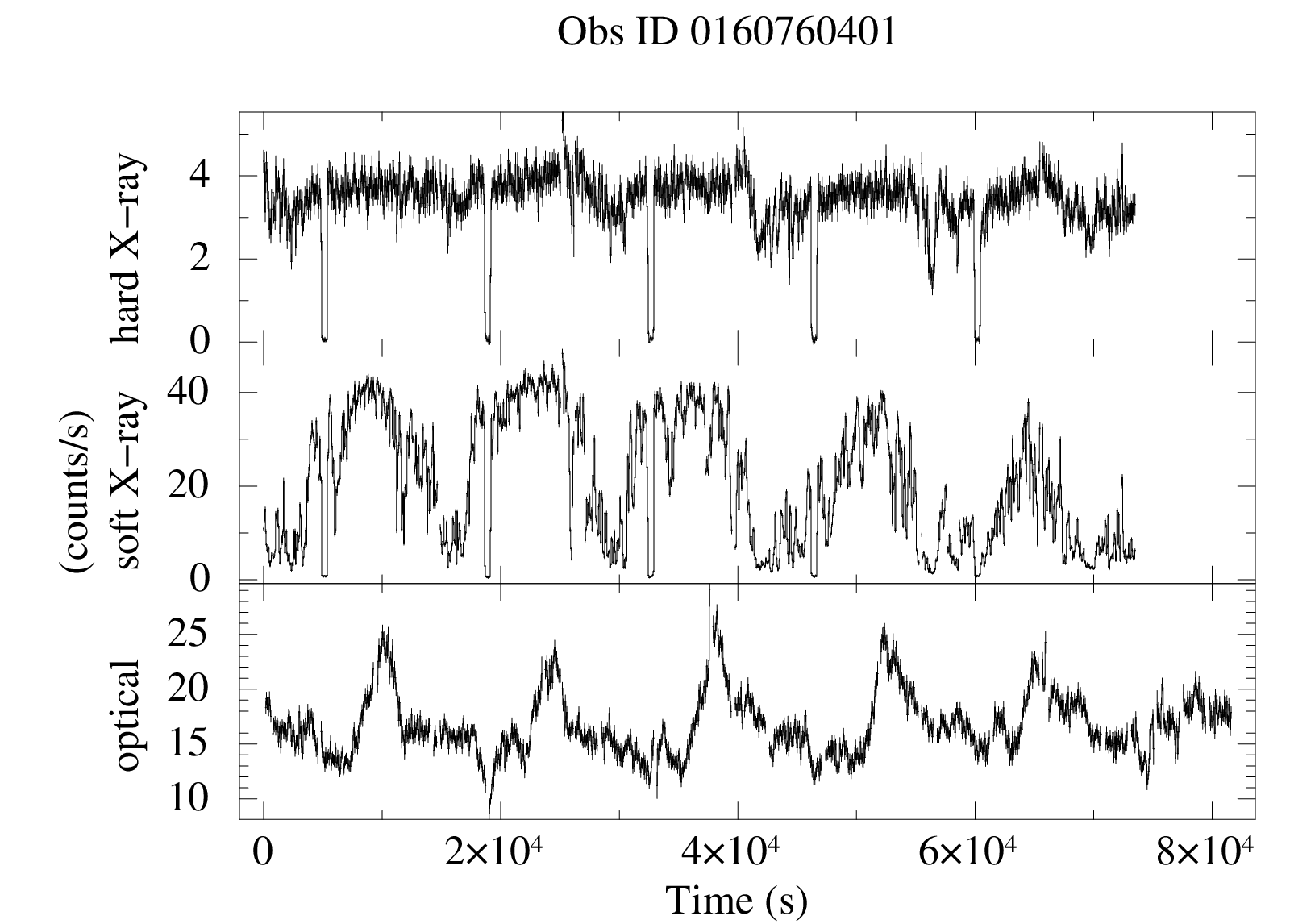}
       \includegraphics[scale=0.22,angle=-90]{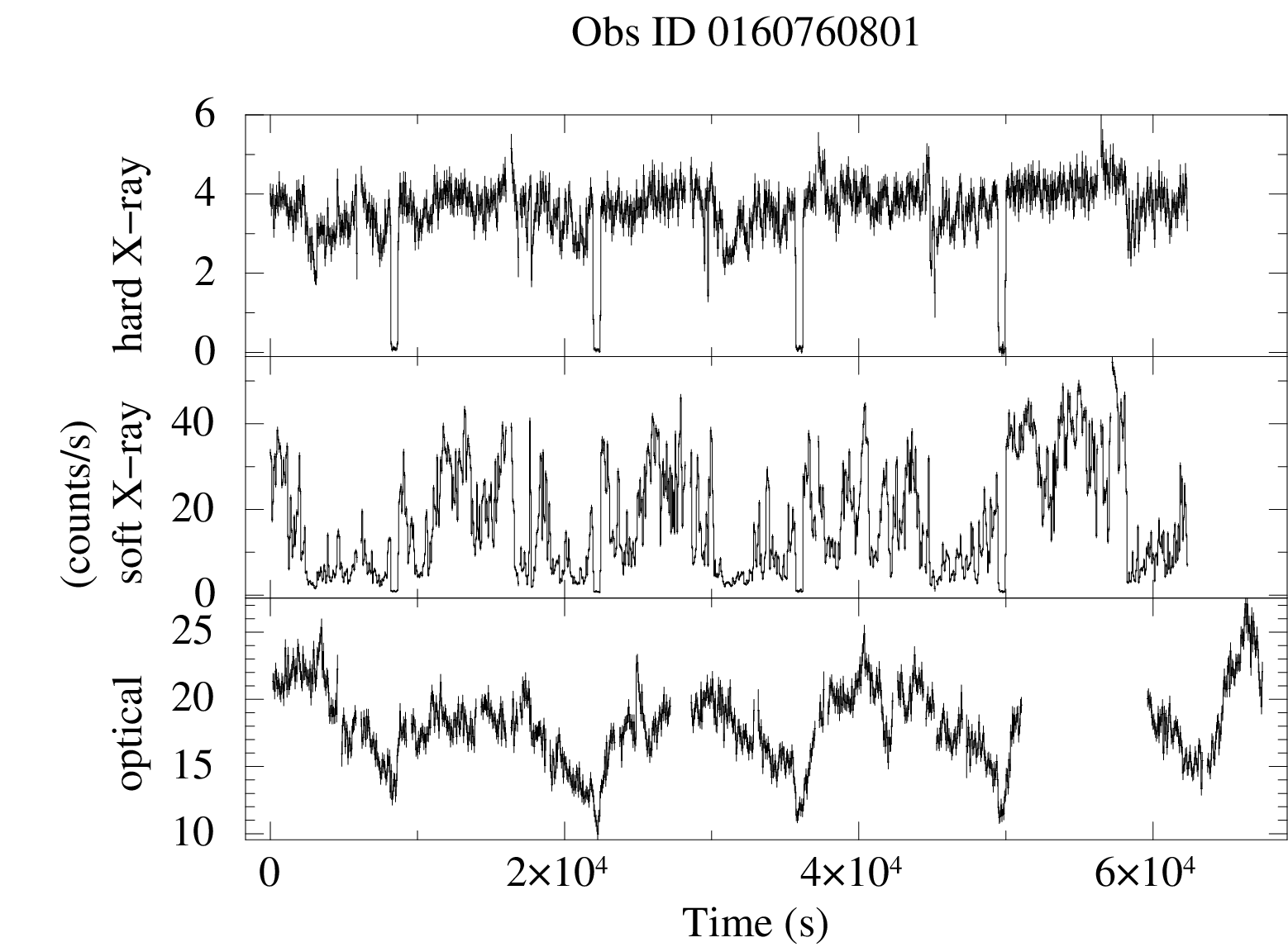}
       
    \end{minipage}  
    \begin{minipage}{1.1\linewidth} 
      \centering
         \includegraphics[scale=0.22,angle=-90]{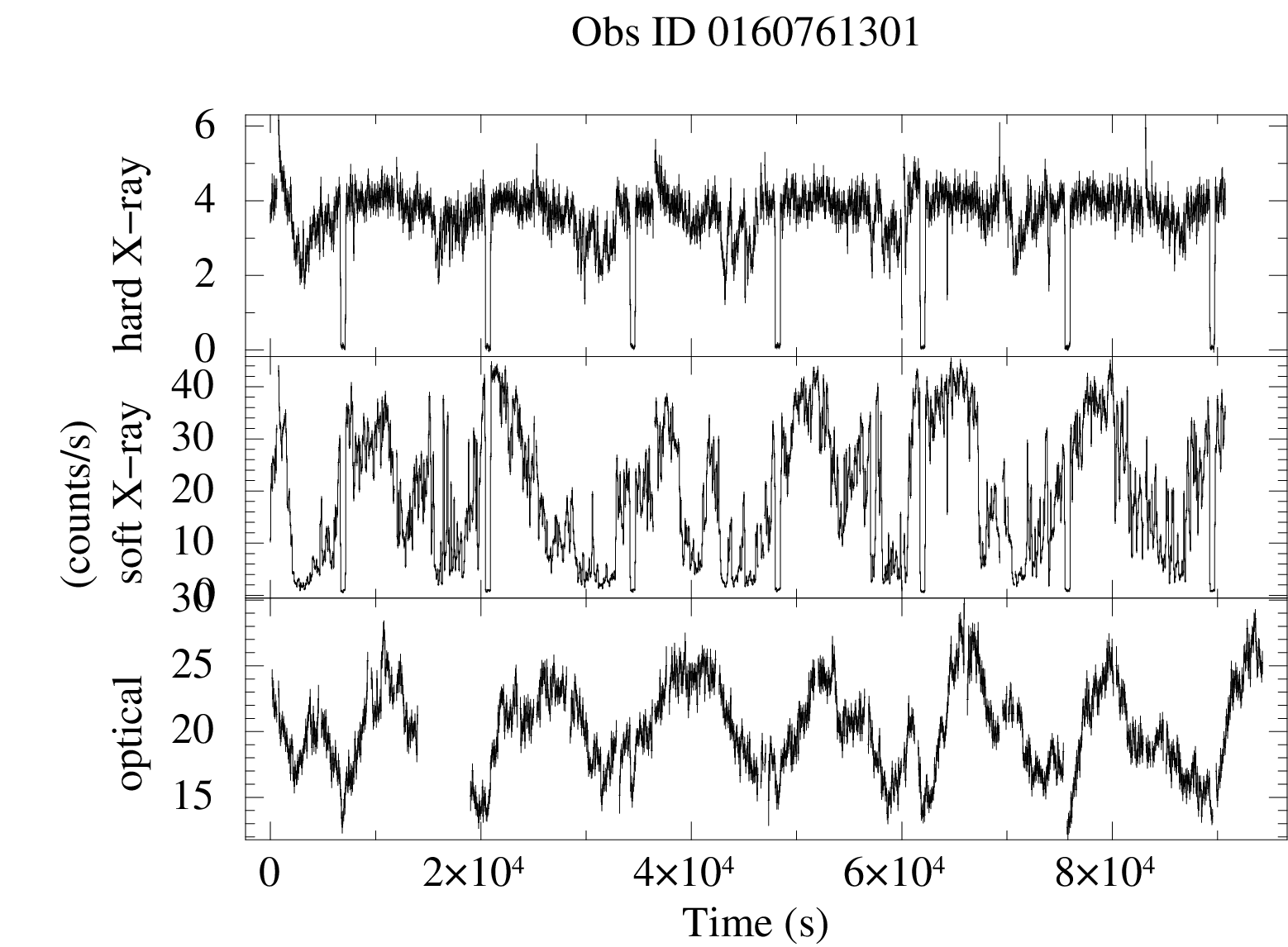}
         
       \end{minipage}      
  
    \caption{Simultaneous hard X-ray (5 - 10 keV), soft X-ray (0.3 - 5.0 keV), and optical light curves 
    for all the \obsids are shown with a filter applied to remove the thermonuclear burst sections. The non-burst X-ray and optical 
    light curves display significant intensity modulation during all the observations. }
\end{figure*}

 \begin{figure*}
\begin{minipage}{1.0\linewidth}
\centering
      \includegraphics[scale=0.23,angle=-90]{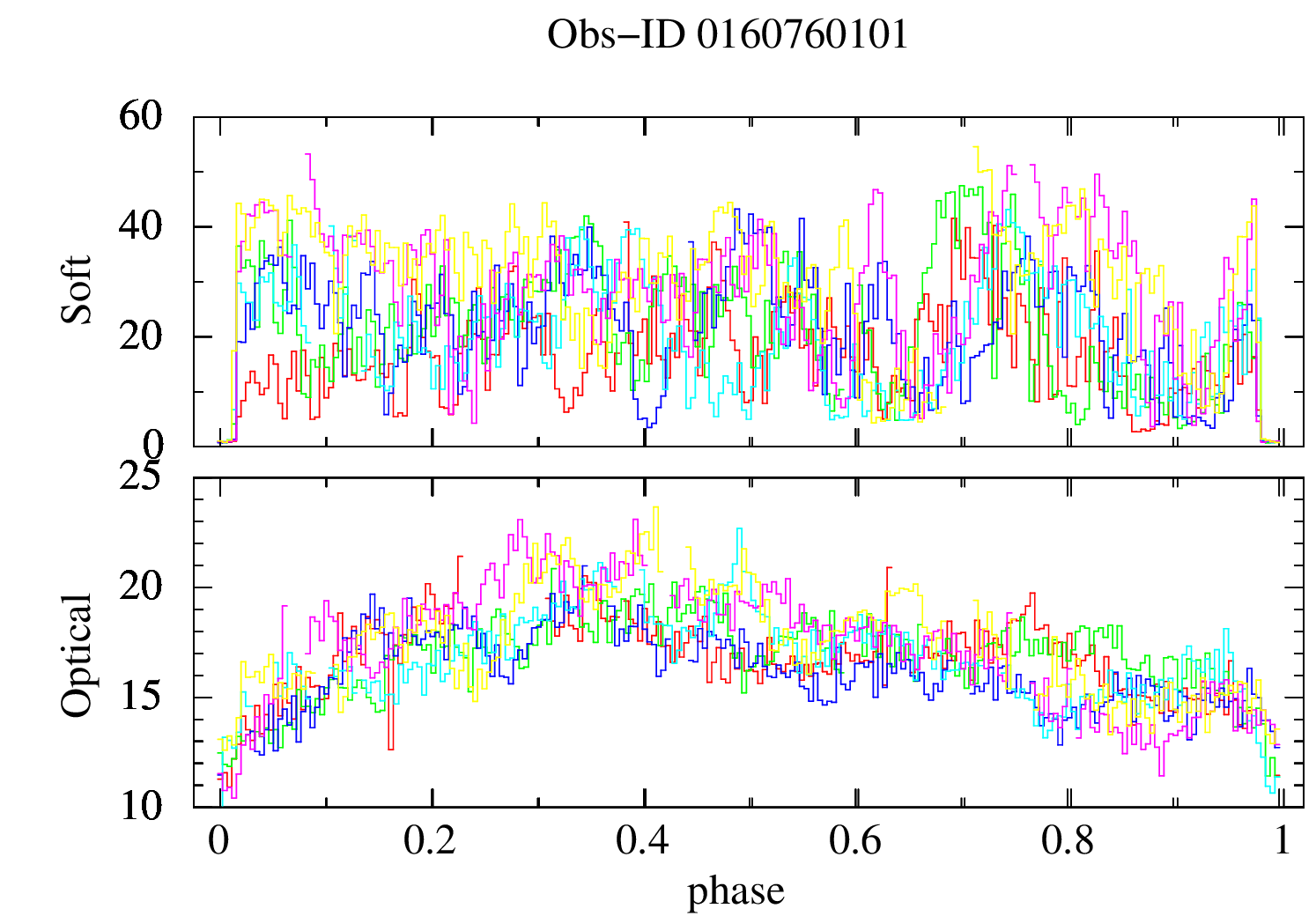}
      \includegraphics[scale=0.23,angle=-90]{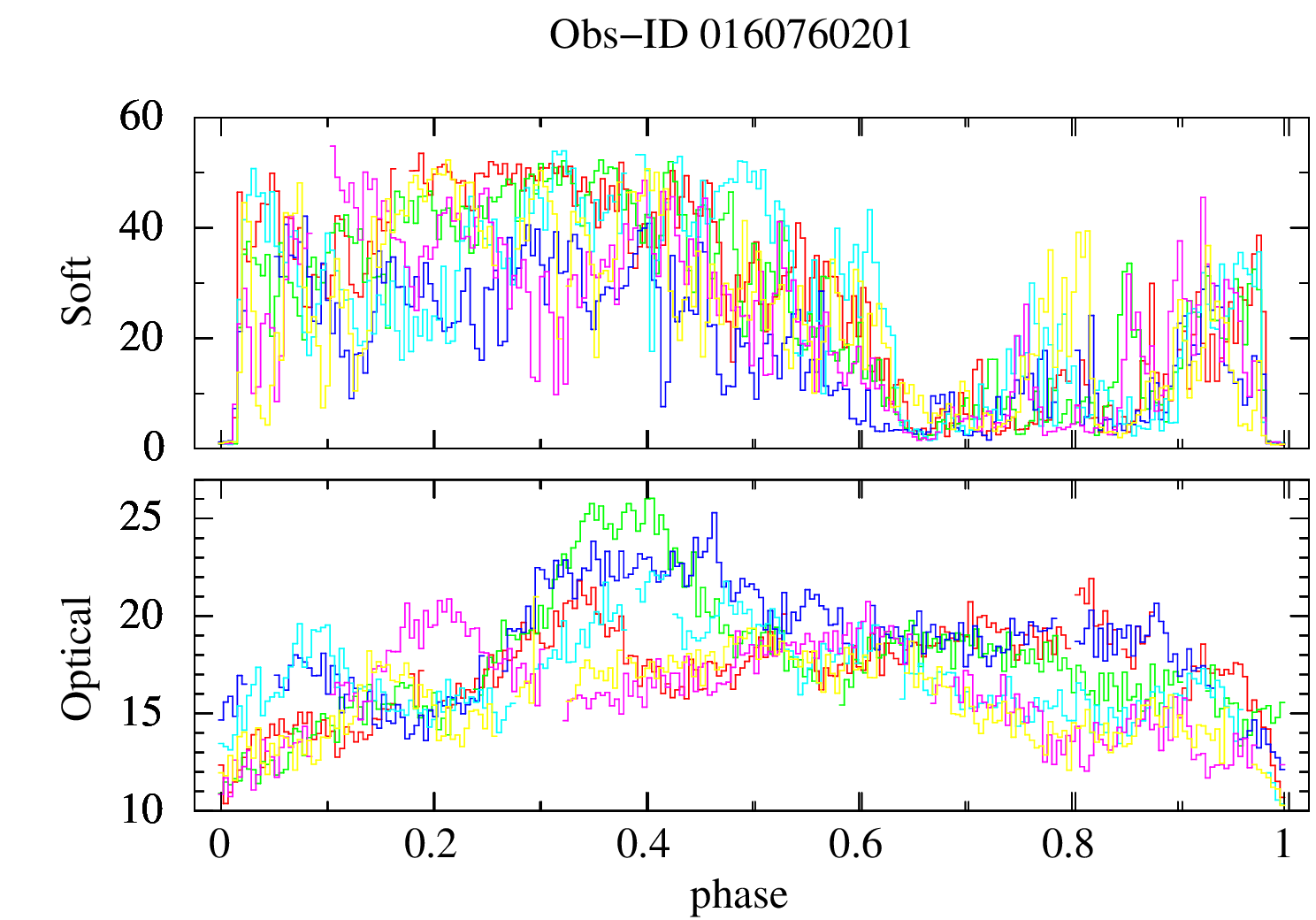}
          
          \end{minipage}     
   \hspace{0.8cm}    
\begin{minipage}{1.0\linewidth}    
\centering
      \includegraphics[scale=0.23, angle=-90]{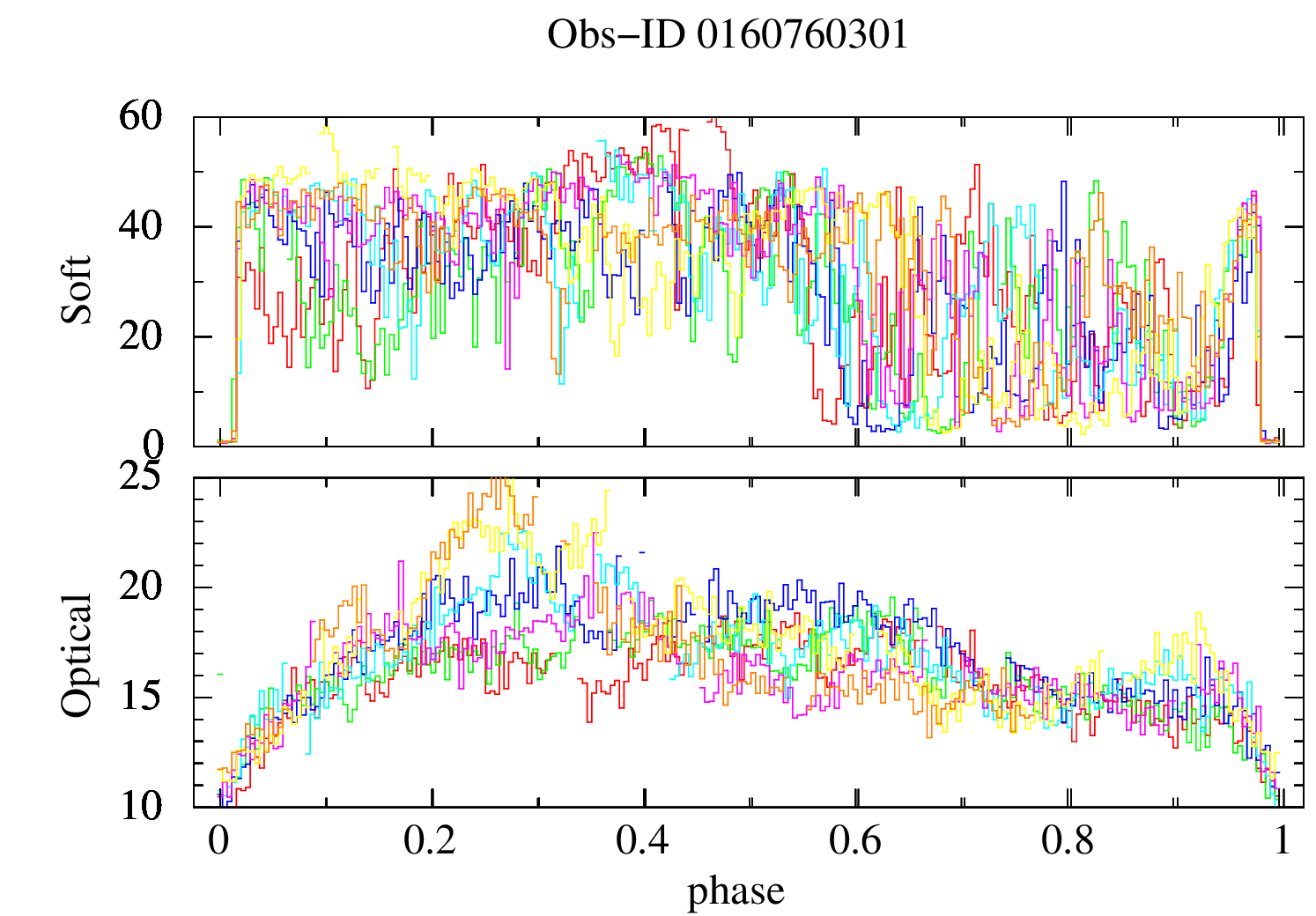}
       \includegraphics[scale=0.23,angle=-90]{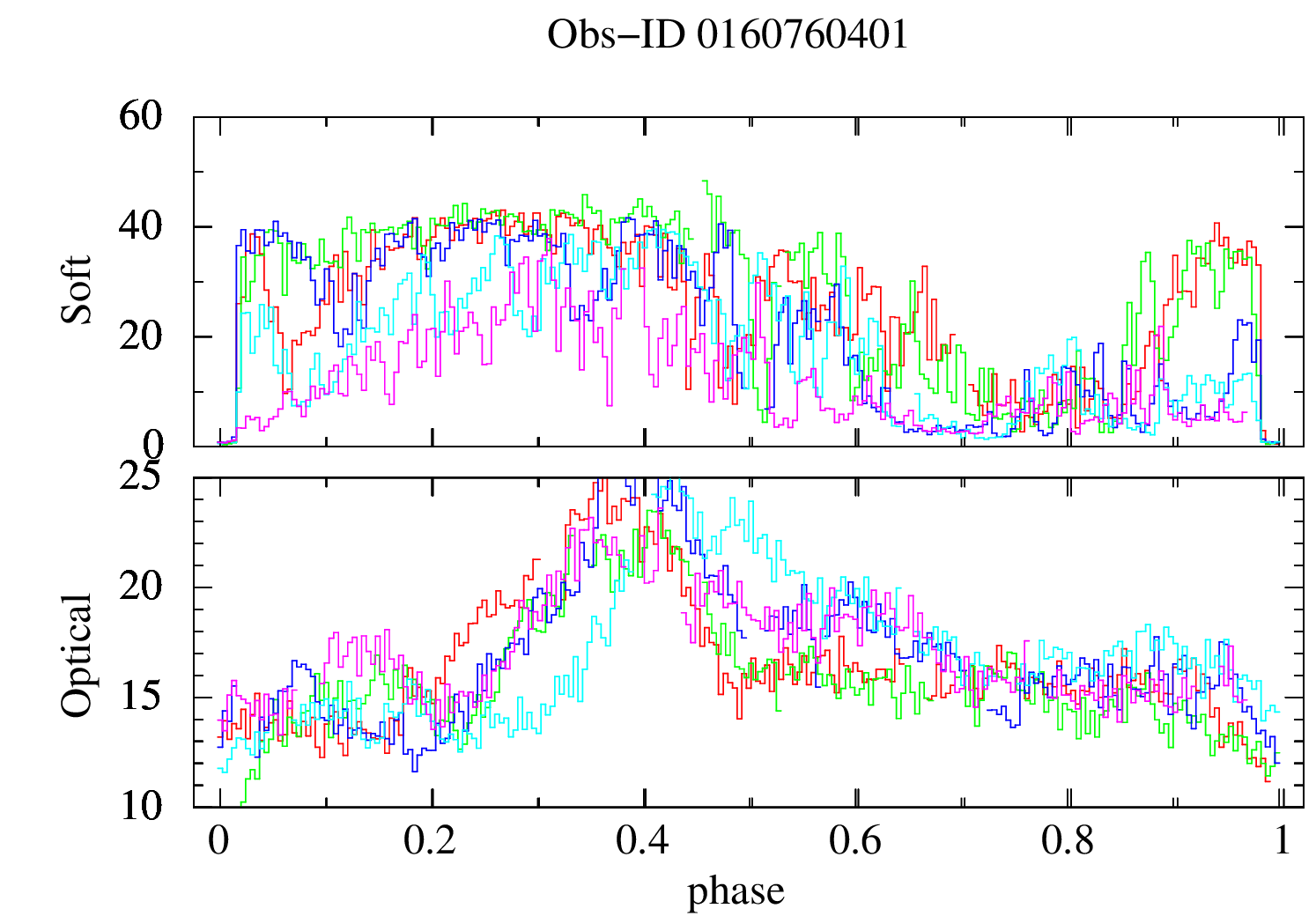}
              
         \end{minipage}     
   \hspace{0.8cm}    
\begin{minipage}{1.0\linewidth}    
\centering
         \includegraphics[scale=0.23, angle=-90]{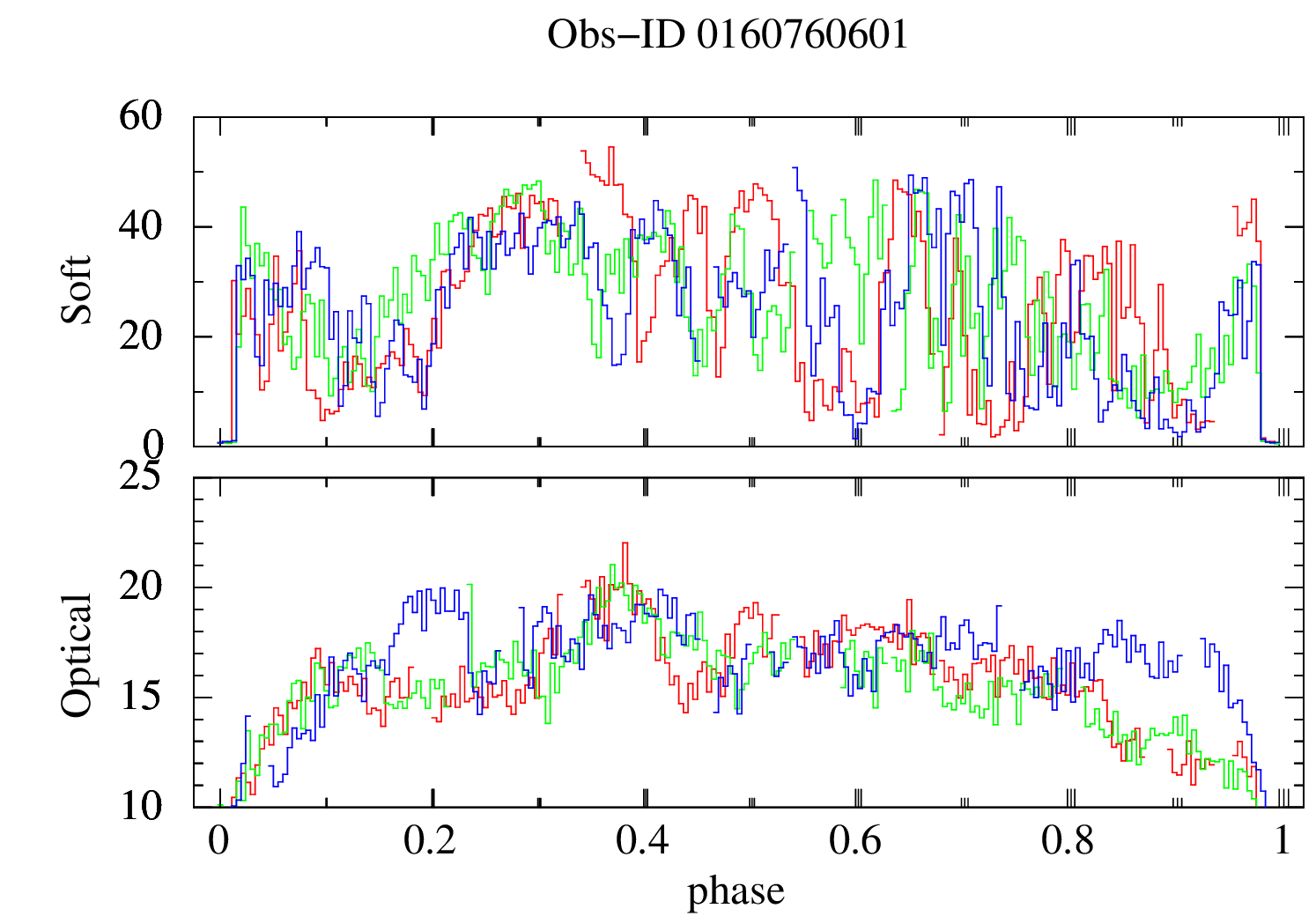}
         \includegraphics[scale=0.23,angle=-90]{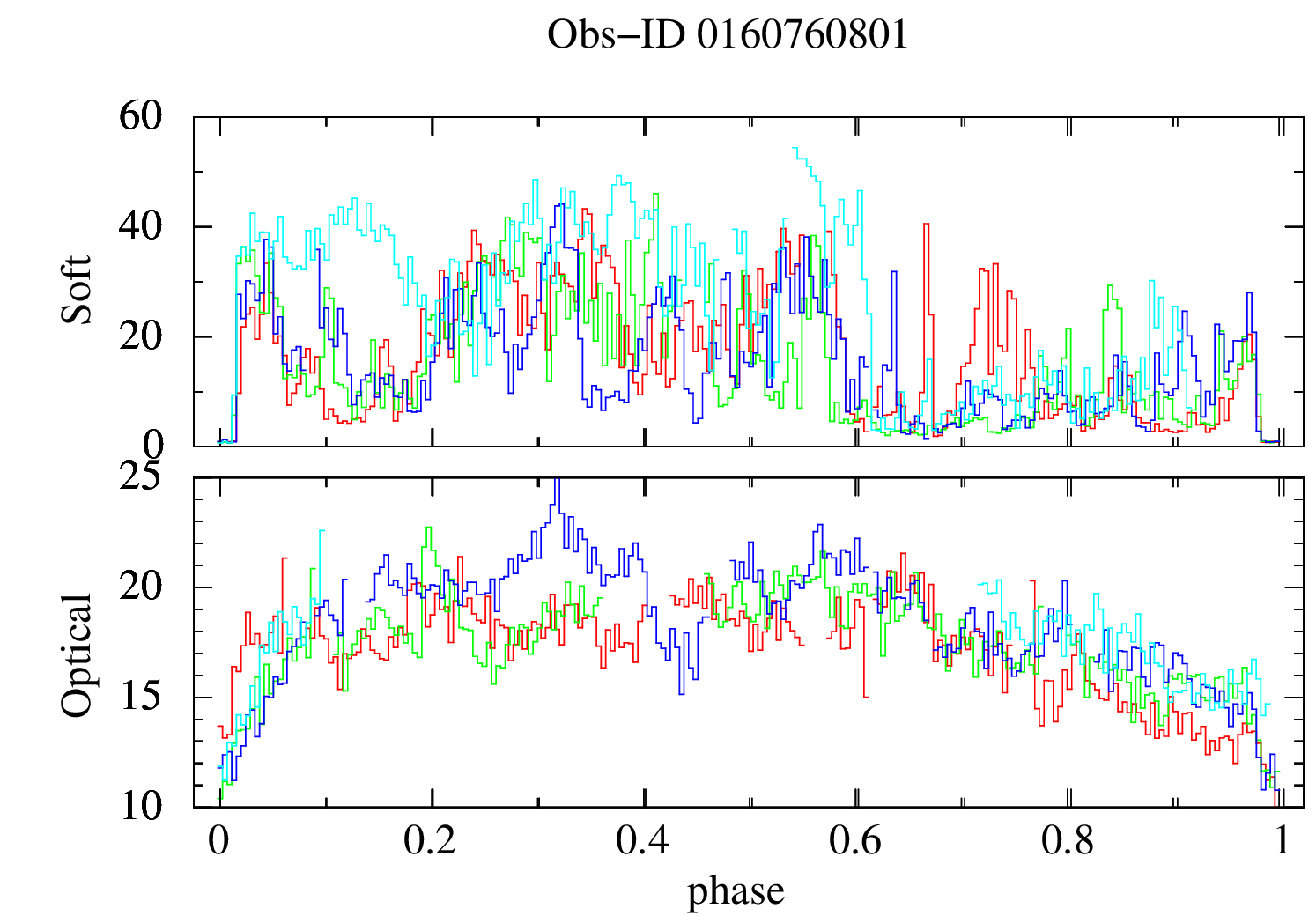}
                  
           \end{minipage}      
 \begin{minipage}{1.1\linewidth}    
\centering
    \includegraphics[scale=0.23,angle=-90]{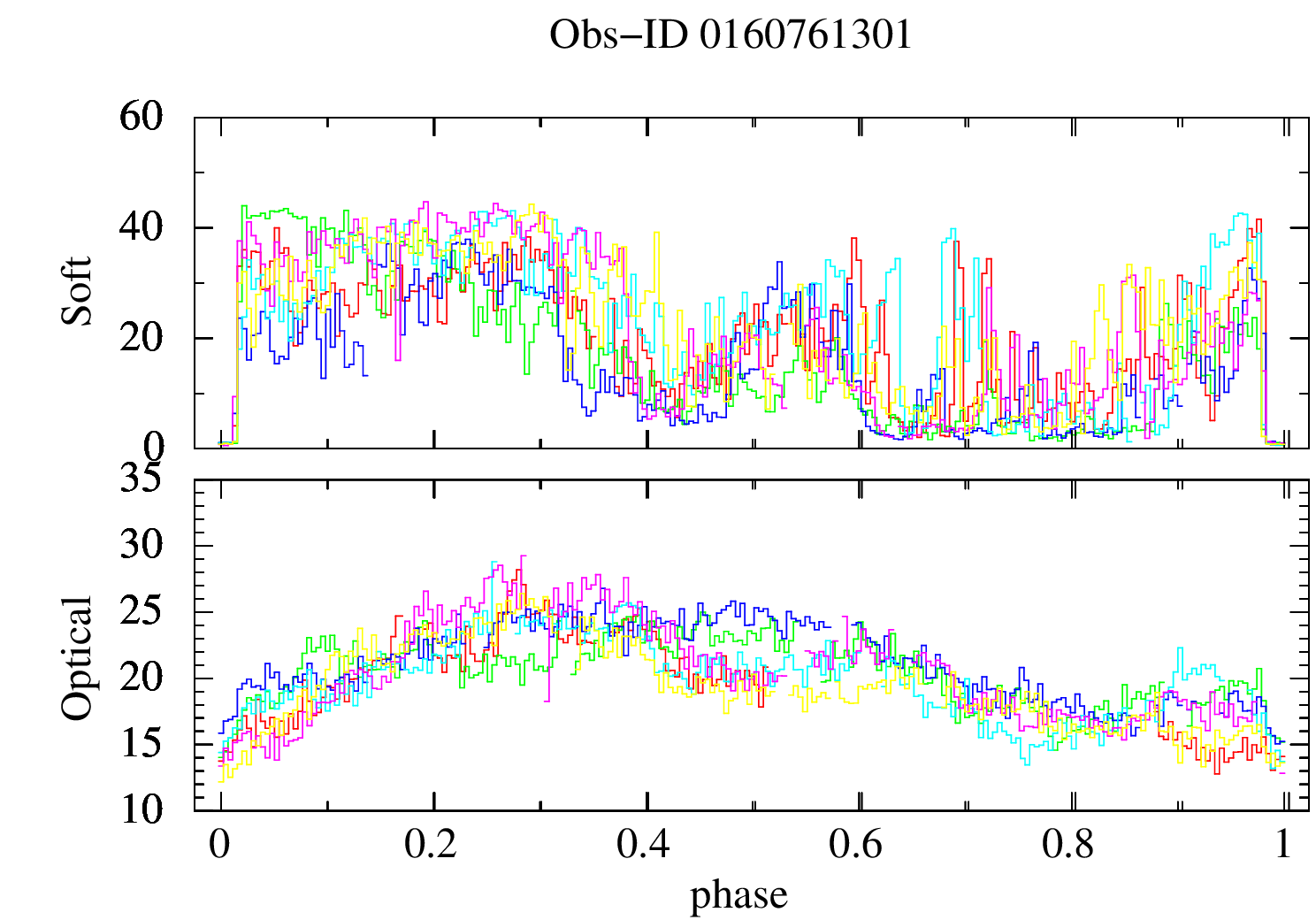}
    
      \end{minipage}
    \caption{ Simultanesou soft X-ray and optical orbit to orbit intensity modulations are shown for all the \obsids.
    Each binary orbit is represented by a different color. The orbital light curve intensity is seen to vary enormously between successive 
    binary orbits both in the soft X-rays and optical.  }
\end{figure*}

 \begin{figure*}
\begin{minipage}{1.0\linewidth}
\centering
 \includegraphics[scale=0.23,angle=-90]{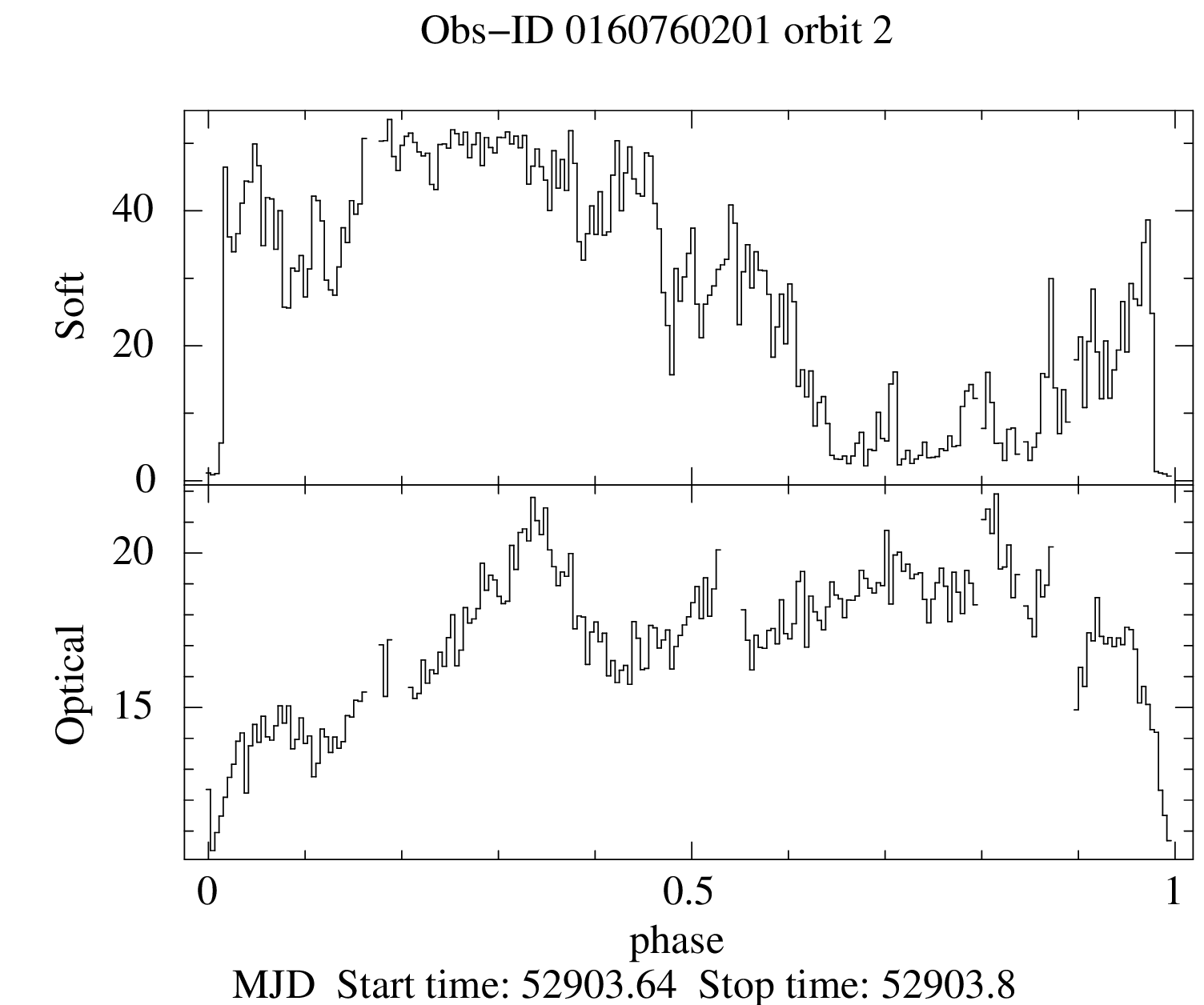}
  \includegraphics[scale=0.23,angle=-90]{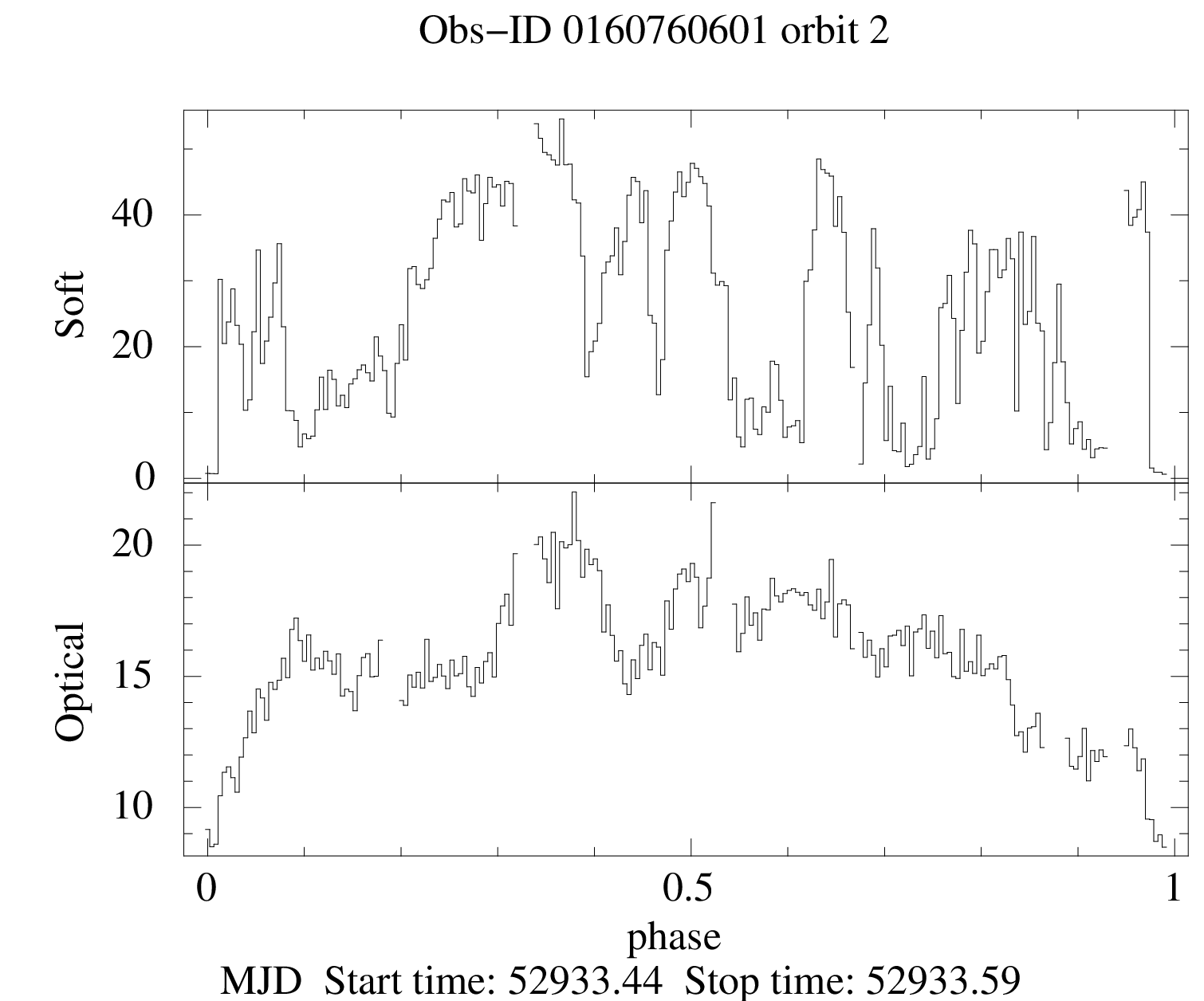}
  
  \end{minipage}  
 \begin{minipage}{1.1\linewidth} 
 \centering
 \includegraphics[scale=0.23,angle=-90]{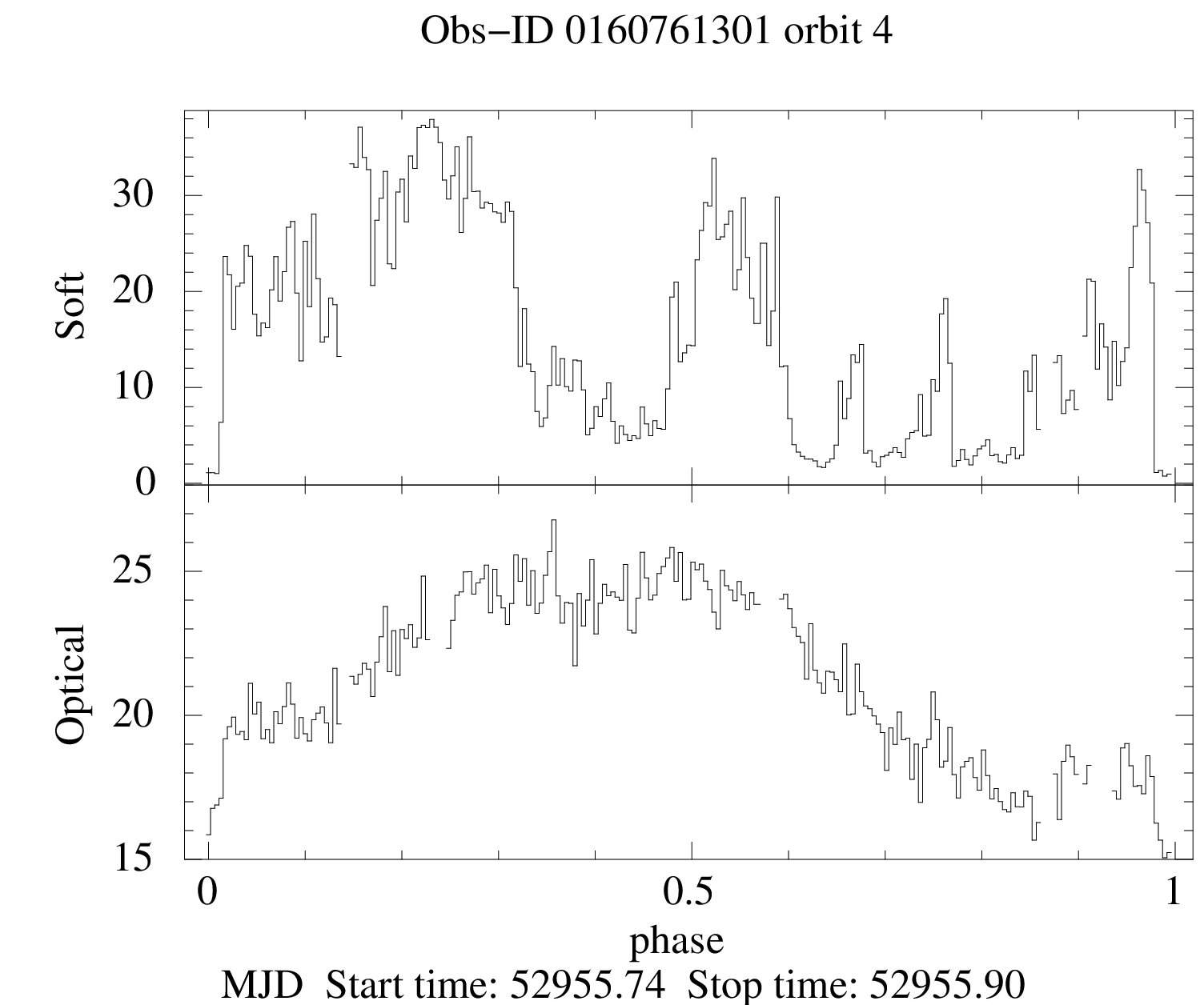}
   
  \end{minipage}  
 \begin{minipage}{1.0\linewidth} 
 \centering
    \includegraphics[scale=0.23,angle=-90]{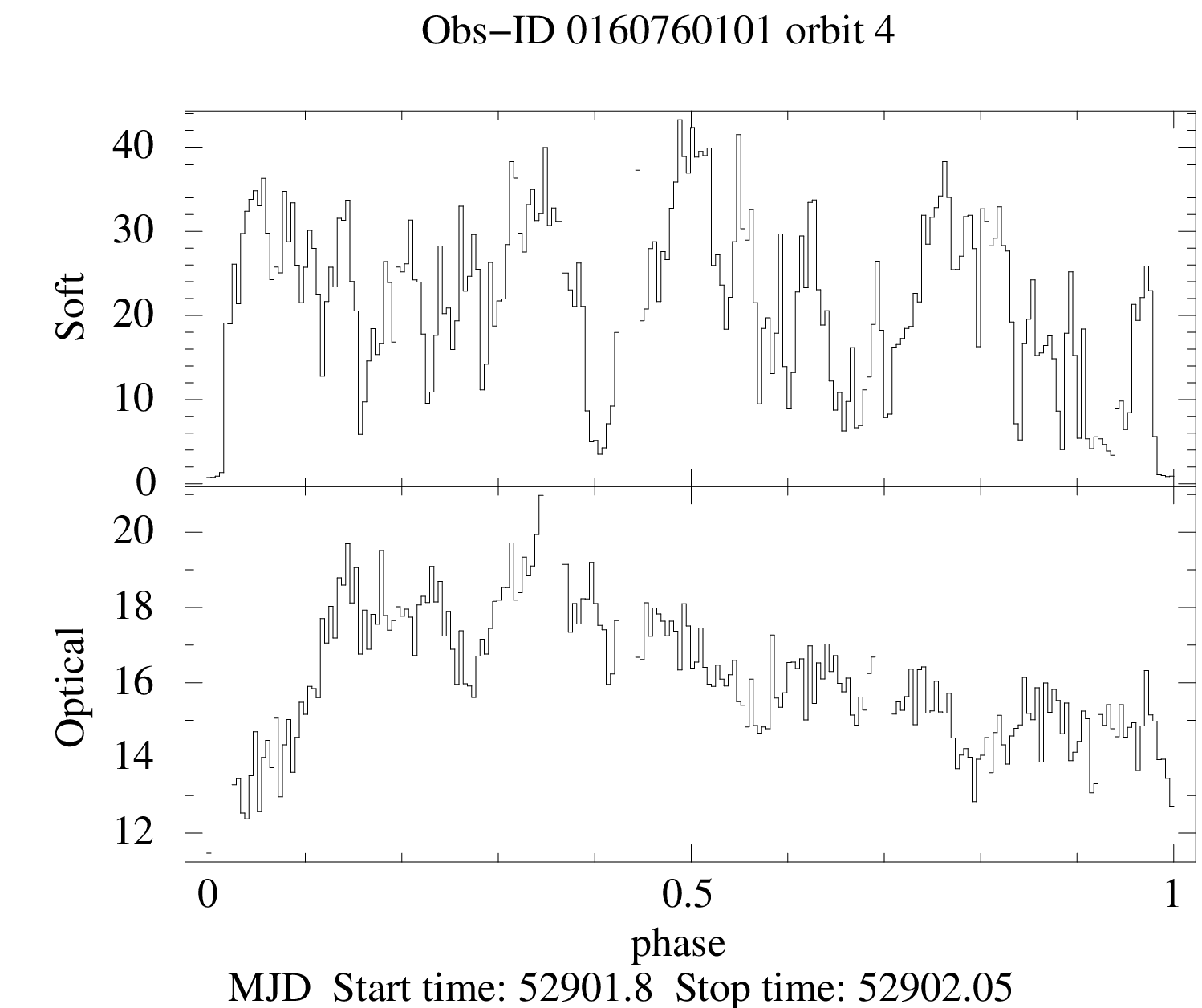}
   \includegraphics[scale=0.23,angle=-90]{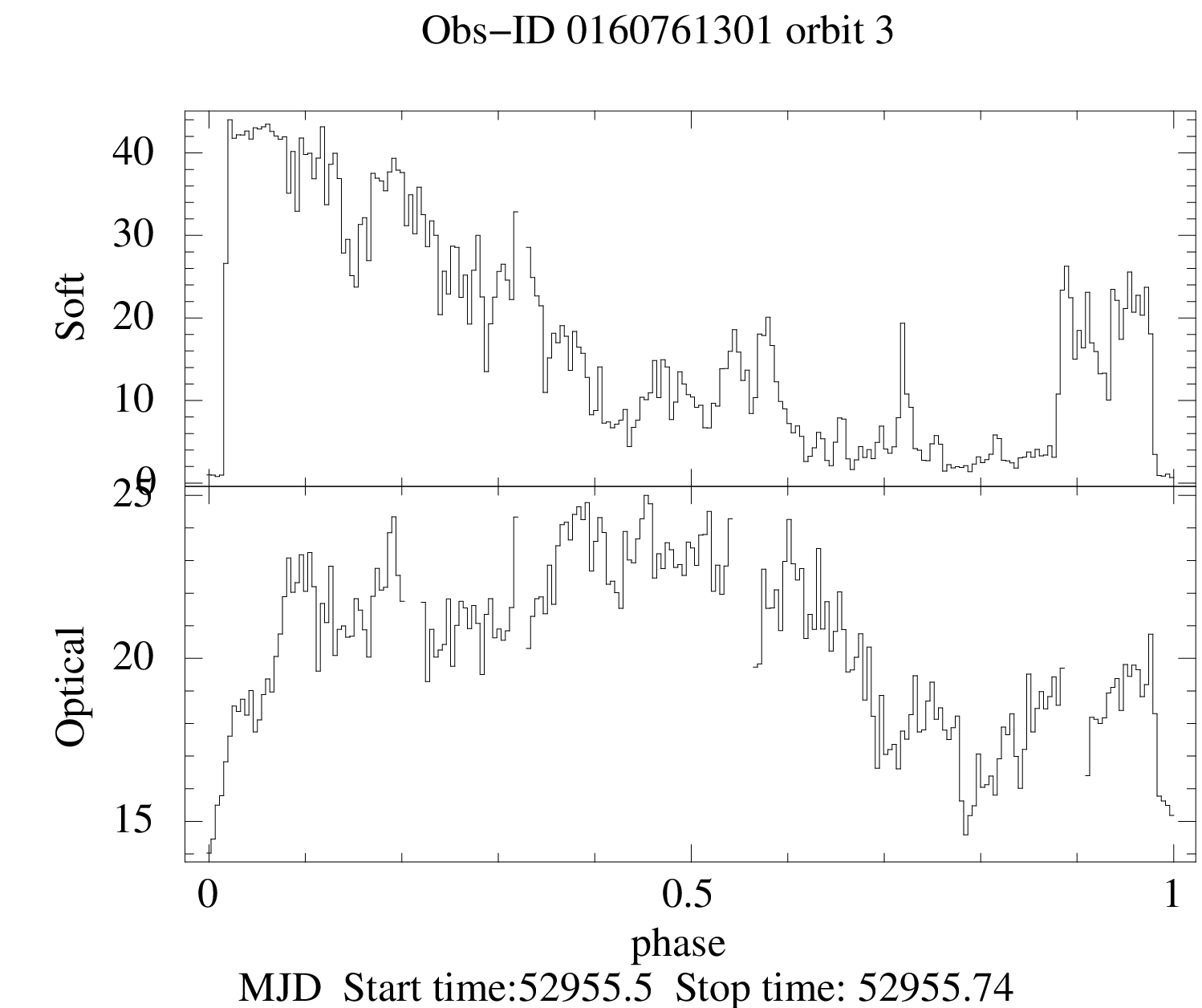}
   
  \end{minipage}  
  \caption{Soft X-ray and corresponding optical variations are classified into five broad classes of orbital profiles.
 1)  A broad pre-eclipse X-ray absorption between phases 0.6-0.9 (top-left panel).
 2) A broad absorption feature around phase 0.2 (top-right panel).
 3) A broad absorption feature around phase 0.4 (middle panel).
 4) Multiple narrow less intense dips at different orbital phases (bottom-left panel). 
 5) Intense absorption during significant portion of the orbit (bottom-right panel). }
\end{figure*}

 \subsection{Cross-correlation analysis}

We define three time-scales of variability to assess the correlation between soft, hard X-rays and optical emission. 
Reprocessing happens at extremely short time-scales, which are insignificant compared to the orbital time-scales. 
Probing such short time-scales is out of the scope of this work. The size of the accretion disk and the binary separation in 
EXO 0748-676 is about 3 light seconds \citep{Hynes2006}.
However, other than the thermonuclear bursts, which have been removed from the current analysis, there is not much short 
time-scale (of the order of a few seconds) X-ray variability. Hence, we do not aim to investigate reprocessing at 
these time scales too.\\
The third is the long term variability at time-scales longer than the above two. To probe this, each orbital segment from the 
soft and hard X-ray light curve, measured from eclipse to eclipse, was cross-correlated with the simultaneous optical orbital 
light curve segment with a bin time of 60 s. The soft and hard X-rays were also cross-correlated similarly. Data gaps have 
been ignored. The thermonuclear X-ray bursts, reprocessed optical bursts and the eclipses were also not included in the cross 
correlation analysis.\\
The time delay between the various light curve segments in the three bands were computed using the XRONOS tool 
\textit{crosscor} which uses the FFT algorithm. The cross correlation profiles have been classified broadly into 
five classes solely on the basis of their delays and profile shapes and do not correspond to the variable dipping 
classification of the soft X-ray light curves, indicated in the previous section. The cross-correlations have 
been computed for sample light curves, one, from each of the five classes. The CCF profiles (Figure 5, left) are plotted 
along with the corresponding orbital light curve segment for the hard X-ray, soft X-ray and optical (Figure 5, right).
 We find that there is little or no correlation between the optical and X-rays at short timescales, and the 
repeating patterns of the CCF are independently related to the orbital period (a few 1000s of seconds), being evident from the CCF peaks.

\begin{figure*}
   
   \begin{minipage}{1.0\linewidth}  
   \centering
 \includegraphics[scale=0.23,angle=-90]{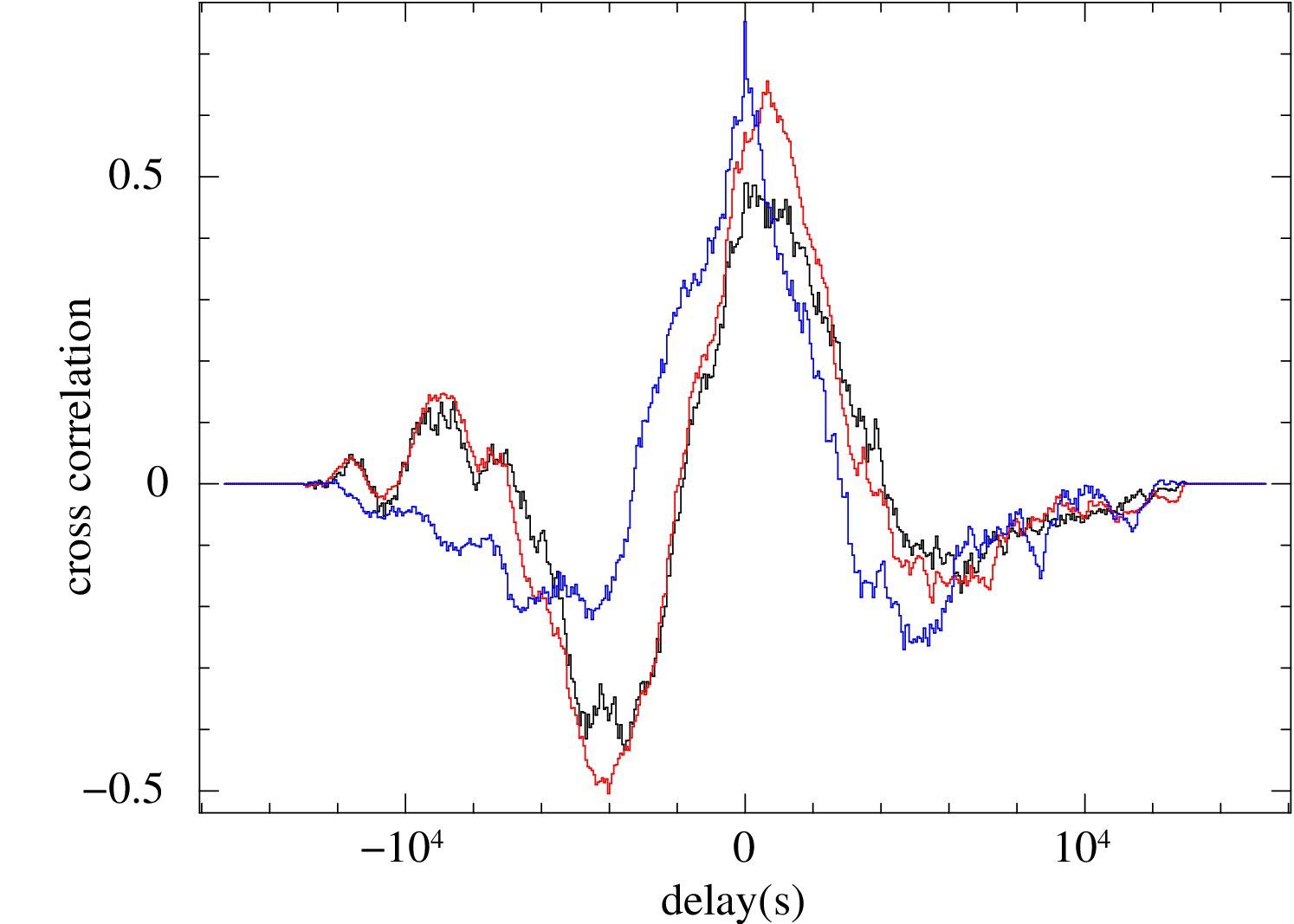}
 \includegraphics[scale=0.22,angle=-90]{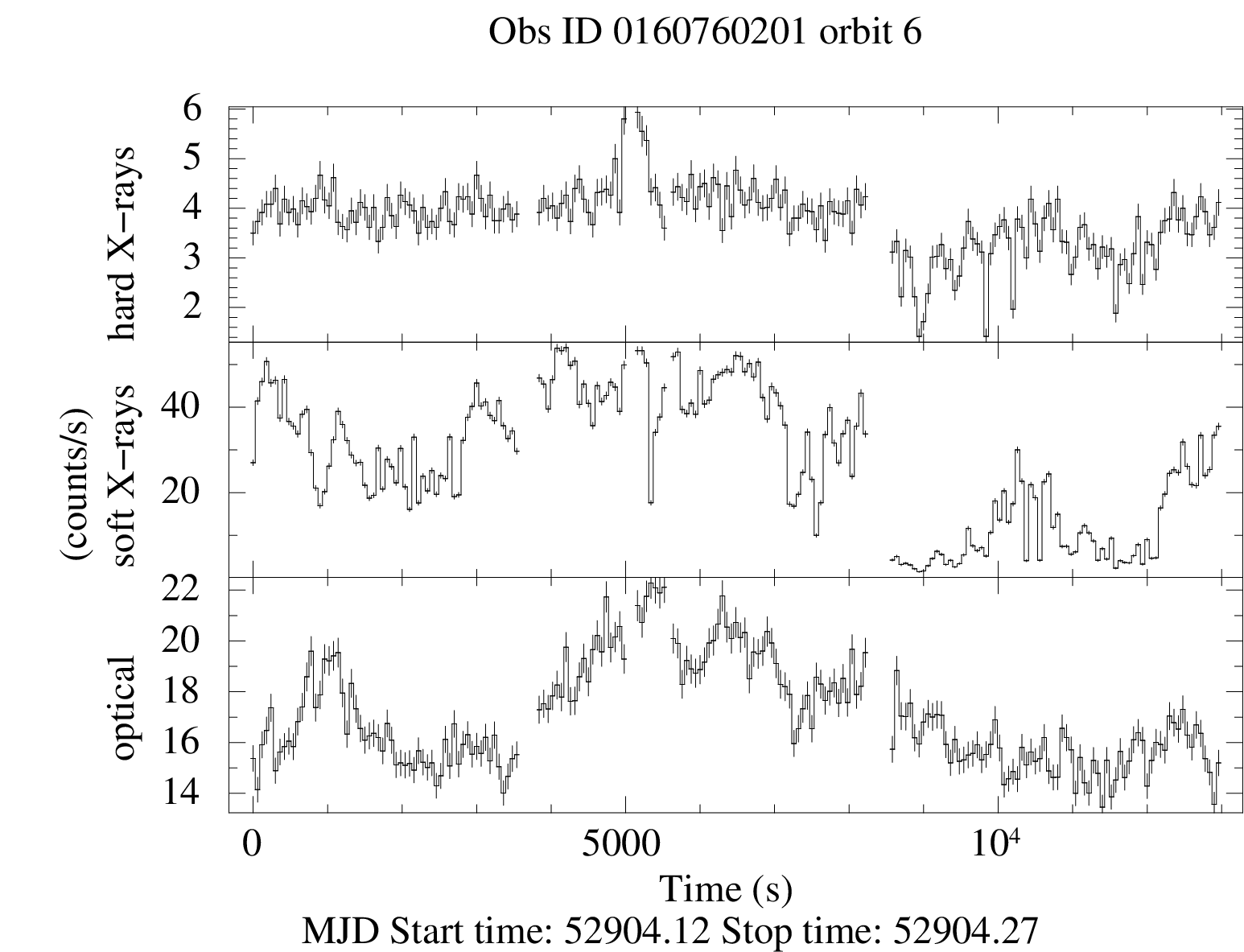}
  \end{minipage}     
   \hspace{0.9cm}    
\begin{minipage}{1.0\linewidth}   
  \centering
 \includegraphics[scale=0.23,angle=-90]{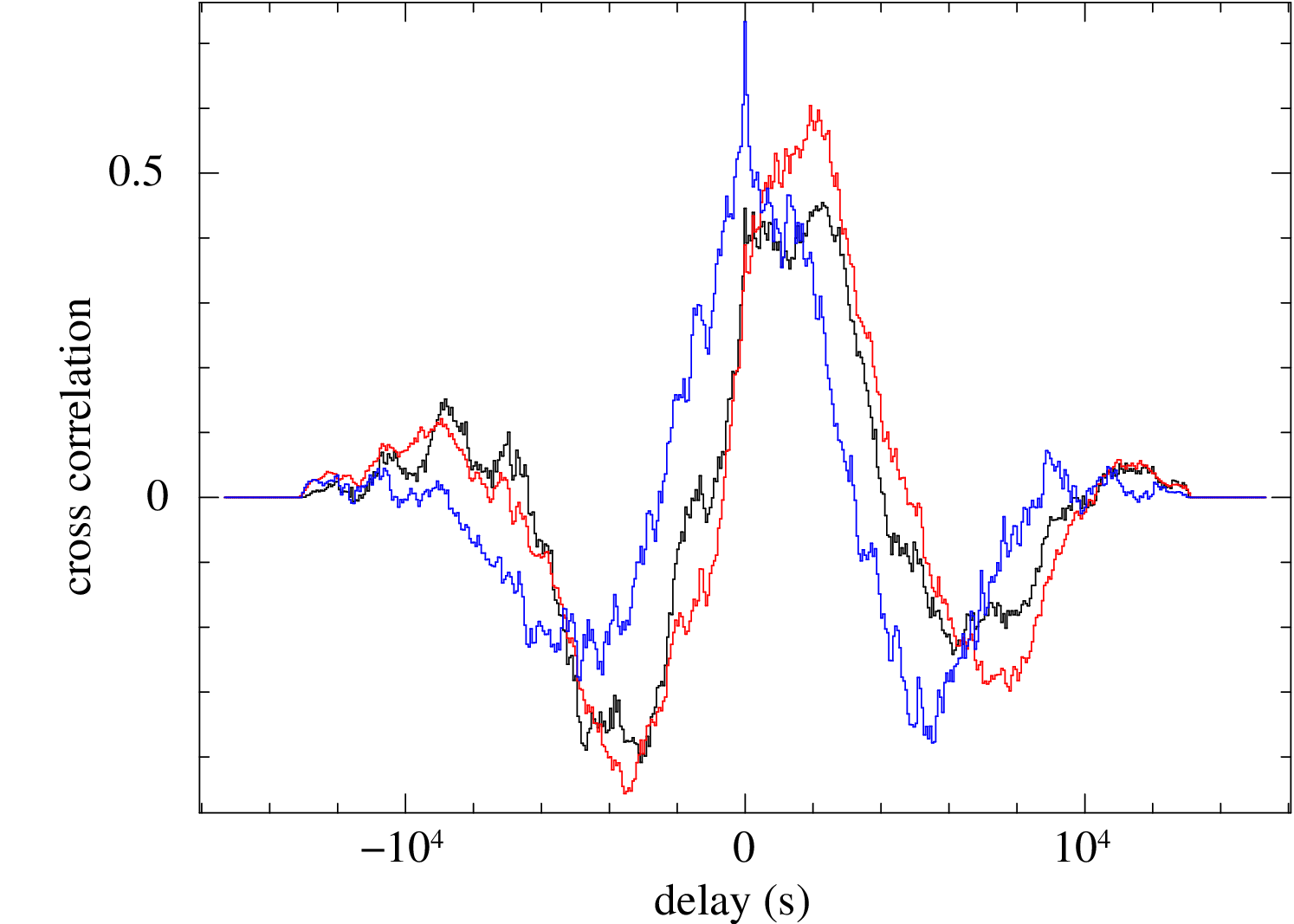}
 \includegraphics[scale=0.22,angle=-90]{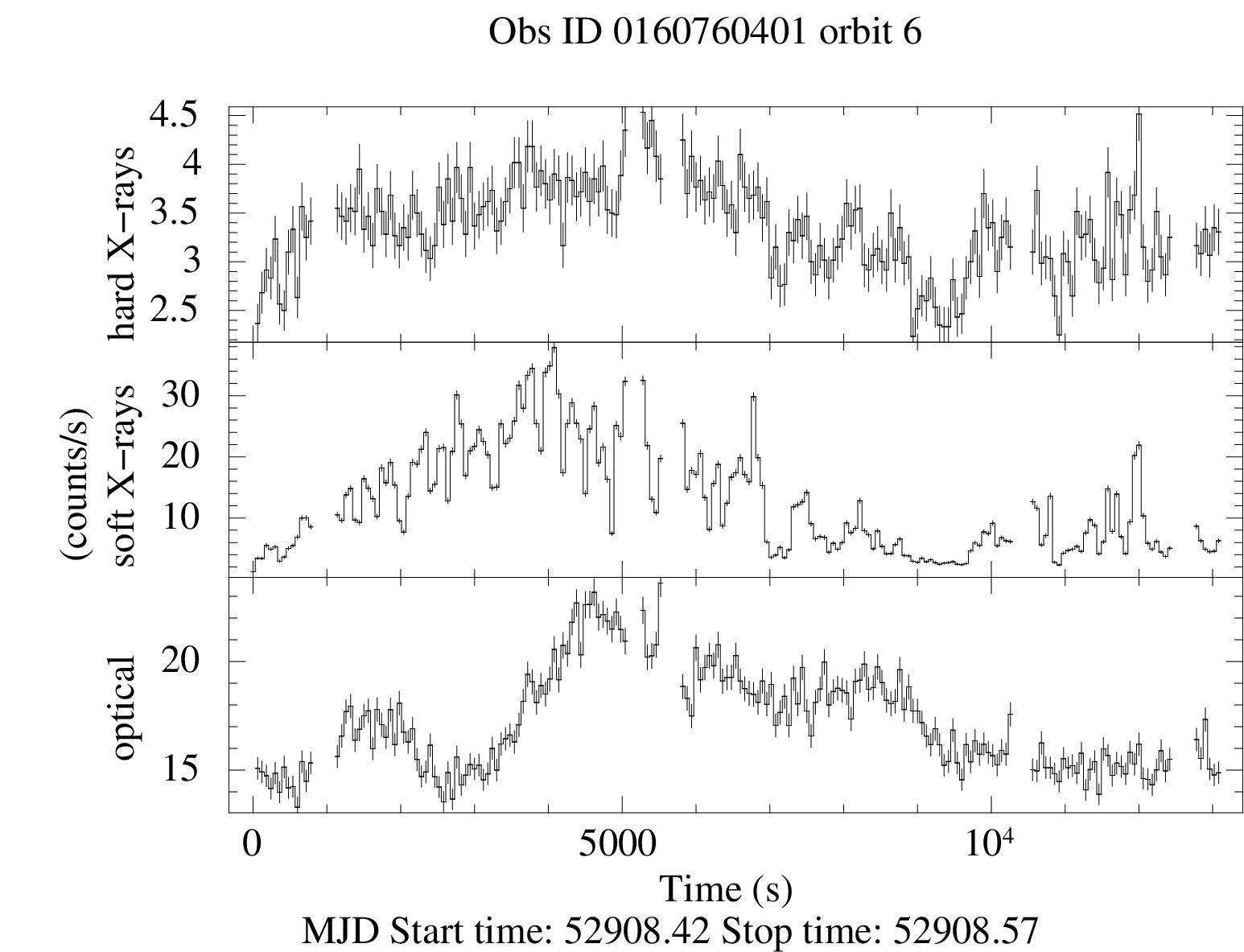}
  \end{minipage}     
   \hspace{0.9cm}    
\begin{minipage}{1.0\linewidth}   
  \centering
 \includegraphics[scale=0.23,angle=-90]{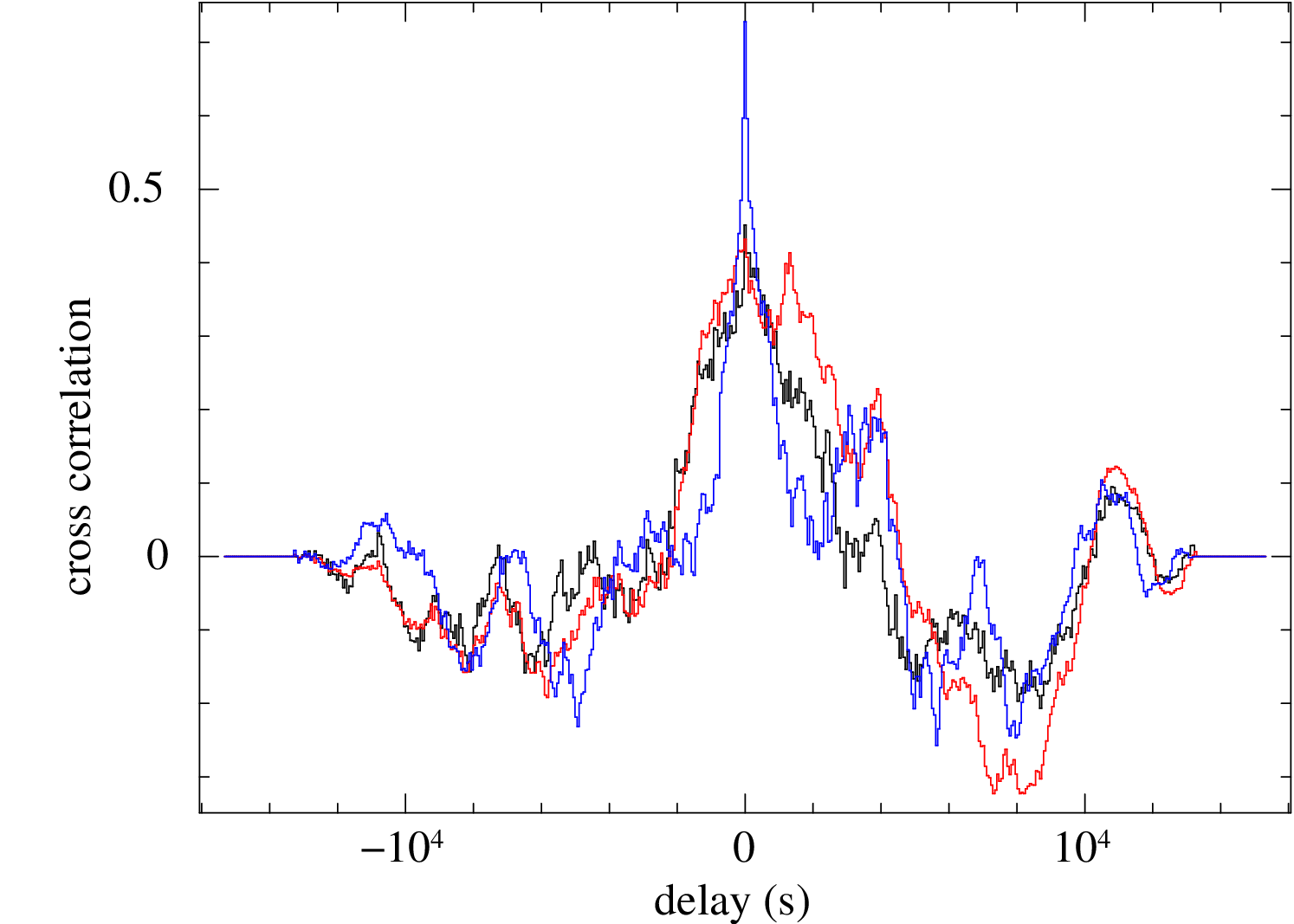}
 \includegraphics[scale=0.22,angle=-90]{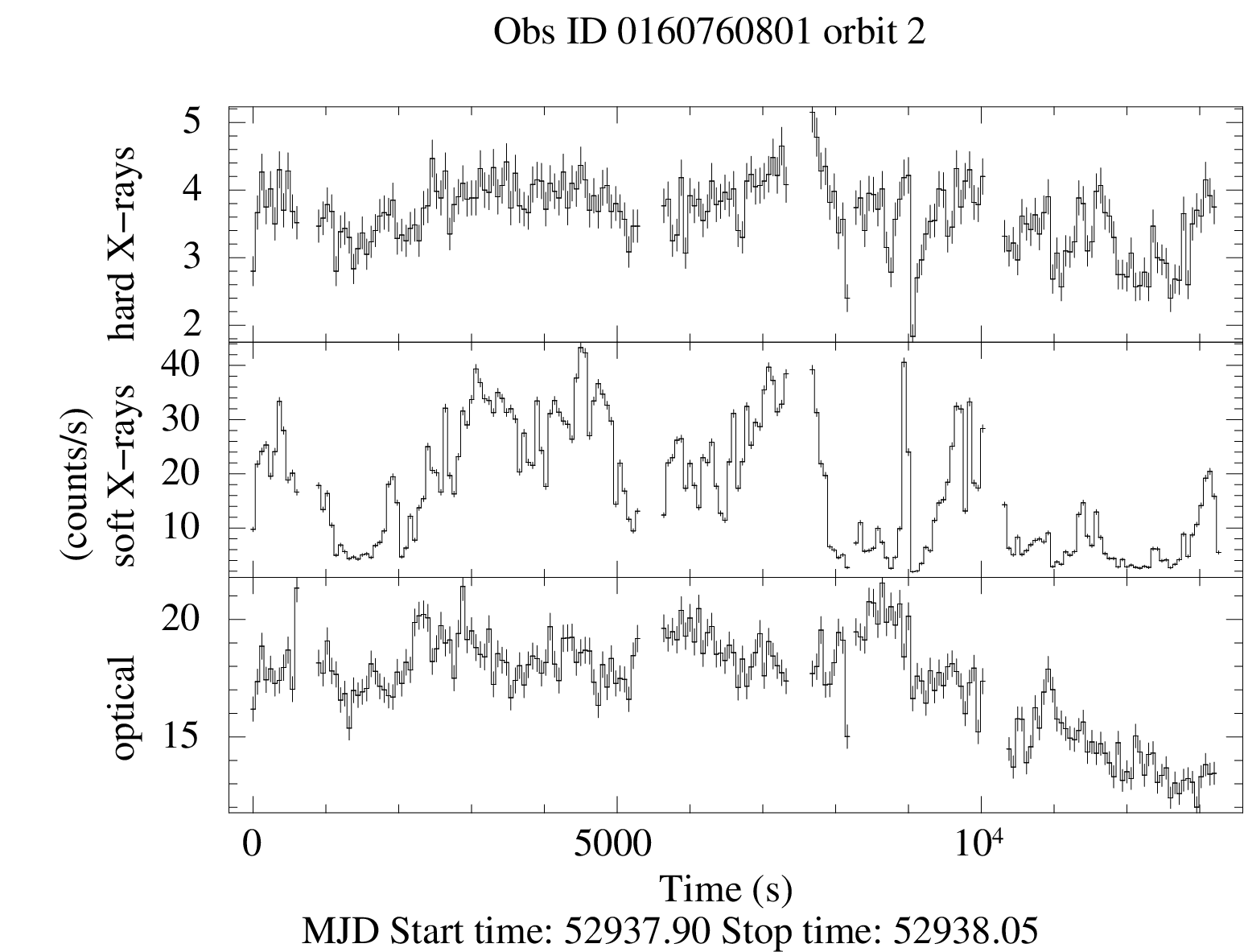}
  \end{minipage}     
   \hspace{0.9cm}    
\begin{minipage}{1.0\linewidth}   
 \centering
 \includegraphics[scale=0.23,angle=-90]{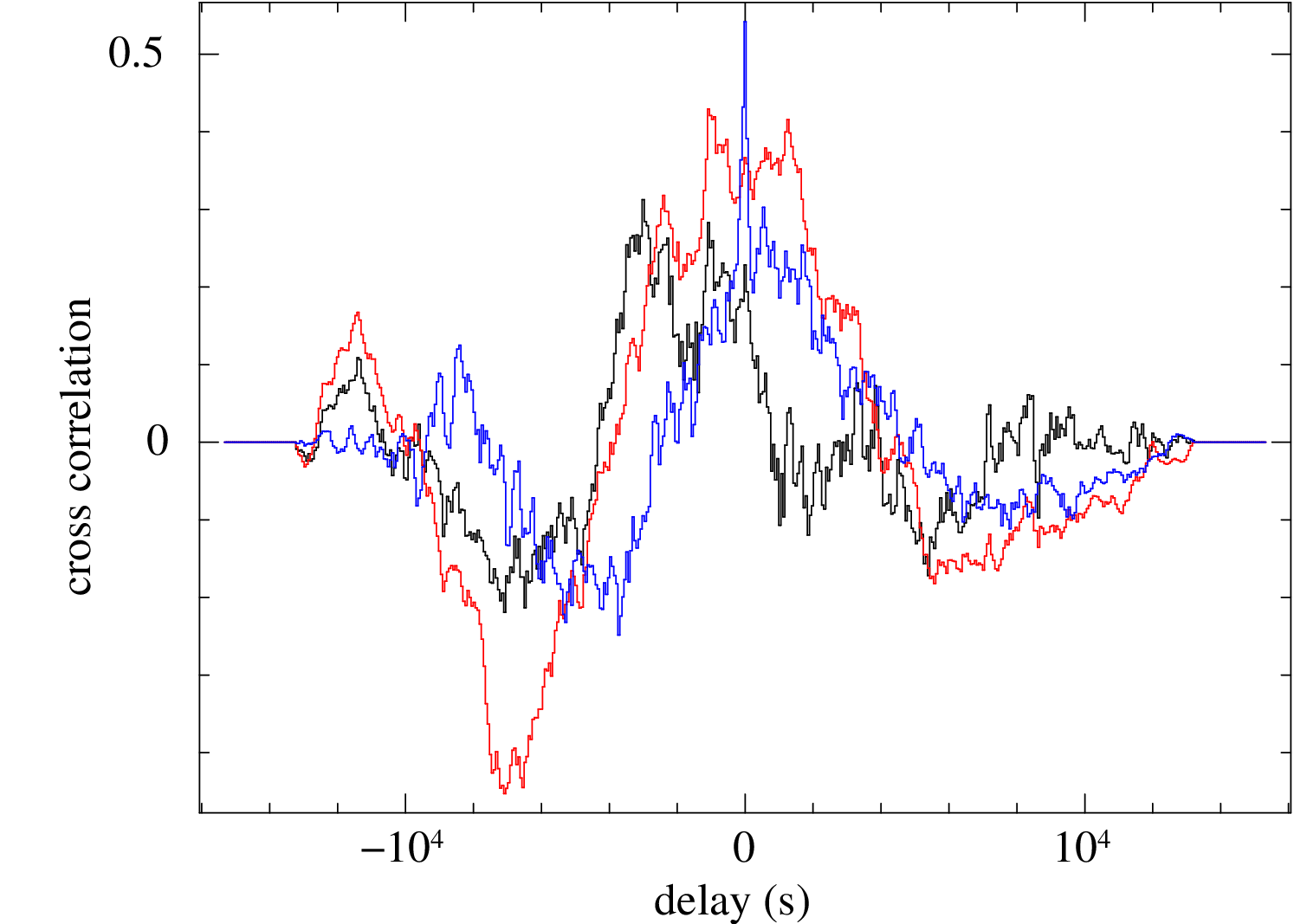}
 \includegraphics[scale=0.22,angle=-90]{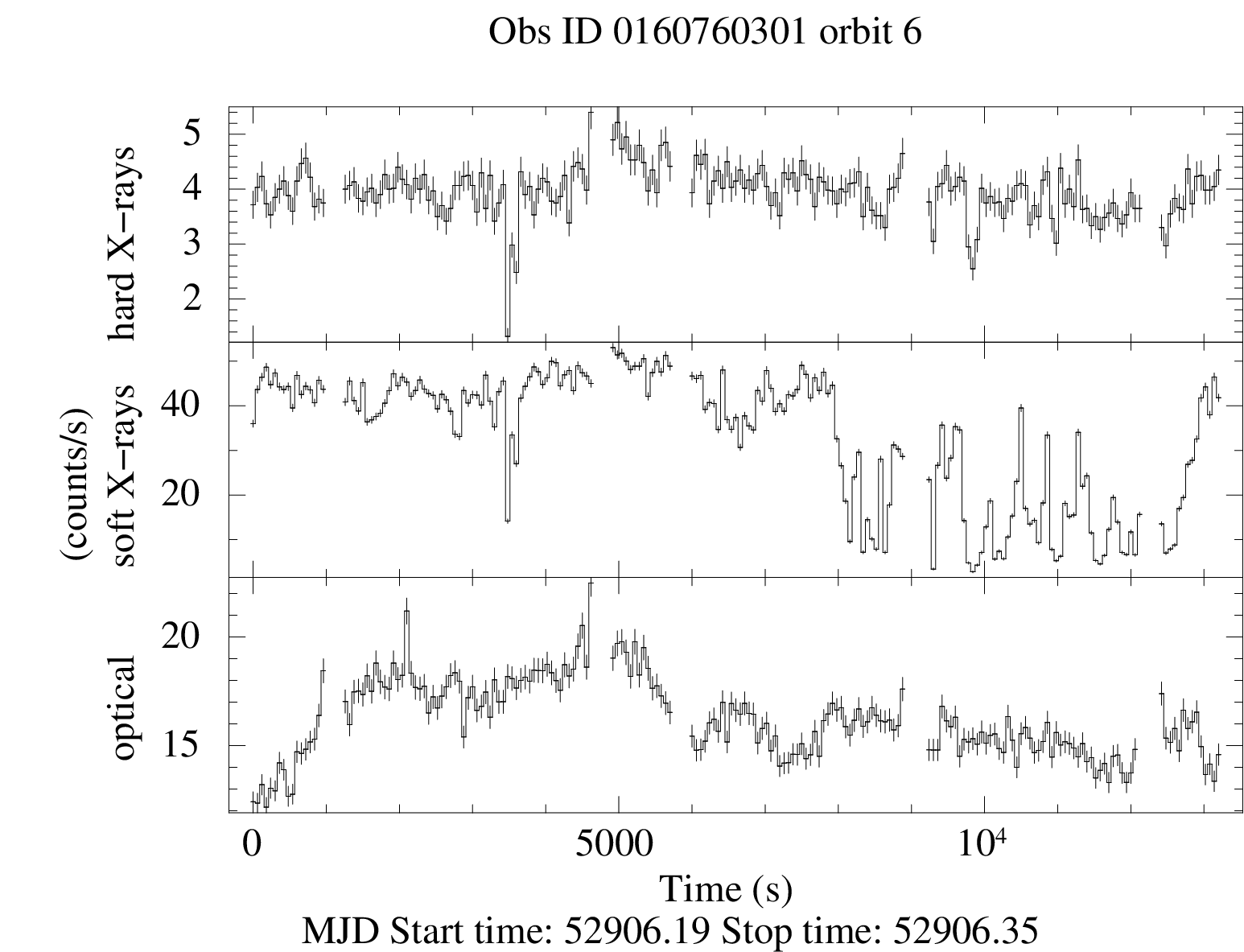}
  \end{minipage}     
   \hspace{0.9cm}    
\begin{minipage}{1.0\linewidth}   
 \centering
 \includegraphics[scale=0.23,angle=-90]{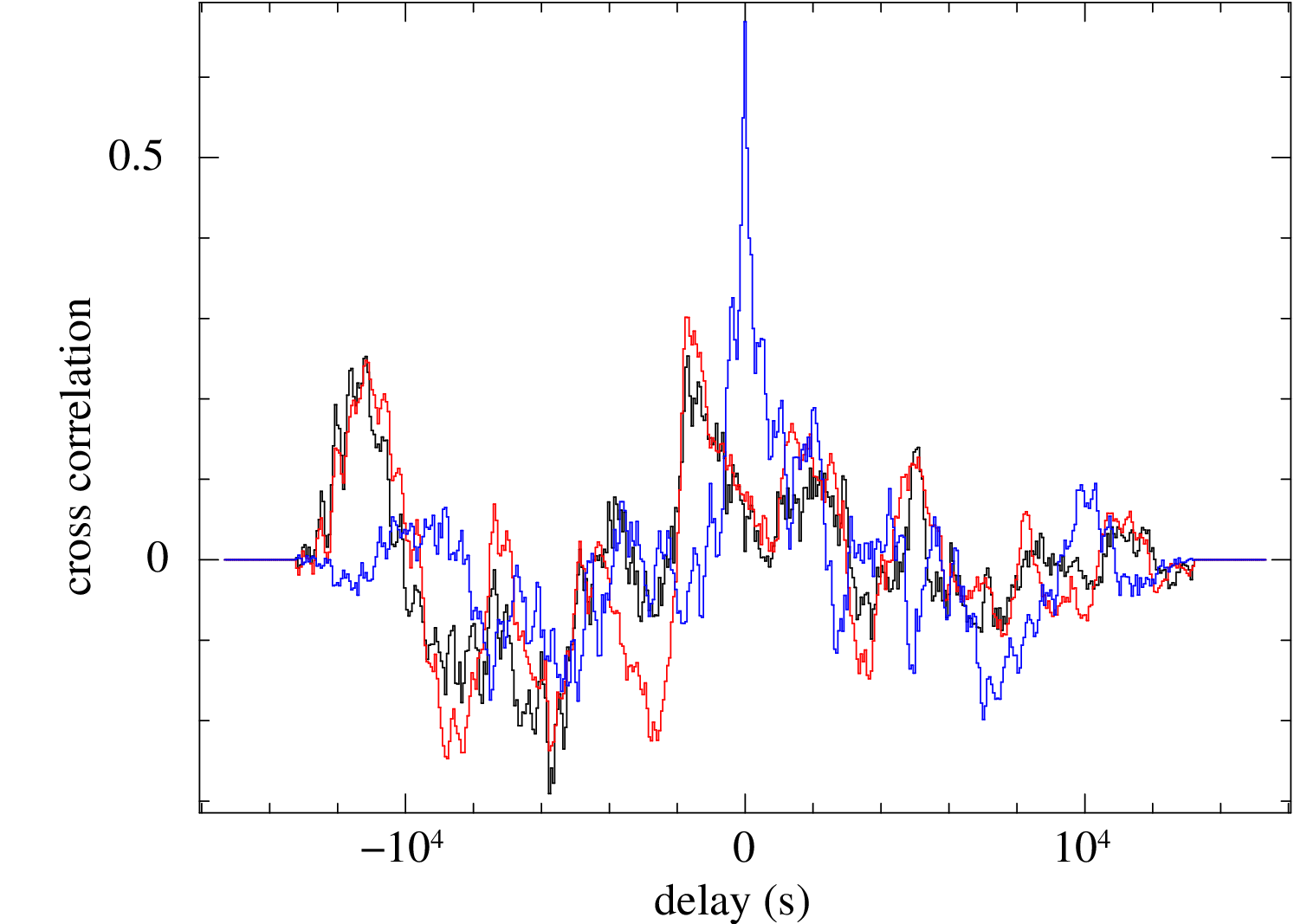}
 \includegraphics[scale=0.22,angle=-90]{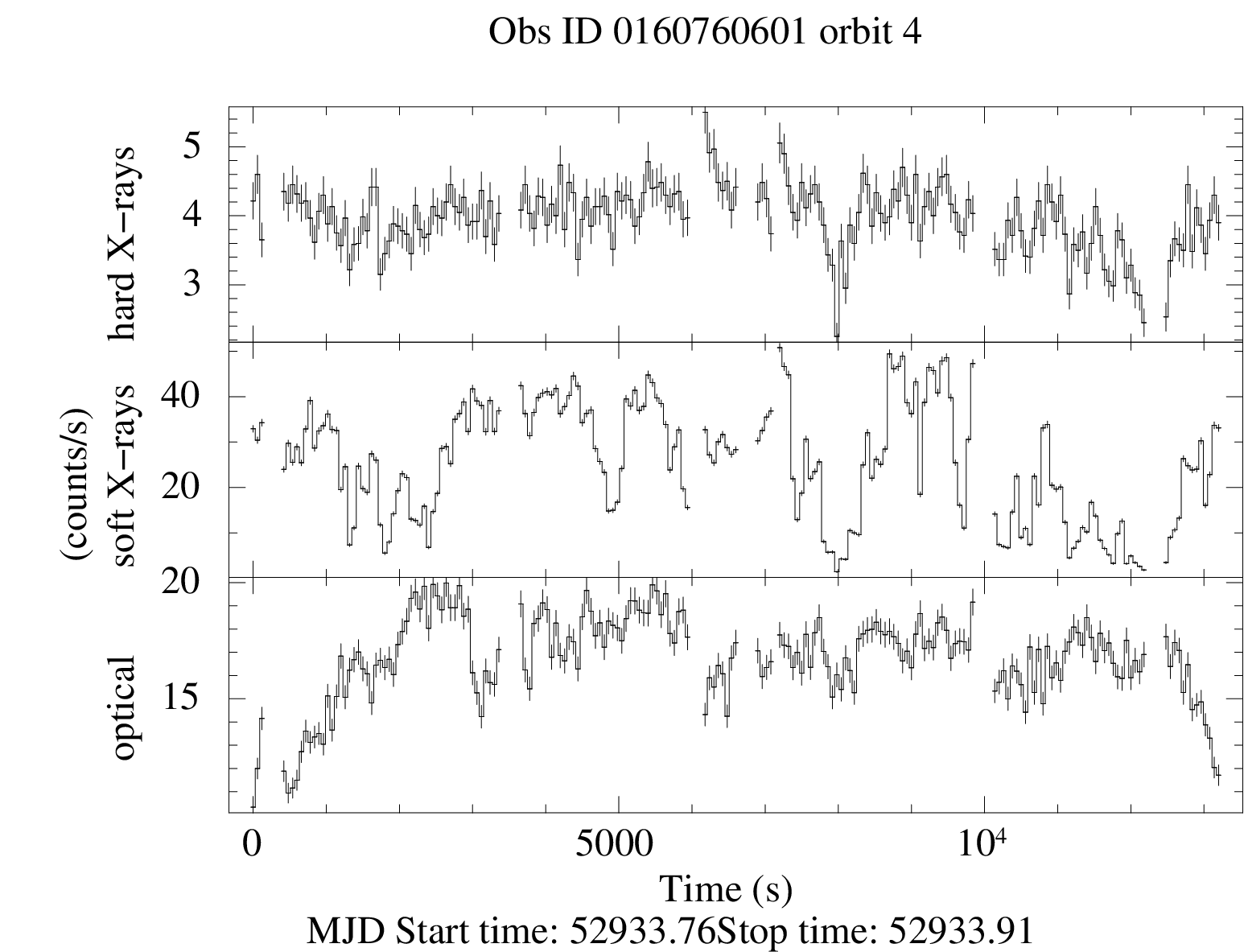}
 
\caption{Five different classes of cross correlation analysis patterns between the soft X-ray and optical (red),
the hard X-ray and optical (black) and the soft and hard X-ray (blue) have been identified and are shown in the left segment
along with the corresponding light curve segments on the right.}
\end{minipage}
 \end{figure*}

%
%
%

     \begin{figure*}[h]
    \centering
    \includegraphics[scale=0.35,angle=-90]{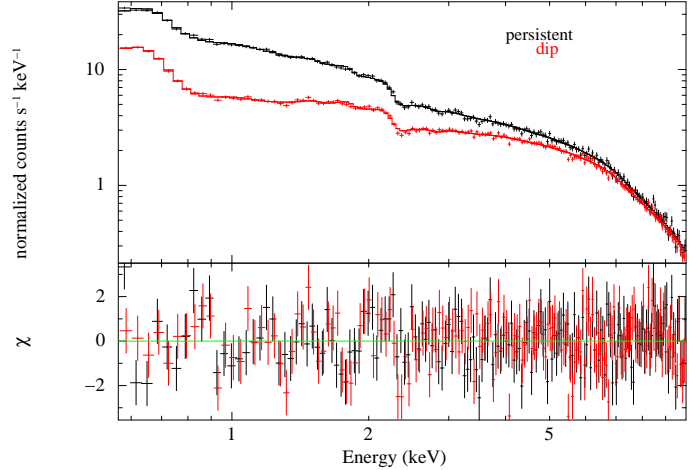}
    \caption{Simultaneous persistent and dipping spectra in the 0.5-10 keV band are shown for the second orbital cycle of \obsid 0160761301.}
    \end{figure*}

 \subsection{Spectral analysis}

      We have also carried out a representative study of intensity resolved spectra of the second orbital cycle from the \obsid 0160761301 
      (Figure 6). We refer to \citet{Diaz} for a more detailed intensity resolved spectral analysis.
      We extracted the EPIC-PN spectra in the 0.5 - 10 keV energy band from a region of 40'' centered on 
      the source. Segments from the light curve with a count rate less than 20 counts/sec in the 0.5 - 10 keV band were identified as 
      dipping and the remaining nearly-steady segment was considered as persistent emission. The responses were generated using 
      the \software{SAS} task \textit{arfgen}. Spectral fitting was performed using \textit{XSPEC} version 12.8.1. \\
      We simultaneously fit the persistent and dip spectra with a power-law model along with neutral hydrogen absorber and 
      a partial covering ionized absorber (\textit{zxipcf} in XSPEC). To avoid the nH parameter taking unreasonably low values, we 
      fixed it to 0.11 x 10$^{22}$ cm$^{-2}$, as similarly considered by \citet{Diaz}. Strong residuals at 0.569 keV and 0.915 keV were
      observed, which we then fit using two gaussians. The \textit{phabs*zxipcf*po + phabs*(gaus+gaus)} model in \textit{XSPEC} gave a 
      reduced $\chi^2$ of 1.22 with 319 dof. Table 3 details the best fit parameters for the spectra. We observe an increase in the 
      covering fraction of the ionized absorber and a decrease in the ionization parameter as we move from persistent emission to 
      dipping, as was also reported by \citet{Diaz}.

 \begin{table*}
\centering
\caption{Best fit spectral parameters for the second orbital cycle of the \obsid 0160761301. }
\resizebox{0.6\linewidth}{!}
{\begin{tabular}{ccccc}
 \hline
 Parameter & model & Persistent & & Dip \\
 \hline
 \hline
 &&&\\
nH (x10$^{22}$ cm$^{-2}$) & \textit{phabs} & &0.11 (frozen)  &  \\
nH (x10$^{22}$ cm$^{-2}$) & \textit{zxipcf} & 4.33$_{-0.97}^{+0.43}$ &  & 4.53$_{-0.57}^{+0.45}$ \\
log($\xi$) & \textit{zxipcf} & 1.39$\pm$0.2  & &  1.09$_{-0.29}^{+0.14}$ \\
cov. f & \textit{zxipcf} & 0.55$_{-0.07}^{+0.03}$ & & 0.78$\pm$0.02 \\
$\alpha$ & \textit{power law} & 1.52$_{-0.09}^{+0.04}$ & & 1.38$\pm$0.05  \\
E (keV) & \textit{gaus} & &0.569 (frozen) &  \\
EW (eV) & \textit{gaus} & 437.2$\pm$0.001 & &666.9$\pm$0.0002 \\
E (keV) & \textit{gaus} & &0.915 (frozen) &  \\
EW (eV) & \textit{gaus} & 70.5$\pm$0.0005 & &49.6$\pm$0.0002 \\
\hline
Reduced $\chi^{2}$  &  & &1.22 for 319 d.o.f  & \\
\hline
 \end{tabular}}
 \end{table*}
 
 \section{Discussion}
 
 \subsection{Summary of results}
  We have studied simultaneous X-ray and optical light curves of EXO 0748-676 from the EPIC-PN detector and the OM from a set of 
  \textit{XMM-Newton} observations carried out during a hard spectral state of the source. A total of 36 simultaneous and complete, 
  X-ray and optical orbital light curves have been studied. The X-ray and optical dips occur at many different orbital phases with 
  varying depths and widths throughout all the 7 observations. The soft X-ray and the optical emission are seen to vary 
  enormously at sub-orbital time scale and also between successive binary orbital cycles. The non-burst, non-eclipsing X-rays 
  (soft and hard) and optical emission do not show any co-variability at the reprocessing time-scales, but show a weak 
  correlation at time-scales of a few 1000s of seconds. EXO 0748-676 has been extensively studied in both X-rays and optical in 
  the past. The nature of its X-ray and optical co-variability has been known to change over the years. Our results provide further 
  indications to the evolving nature of the accretion disk in this source.\\

  \begin{figure*}[h]
  \centering
  \includegraphics[scale=0.45, angle=-90]{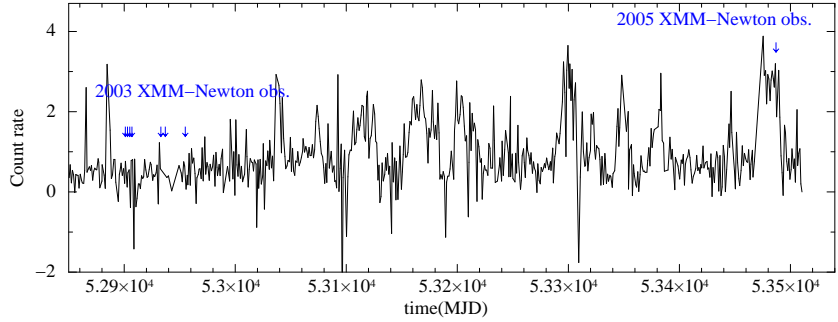}
 \caption{The \textit{RXTE}-ASM light curve of EXO 0748-676 showing the high-intensity 2005 soft state and low-intensity 2003 
 (our current dataset) hard state, marked with arrows. These very different X-ray states also show up in the spectral properties 
 of the source.}
 \end{figure*}
  
\subsection{Variability In X-rays (soft and hard)}

  During the observations reported here, EXO 0748-676 had an average unabsorbed X-ray flux of 2.25x10$^{-10}$ erg cm$^{-2}$s$^{-1}$
  \citep{Boirin}, which is nearly 1$\%$ L$_{Edd}$. This persistent flux, representative of the mass accretion rate of the neutron star,
 has been one of the lowest values ever measured for this system before it went into quiescence \citep{Hynes_2009}.
  The system was in an X-ray ``faint-hard'' state throughout all the 7 observations reported here. However, the soft X-ray light 
  curves, excluding the thermonuclear bursts and residual eclipse emission, have a large rms in their variability unlike the hard 
  X-ray light curve variability, which is lower, indicating an unvarying accretion rate and persistent hard X-ray emission from the 
  neutron star.\\
   
   The soft X-ray light curves, in addition to the eclipses and the bursts, showed intense dipping over a range of orbital phases. 
   The dipping behavior, in it's depth, continuity, orbital phase range, etc. changed considerably from orbit to orbit. 
   A few hard X-ray dips are coincident with strong soft X-ray dips; Also, the absorbing efficiency of the dips is more pronounced at
   lower energies. The duration of the X-ray dips allow us to roughly estimate the azimuthal angular extent of the absorbing material.
   As can be seen in Figures 2, 3 and 4, the duration of the dips have a large variation: multiple very narrow dips 
   (as short as 1\% of the orbital period) to one or two wide dips (as long as $\sim$40\% of the orbital period). The soft X-ray dips 
   therefore indicate simultaneous presence of multiple vertical structures with narrow (a few degree) to broad ($\sim$150$^\circ$) 
   azimuthal extent.\\

   EXO 0748-676 has shown a bimodal behavior of transitioning between the hard and soft state in X-rays, while having 
   low-to-moderate accretion rates (0.01-0.5 L$_{edd}$). Figure 7 shows the RXTE/ASM light curve indicating
   the epochs of two \textit{XMM-Newton} observations, one during a bright soft spectral state in 2005 and 
   another during a faint 2003 (this work) hard spectral state.  
   Excluding the thermonuclear bursts, the 2003 observation shows an average of 25 counts/s,  while the soft state 2005 
   observation has a count rate that is about an order of magnitude higher. \citet{Ponti2014} showed that the dipping phenomenon in 
   the 0.5 - 5 keV band appears less intense in soft state compared to the hard state. The dips in the soft state are shallow, being 
   almost 75\% of the average count rate. The hard state dips on the other extend much deeper, going down to $\sim$33\% of the average 
   count rate (see Figure 8).\\
   
   Interestingly, light curves from an \textit{XMM-Newton} observation in 2003, studied earlier by \citet{Diaz} and \citet{Boirin}, 
   which have also been analyzed in the present paper, show that the dips have widely different characteristics, including some 
   appearing after the eclipse (\obsids 0160760601 and 0160760801, Figures 2, 3 and 4), even when the source is in the same hard 
   spectral state. \\

\begin{figure*}
\centering
 \includegraphics[scale=0.3, angle=-90]{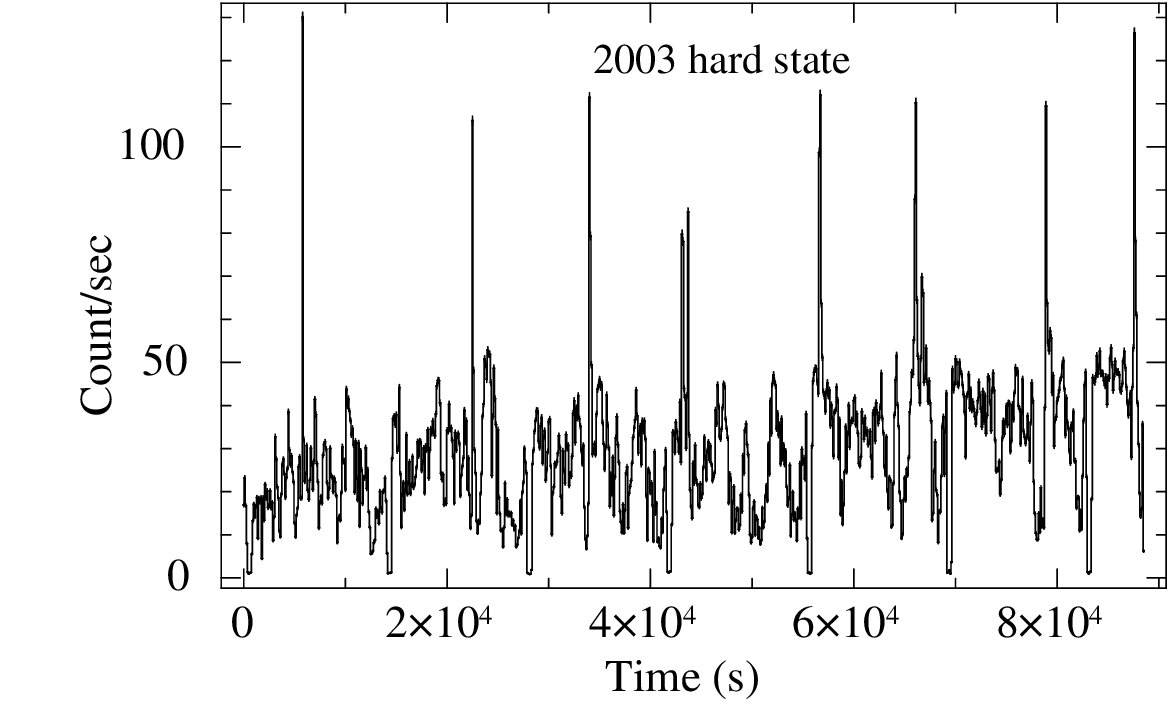}
 \includegraphics[scale=0.3, angle=-90]{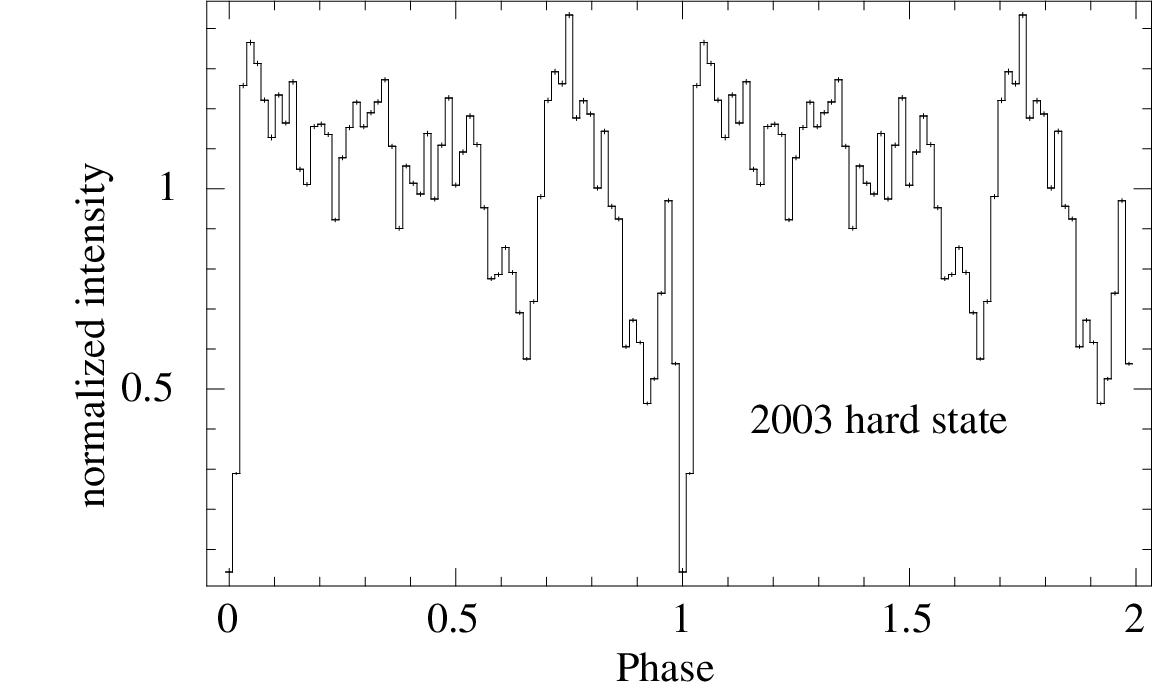}
 
 \includegraphics[scale=0.3, angle=-90]{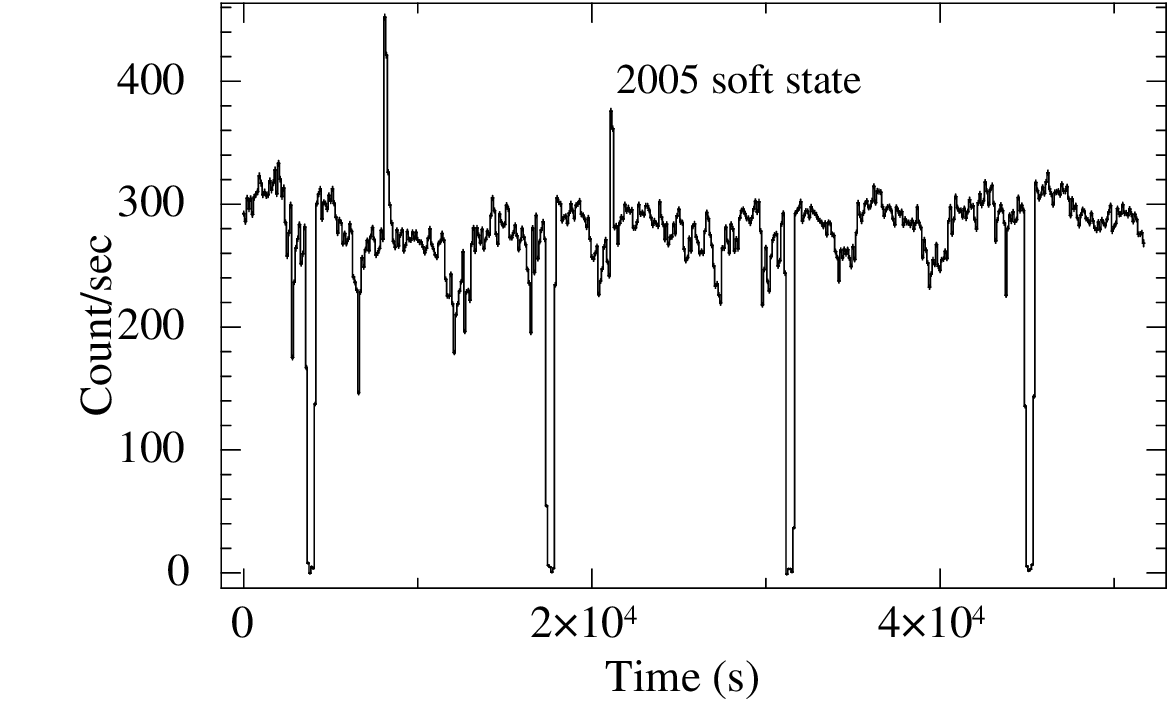}  
 \includegraphics[scale=0.3, angle=-90]{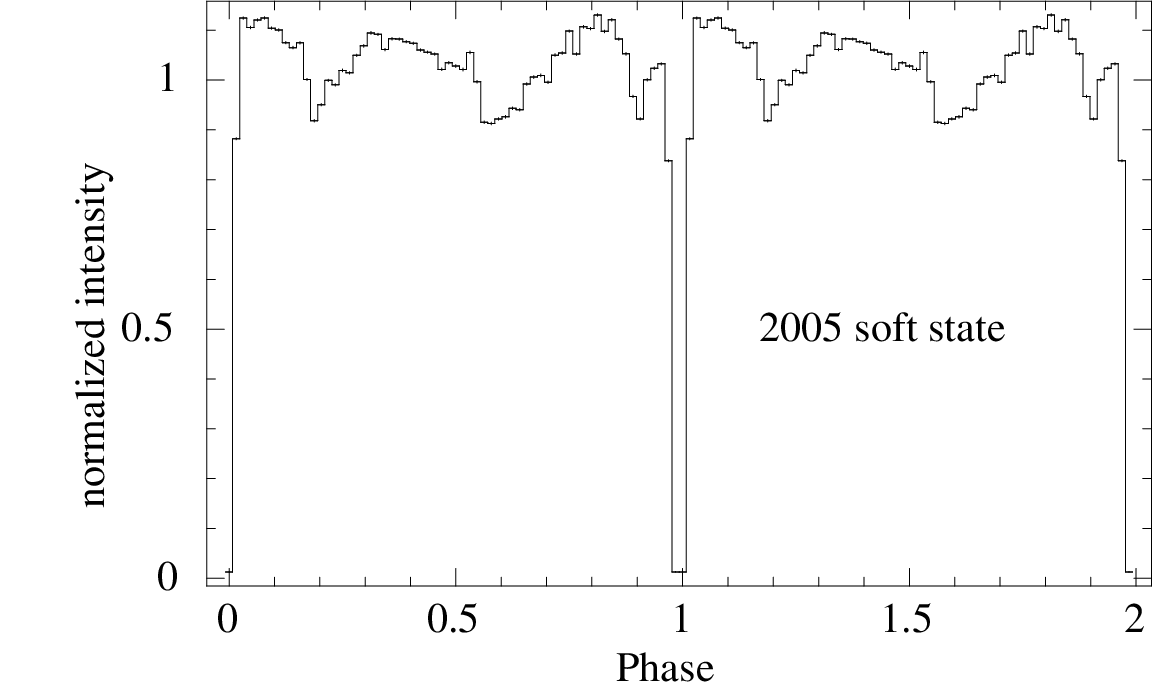}
  
   \caption{ \textit{XMM-Newton} light curves (2.5 - 10.0 keV) with a binning of 100 s and normalized average orbital profiles for the 
   2003 \obsid 0160760101 (top) and the 2005 \obsid 0212480501 (bottom) are shown. The 2005 soft 
   state folded profile shows less intense X-ray dips as compared to the more intense dips when the source was in a hard state in 2003. }
 
\end{figure*}

   The soft X-ray photons, get absorbed by disk structures at different azimuthal angles, due to which they display 
   phase-dependent dipping behavior. The structures in the accretion disk that cause the dips can have various origins like irradiation
   induced vertical structure or warps, stream-impact bulge, ionized absorbers etc., which we describe in detail below. 
  
  \subsubsection{Irradiation induced vertical structures}
  Irradiation effects are generally invoked in disk accreting LMXB systems \citep{deJong} like 4U 1822-371 
  \citep{MasonCordova} and 4U 1916-05 \citep{Callanan,Homer}. These effects could lead to large scale height and non-axisymmetric 
  vertical structures. These structures can explain the soft X-ray modulations seen at different orbital phases \citep{Hakala}. 
  Radiation-driven accretion disk warps can lead to instabilities that would produce many of the observed modulations 
  for LMXBs \citep{Pringle}. Warped, precessing accretion disks enable irradiation of outer disk portions in an azimuthally 
  asymmetric way \citep{Clarkson}. For Neutron star X-ray binaries, warping effects get prominent after $\sim$3x$10^8$ cm and 
  thus affect the outer regions of accretion disks \citep{Pringle}. Although \citet{Ogilvie} showed that not all LMXBs are likely
  to experience warping, there is strong observational evidence in our case for some form of disk structures that must be present 
  above the disk plane. \citet{Kotze} analyzed the long term RXTE ASM light curve of EXO 0748-676 and carried out a period 
  search using the Lomb Scargle method and reported a super-orbital period corresponding to 180.8$\pm$0.3 d during the ``active'' 
  state. They interpret this super-orbital periodicity as a consequence of coupled precessional and warping effects. 
  The current dataset is obtained from a  nearly 60 day long observing campaign and hence is not sufficient to look for super-orbital 
  period signatures. Furthermore, \citet{Homan_2015} state that origin of the 1 mHz low frequency QPO in EXO 0748-676 could be 
  geometrical in nature, possibly a misaligned, precessing inner accretion flow.

  \subsubsection{Stream-impact bulge at the outer accretion disk} 
  Changing aspects of the X-ray heated regions of the disk like the stream-impact plasma bulge at the outer 
  disk and/or the ionized plasma above the disk may also contribute to the soft X-ray variability at different orbital phases.
  They could be responsible for blocking the direct view of the X-ray emitting region, which could cause X-ray dips at 
  different orbital phases. 
  
   \subsubsection{Ionized absorber} 
   \citet{Ponti2014} showed that the presence of a large ionized absorbing component, reduces the opacity,
   which causes less intense X-ray dips in the soft state compared to the hard state (Figure 8). To explain the nature of 
   the ionized absorber responsible for the dips, two strikingly contrasting models, exist. One involves complex continuum models
   \citep{church1998, church_1995} and the other considers simple ionized absorbers \citep{Diaz}. 
   The first approach by \citet{church_1995} assumes the X-ray emission originates from a point-like blackbody together with an extended 
   corona, where the dipping spectra are modeled by partial or progressive covering of the extended corona.
   However, \citet{Diaz} self-consistently demonstrated that changes in the properties of an ionized absorber 
   provide an alternative and satisfactory explanation for the overall spectral changes during dips for EXO 0748-676 using the current 
   XMM-\textit{Newton} dataset Obs ID 0160761301, without having to invoke partial or progressive covering models. \\
   
      A detailed spectral study of this observation, considering all the dips, by \citet{Diaz} revealed that the changes in 
      the spectral continuum from persistent to dipping intervals, reflected in a significant increase in the absorption column density 
      of the ionized absorber upto a factor of 5 and a decrease in the ionization parameter log($\xi$) by $\sim$0.2. Their study indicates 
      that, in EXO 0748-676, which shows sharp and deep dips, the absorber is significantly less ionized and relatively cool 
      (log($\xi$)$<$2.5) unlike other LMXB dipping sources with more ionized absorbers.

  \subsection{Lack of correlated variability between X-rays and optical}
  \subsubsection{Reprocessing}
  Global optical/NIR and X-ray correlation studies conducted in many low magnetic field (B $\leq$10$^{11}$G) neutron star LMXBs 
  by \citet{Russell} showed that the optical emission can be accounted for by thermal emission due to reprocessing except in 
  cases where the source is in a hard state and its luminosity is very high ($>$10$^{37}$ ergs/s).
  In such a scenario, synchrotron emission from jets contributes heavily to the NIR and optical emission.  
  However, since X-ray reprocessing usually dominates the optical emission for lower luminosities, which is the current case, 
  the X-ray spectral state changes should reflect in the optical emission for this binary. 

  \subsubsection{Burst reprocessing}
  Simultaneous X-ray and optical observations of thermonuclear X-ray bursts in LMXBs are very useful in 
  understanding the X-ray reprocessing phenomena and also to investigate the relative importance of the two processes of optical 
  emission: X-ray heating and viscous dissipation. Burst reprocessing into optical emission  has been observed in many LMXB sources 
  and is an expected consequence of interception of a fraction of the primary burst X-ray photons by the disk and/or companion surface. 
  \citet{Dubus} showed that a point source cannot irradiate a planar (non-warped) stationary disk because self-screening 
  would prevent the outer parts from getting affected by irradiation. A non-irradiated disk model is incapable of explaining 
  the optical emission from a large fraction of persistent LMXBs. A concave disk (a consequence of irradiation), 
  extending to larger scale heights as we move outwards \citep{deJong} would ensure the outer disk intercepts some part
  of the X-rays. In the case of EXO 0748-676, the delays between the X-ray and optical bursts are consistent with the binary 
  separation \citep{Hynes2006,BP}. Smearing in the delays, also indicates a finite extent of this reprocessing region
  which is most likely present at the outer accretion disk. Reprocessing of the thermonuclear X-ray bursts 
  therefore show that since a significant part of the burst X-ray emission is reprocessed in the outer accretion disk, same may hold true
  for the non-burst emission. 

  \subsubsection{Non-burst reprocessing fraction}
  Considering a conversion factor for the non-burst reprocessing phenomenon,
  similar to the burst emission, roughly 24\% of the total optical emission can be assumed to be due to the reprocessing of X-rays. 
  However, this is a lower limit, and a larger fraction of the optical emission is probably due to X-ray reprocessing. 
  The X-ray emission from LMXBs is anisotropic and is enhanced in the direction perpendicular to the disk plane \citep{Fujimoto, He}. 
  The persistent X-ray emission tends to have a higher degree of anisotropy compared to the burst emission. 
  Thus, for high inclination systems like EXO 0748-676, the observable X-ray to optical conversion ratio of the non-burst emission 
  would be higher than the burst emission. The residual optical emission during the X-ray eclipses is partly from the companion star 
  and partly from the un-eclipsed disk. 
  Considering an average source magnitude of $\sim$17, the optical flux in the V-band is 
  $\sim$4.1 x 10$^{-13}$ erg cm$^{-2}$ s$^{-1}$ and the 1 - 10 keV X-ray flux is 2.25 x 10$^{-10}$ erg cm$^{-2}$ s$^{-1}$. 
  This gives an X-ray to V-band optical flux ratio of $\sim$550 and a reprocessing efficiency from X-ray flux to V-band 
  optical flux of $\sim$0.044\%. Assuming the companion in EXO 0748-676 to be a typical M2V type star with radius of 
  0.44 R$_{\odot}$, a surface temperature of 6000 K \citep{Hynes_2009} and a source distance of 5 kpc, the total 
  optical flux is $\sim$2.7 x 10$^{-13}$ erg cm$^{-2}$ s$^{-1}$. The V-band optical flux of the companion star is 
  thus a small fraction of the total source emission in the V-band mentioned above.
  
  \subsubsection{Optical eclipses/dips correlations with X-rays}
  The optical eclipses of EXO 0748-676 are not total, and are broader compared to the X-ray eclipses. This indicates that a part of 
  the optical emission is from a region that is centered around the X-ray source, but larger, perhaps with a size comparable to the size 
  of the companion star. A non-zero optical intensity during the eclipse indicates a considerable emission from a broader region, perhaps 
  the outer disk and a small contribution from the companion. Although the broad optical eclipses do coincide with the X-ray eclipses, 
  a number of other variabilities (like narrow or wide dips) are not particularly coincident with the soft X-ray dips. The CCFs display 
  variable positive soft X-ray to optical as well as hard X-ray to optical delays at time scales of a few 1000 s 
  for almost all the orbital profiles. The cross-correlation coefficients are small in all the 7 observations. Also, the CCFs are 
  variable from orbit-to-orbit in delays and strength. The lack of correlation at reprocessing time-scales is expected and 
  is consistent with the correlation analysis results. However, the observed weak correlation, seen in the repeating 
  patterns of the CCF, that exists at sub-orbital time-scales of a few 1000s of seconds, is probably related to orbital period.
  This time scale is different than the dynamic or viscous time-scales in the binary and can possibly arise due to 
  variable outer accretion disk structures, discussed in section 4.1.\\
   
  We also note that we were unable to compare the X-ray optical co-variability of the source during the hard spectral state (2003) 
  with the soft spectral state (2005), since the OM data from the \textit{XMM-Newton} observation in 2005 (\obsid 0212480501) 
  was of a poor statistical quality. 
    
  \subsubsection{Large optical variability}
  The input for the reprocessing phenomena, the X-rays from the compact star is likely to be stable since
  the mass accretion rate and luminosity of the source is fairly constant during the entire duration of all the 7 
  observations \citep{Boirin}. The optical emission from a stable accretion disk with fairly constant 
  mass accretion rate must have a smooth orbital modulation. It is unexpected to show changes in shape as shown in
  Figure 3. The soft X-ray variability is due to absorption by structures in the accretion disk. However, emission from
  the outer disk will have variability that is different from the X-rays. Vertical accretion disk structures like the ones addressed in 
  section 4.1 can produce orbital optical intensity variation both due to changing visibility of the structures themselves or the 
  visibility of their X-ray reprocessing faces. 
   
  \subsection{Mean optical intensity and optical states}
  The optical counterpart of EXO 0748-676 is known to exhibit high and low intensity states similar to those defined for the
  X-ray \citep{Motch_exo}. The high state corresponds to a high mean optical intensity which also exhibits a broad minima at
  the phases 0.6 - 1.2. The low state, on the other hand exhibits a lower mean optical intensity with the broad optical dip disappearing 
  and the sinusoidal shape flattening out \citep{Thomas}. In the current dataset, the optical light curves of  
  \obsids 0160760101, 0160760201, 0160760301 and 0160760401 show a broad minima, with some orbits showing peaks, and the shape seen
  in \obsids 0160760601, 0160760801 and 0160761301 is different: flattened sinusoids without any peaks. Within the current set of
  7 observations having similar orbital averaged optical intensity, the shape of the optical modulation varied between what was earlier 
  known to be associated with the optical bright and low intensity states, indicating that the optical orbital intensity profiles are not 
  intrinsically related to the mean optical intensity.
  
  \section{Conclusions}
 
 From simultaneous X-ray and optical observations of EXO 0748-676 in the hard spectral state, we conclude that
there is no co-variability between the X-rays and optical light seen in most of the orbital light curves at reprocessing 
or orbital time-scales, but a weak correlation exists at a few 1000s time-scales. There exist vertical structures in the 
outer accretion disk that evolve at binary orbital time scales of a few hours. Large optical variability 
between successive orbits could be the result of reprocessing occurring at these evolving disk structures. These structures 
could be the result of irradiation effects and/or warping. Modeling these effects would require more sophisticated understanding 
of an irradiated accretion disk and warps, both of which are still at a very nascent stage. More simultaneous X-ray and optical 
data would improve statistics and help analyze these scenarios better.\\

\textit{Acknowledgements} :
This work has made use of archival data obtained from the High Energy Astrophysics Science Archive Research Center (HEASARC),
provided by the NASA Goddard Space Flight Center.

\bibliography{0748_new_draft}
\bibliographystyle{elsarticle-harv}

\begin{appendix}

\renewcommand\thefigure{\thesection.\arabic{figure}}   
\setcounter{figure}{0}

\section{36 complete simultaneous hard X-ray, soft X-ray and optical orbital light curve profiles.}

EXO 0748-676 was observed during September-November, 2003 with \textit{XMM-Newton}. A total of 7 observations were 
carried out covering 36 complete orbital light curve segments, simultaneously in the X-ray and the optical bands.
The simultaneous hard X-ray (5 - 10.0 keV), soft X-ray (0.3 - 5.0 keV) and optical (white band) light curves are shown 
for all the 7 \obsids. The thermonuclear bursts have been excluded in order to highlight the non-burst intensity modulation
between successive orbital light curve segments. Incomplete orbital segments at the start and end of each \obsid are not shown here. 

\setcounter{figure}{0}

\begin{figure*}
 \begin{minipage}{1.0\linewidth}  
  \centering
 \includegraphics[scale=0.23, angle=-90]{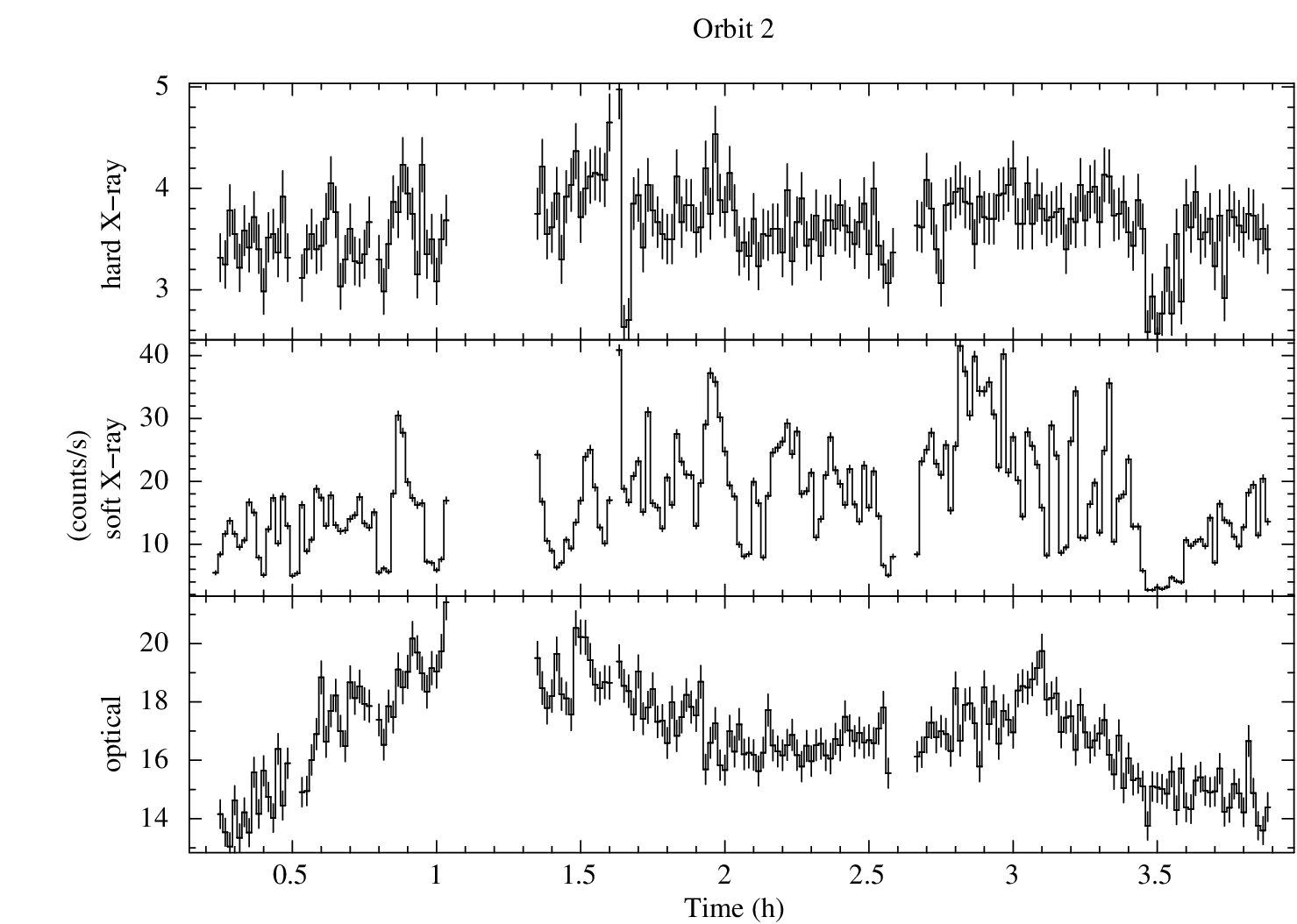}
 \includegraphics[scale=0.23, angle=-90]{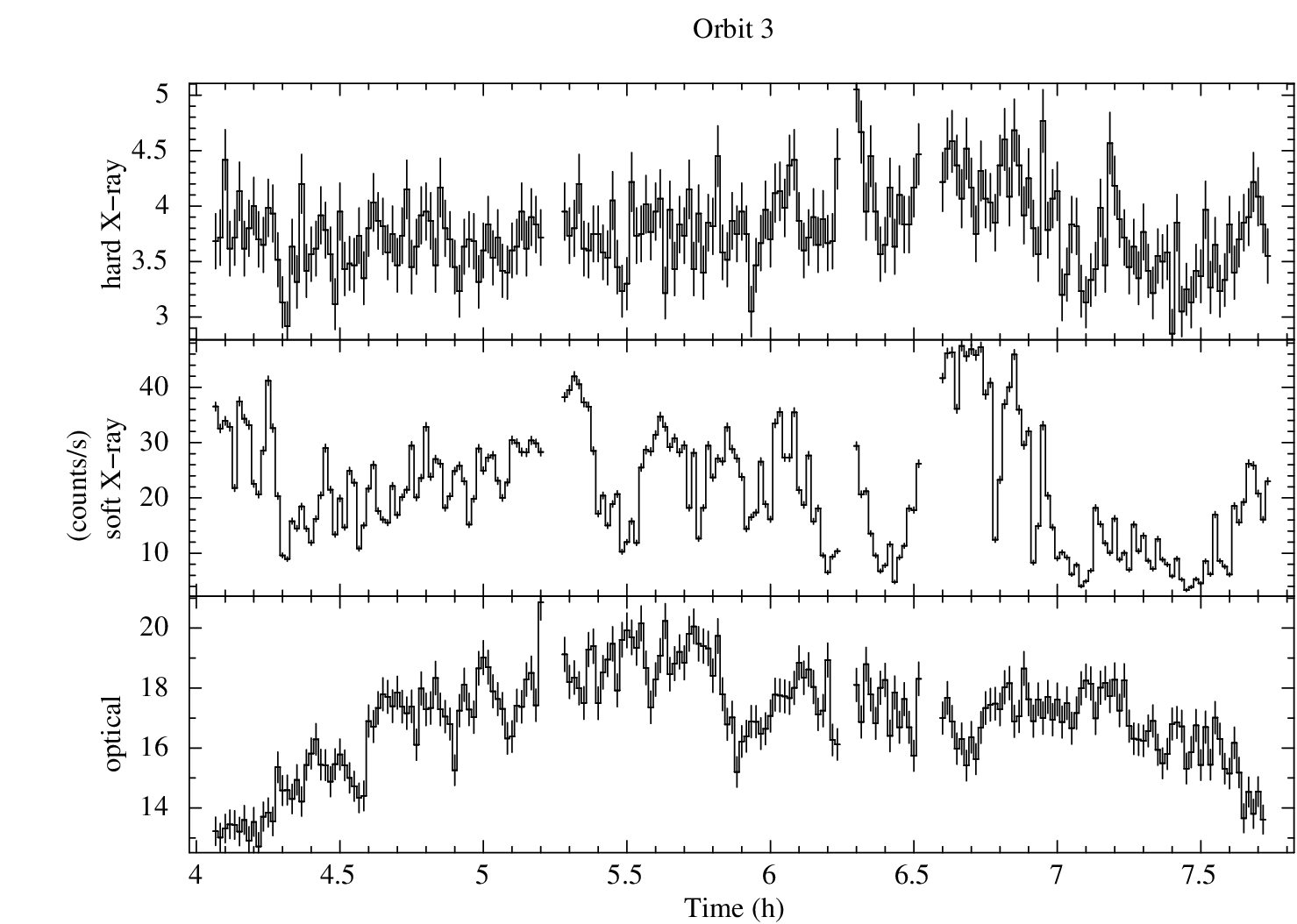}
 
  \end{minipage}     
   \hspace{0.9cm}    
\begin{minipage}{1.0\linewidth}   
  \centering
 \includegraphics[scale=0.23, angle=-90]{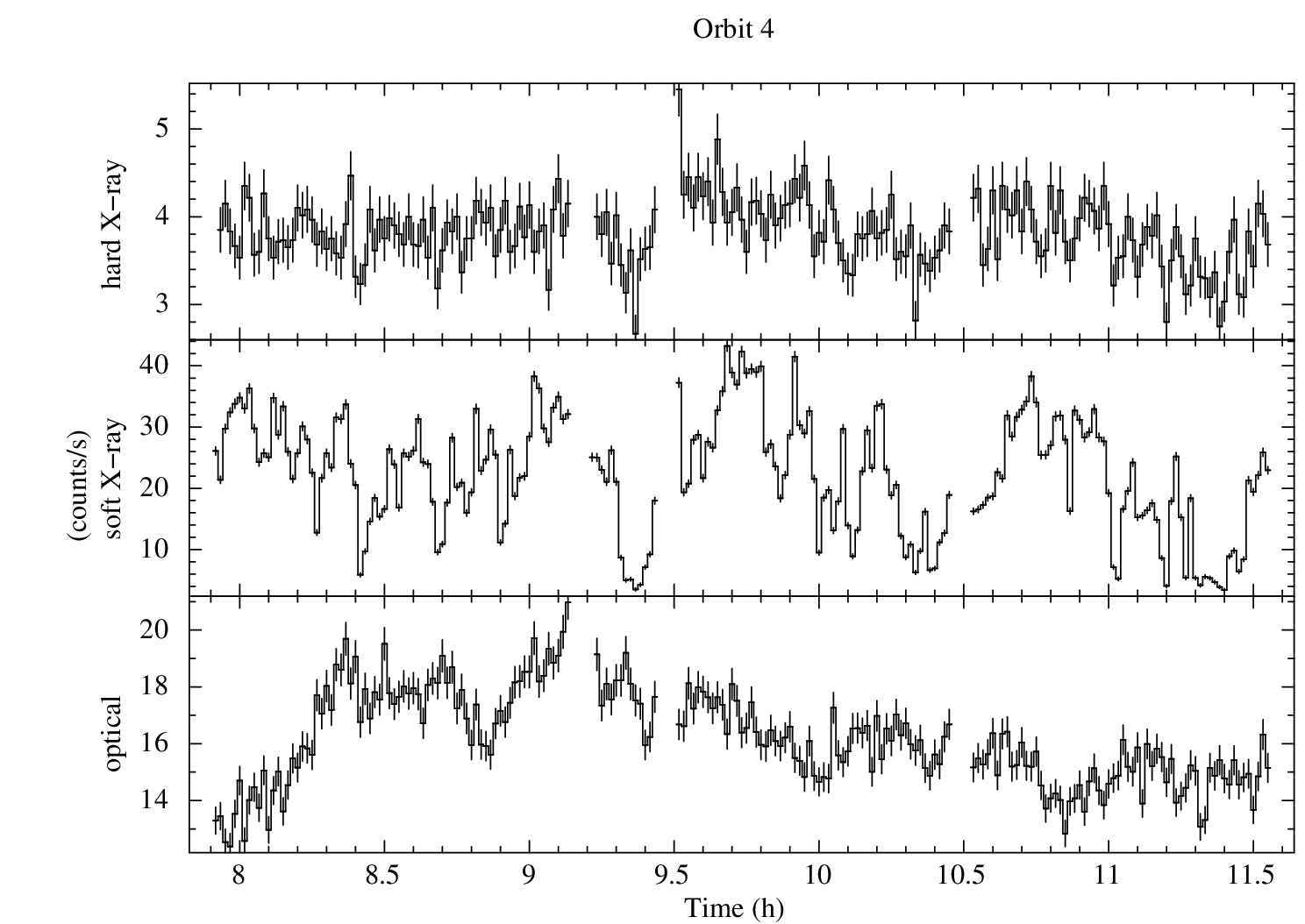}
  \includegraphics[scale=0.23, angle=-90]{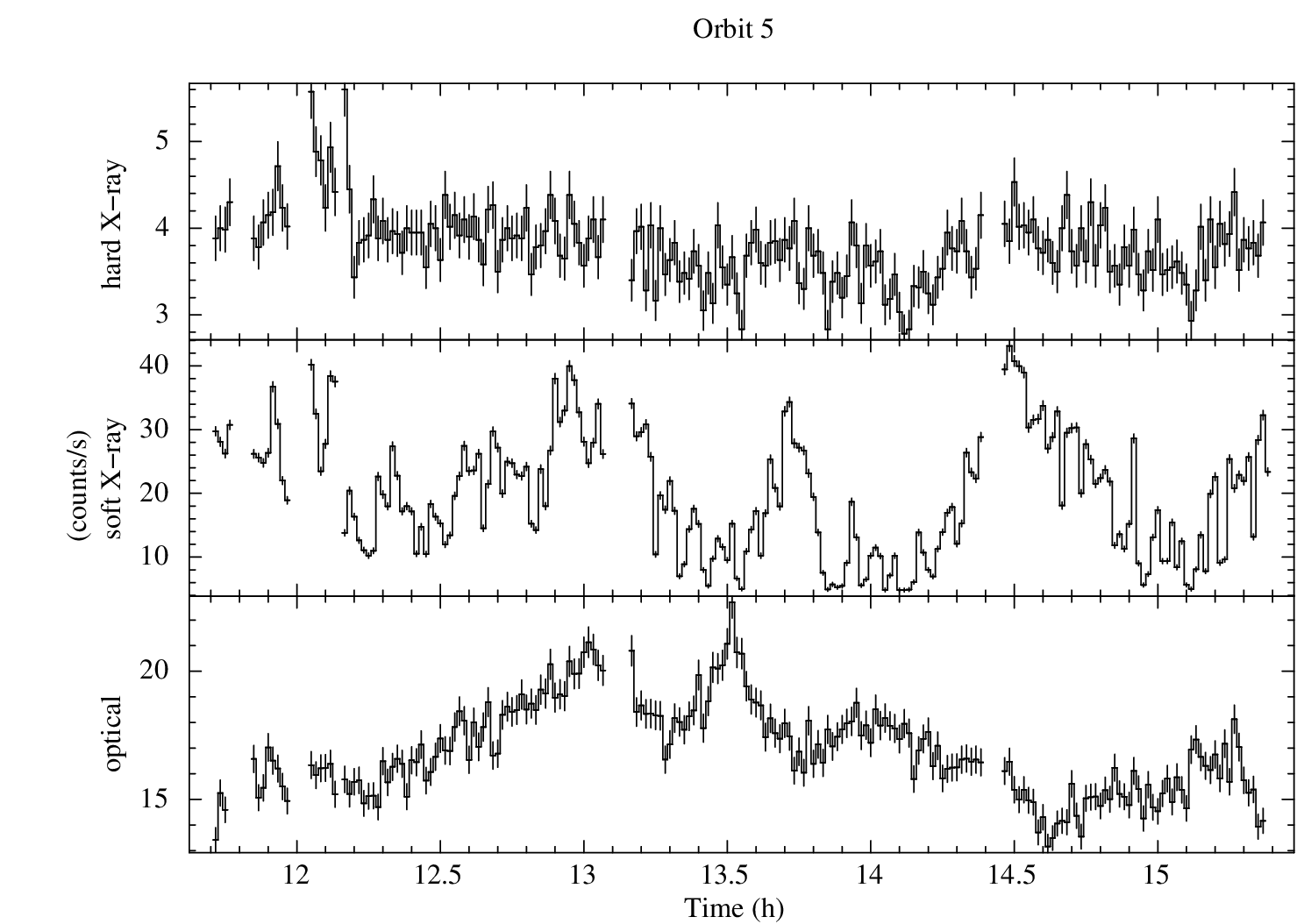}
  
  \end{minipage}     
   \hspace{0.9cm}    
\begin{minipage}{1.0\linewidth}   
  \centering
 \includegraphics[scale=0.23, angle=-90]{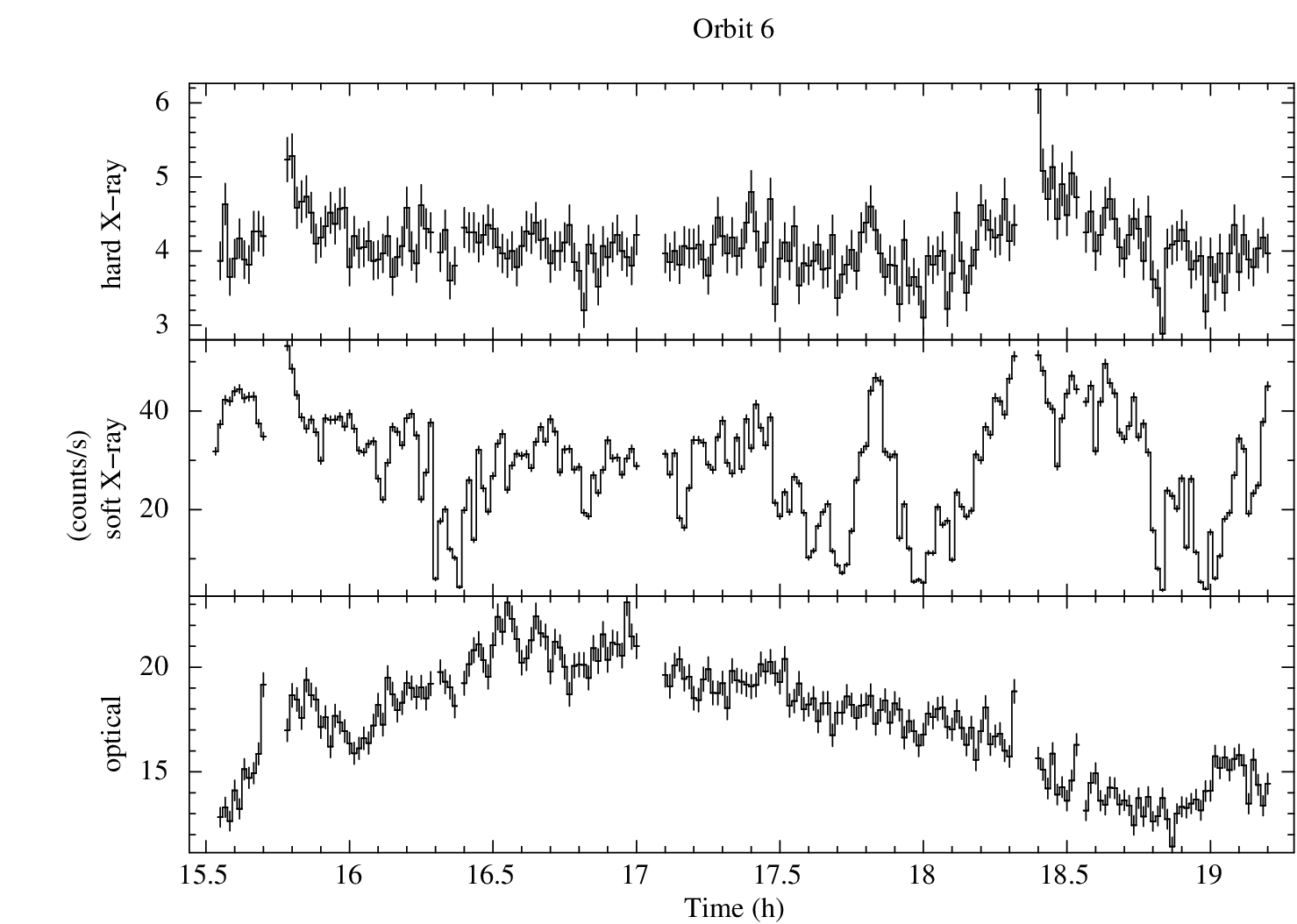}
 \includegraphics[scale=0.23, angle=-90]{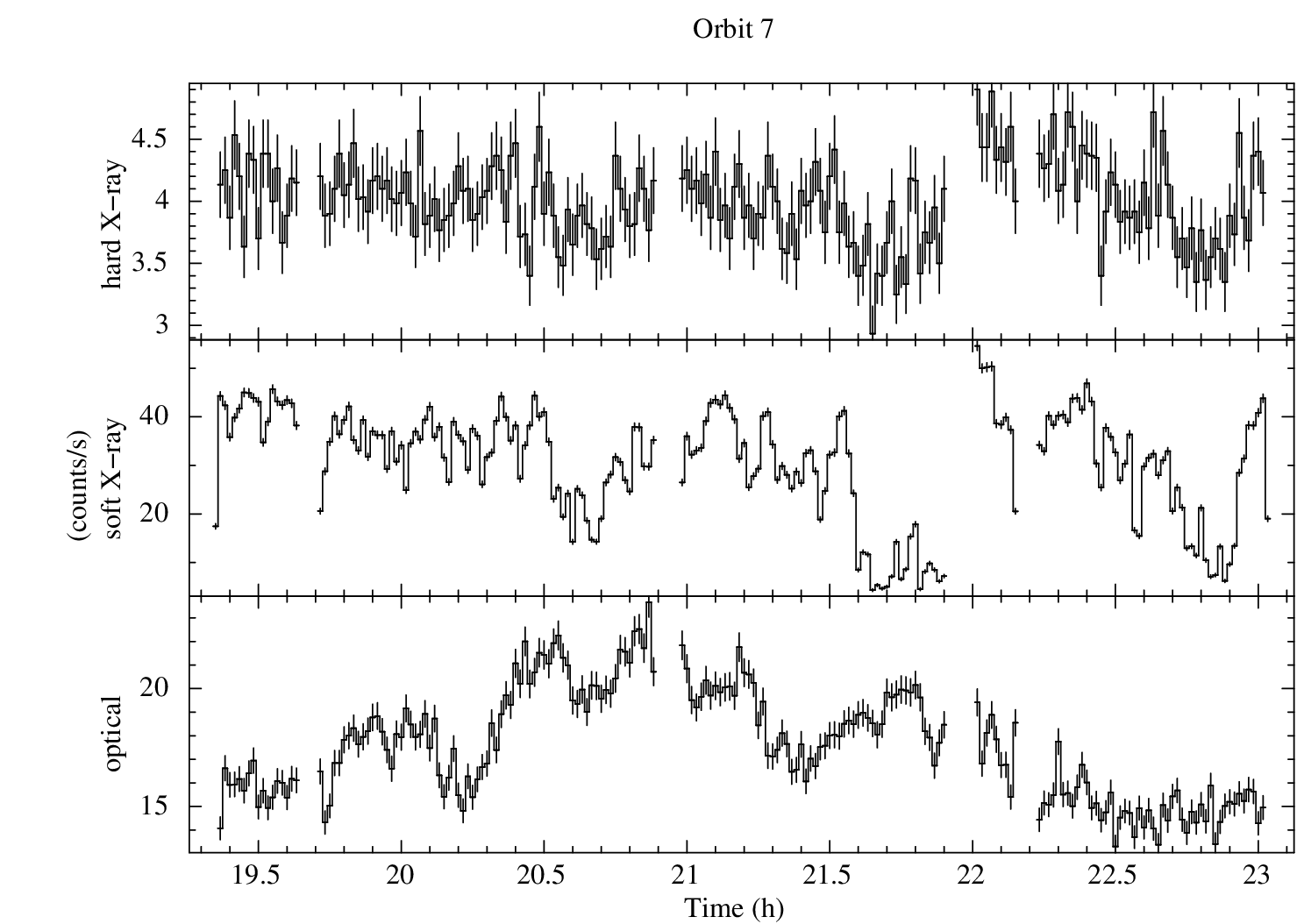}
 
\end{minipage}
\caption{\obsid 0160760101}
 \end{figure*}
 
\begin{figure*}
\begin{minipage}{1.0\linewidth}  
  \centering
  \includegraphics[scale=0.23, angle=-90]{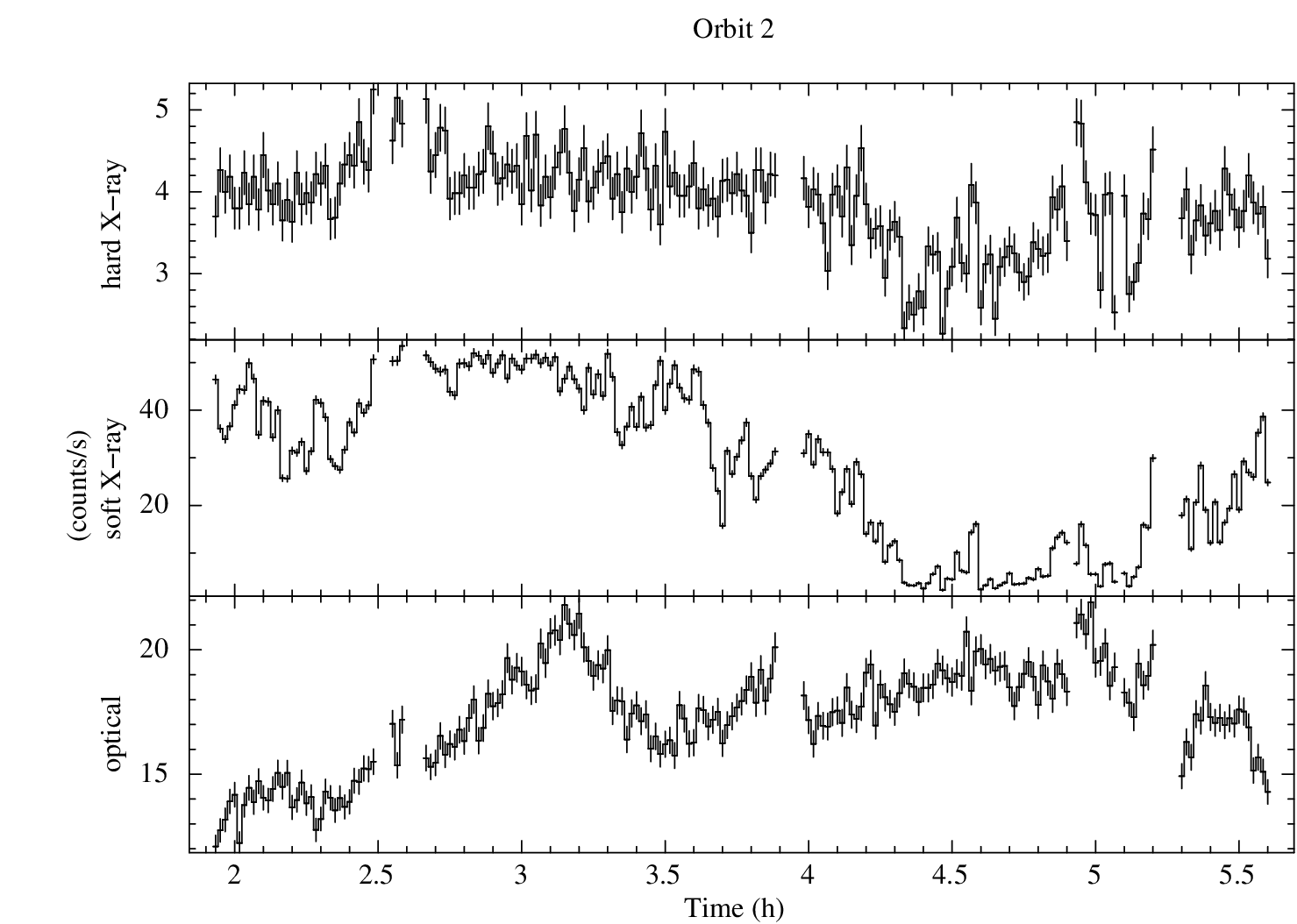}
 \includegraphics[scale=0.23, angle=-90]{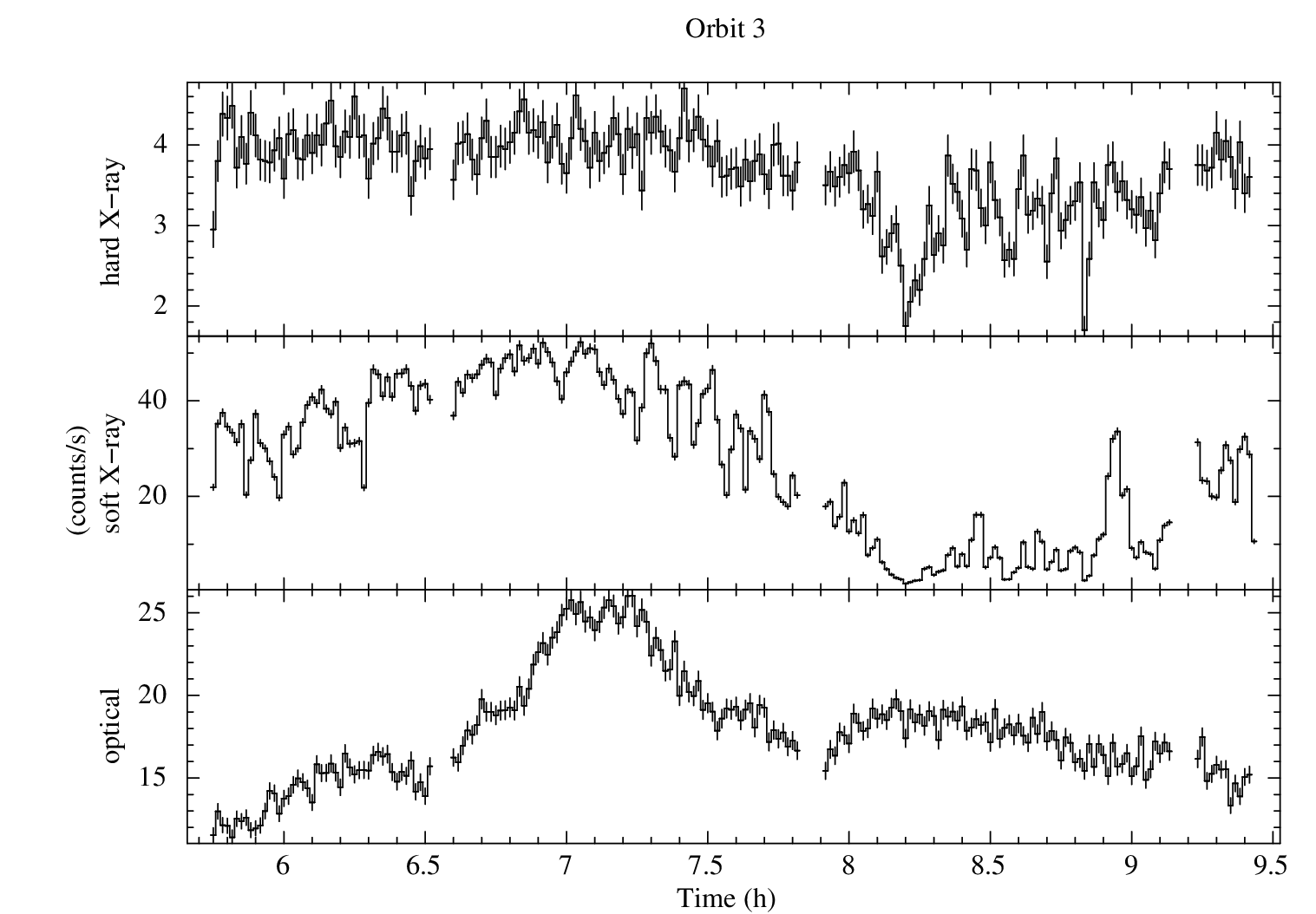}
 
 \end{minipage}     
   \hspace{0.9cm}    
  \begin{minipage}{1.0\linewidth}   
 \centering
 \includegraphics[scale=0.23, angle=-90]{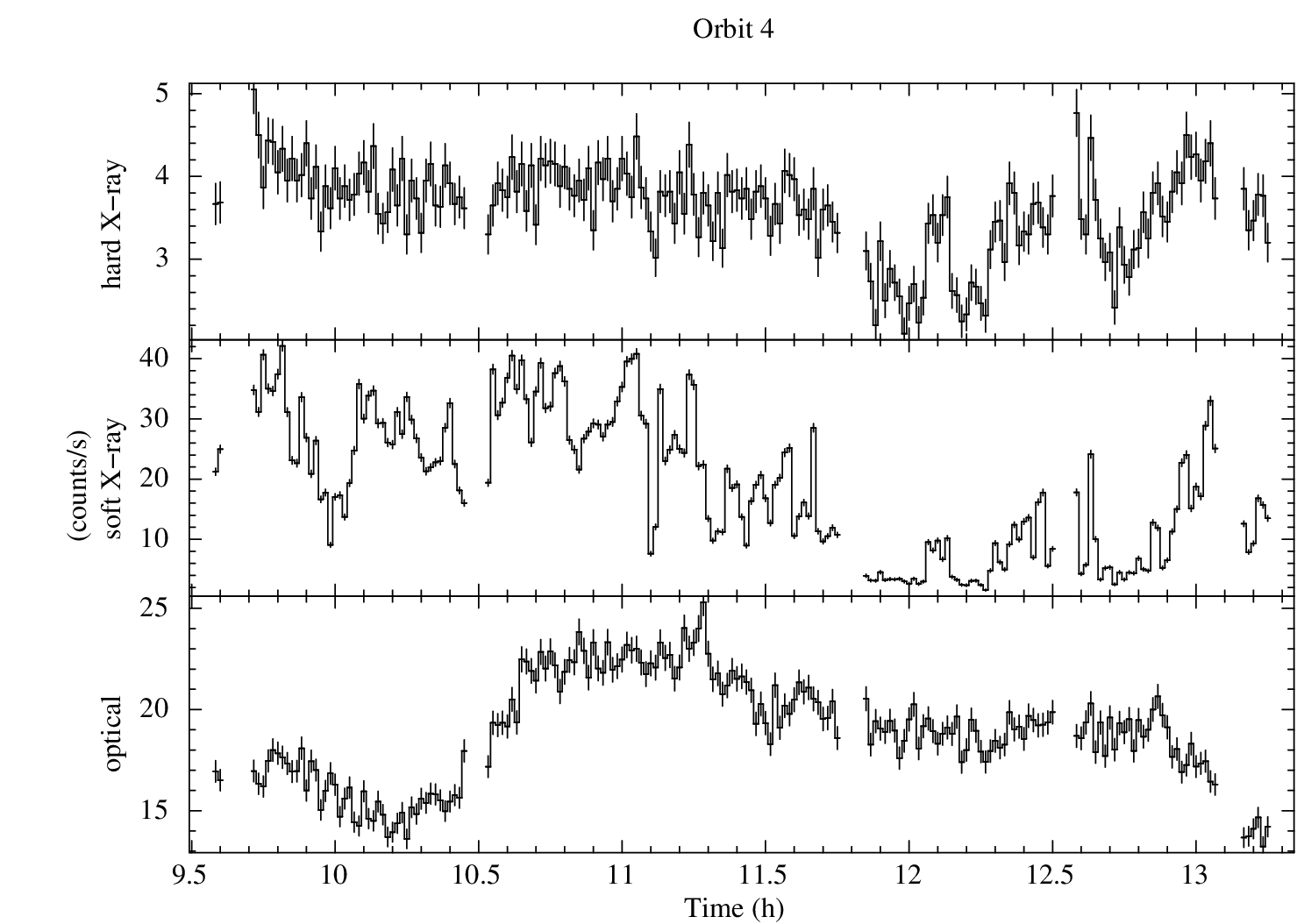}
  \includegraphics[scale=0.23, angle=-90]{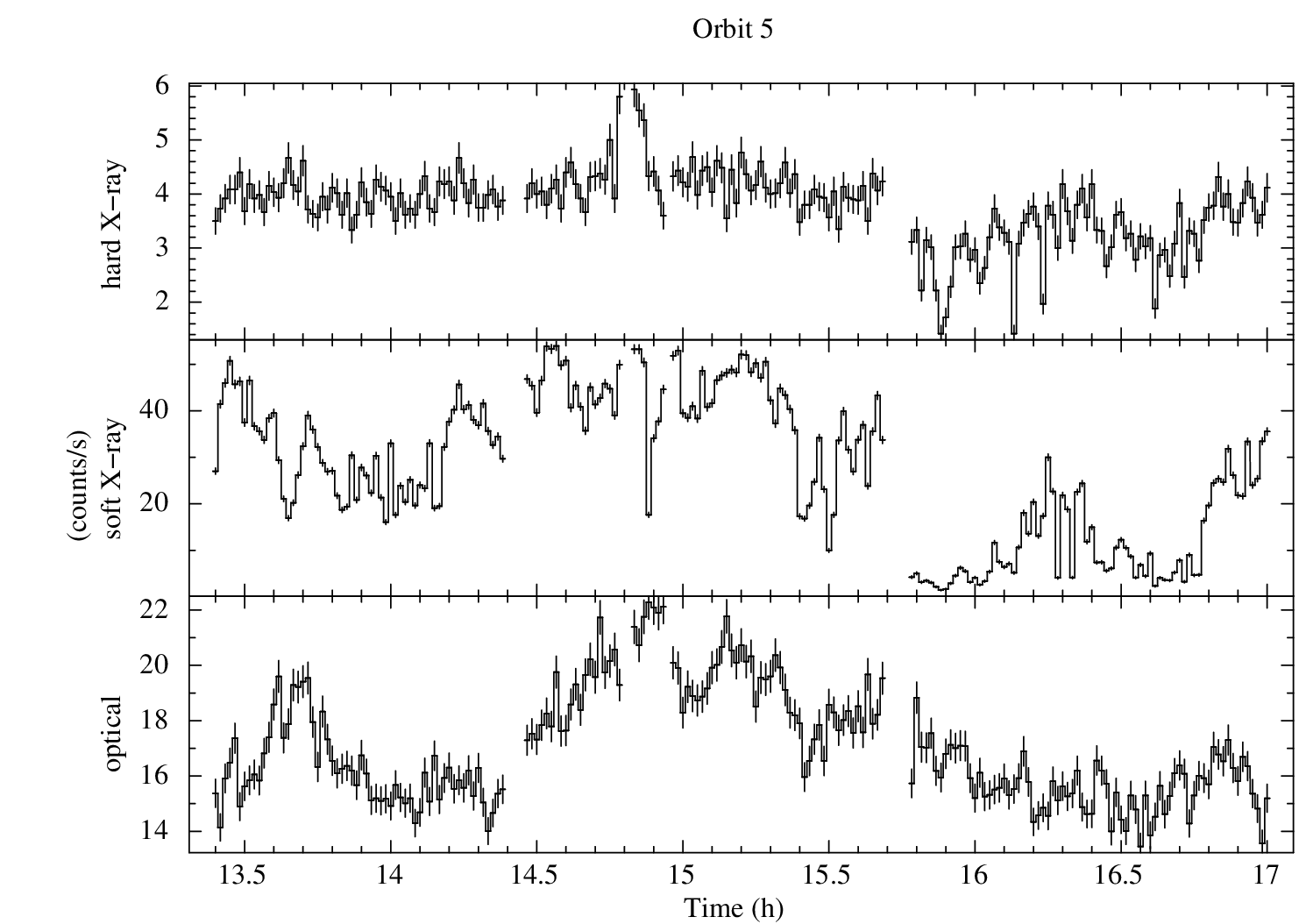}
  
 \end{minipage}     
   \hspace{0.9cm}    
  \begin{minipage}{1.0\linewidth}   
 \centering
 \includegraphics[scale=0.23, angle=-90]{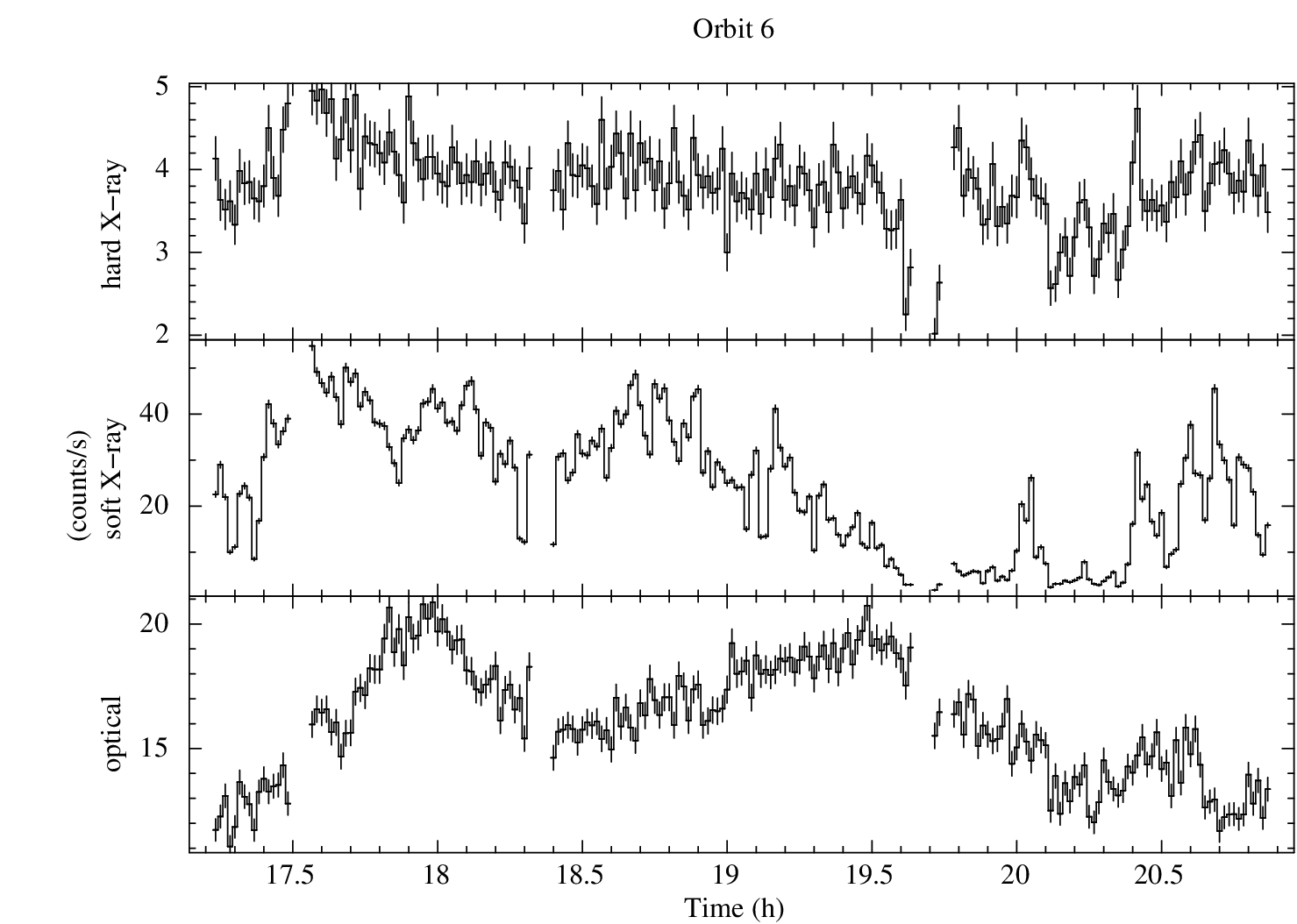} 
 \includegraphics[scale=0.23, angle=-90]{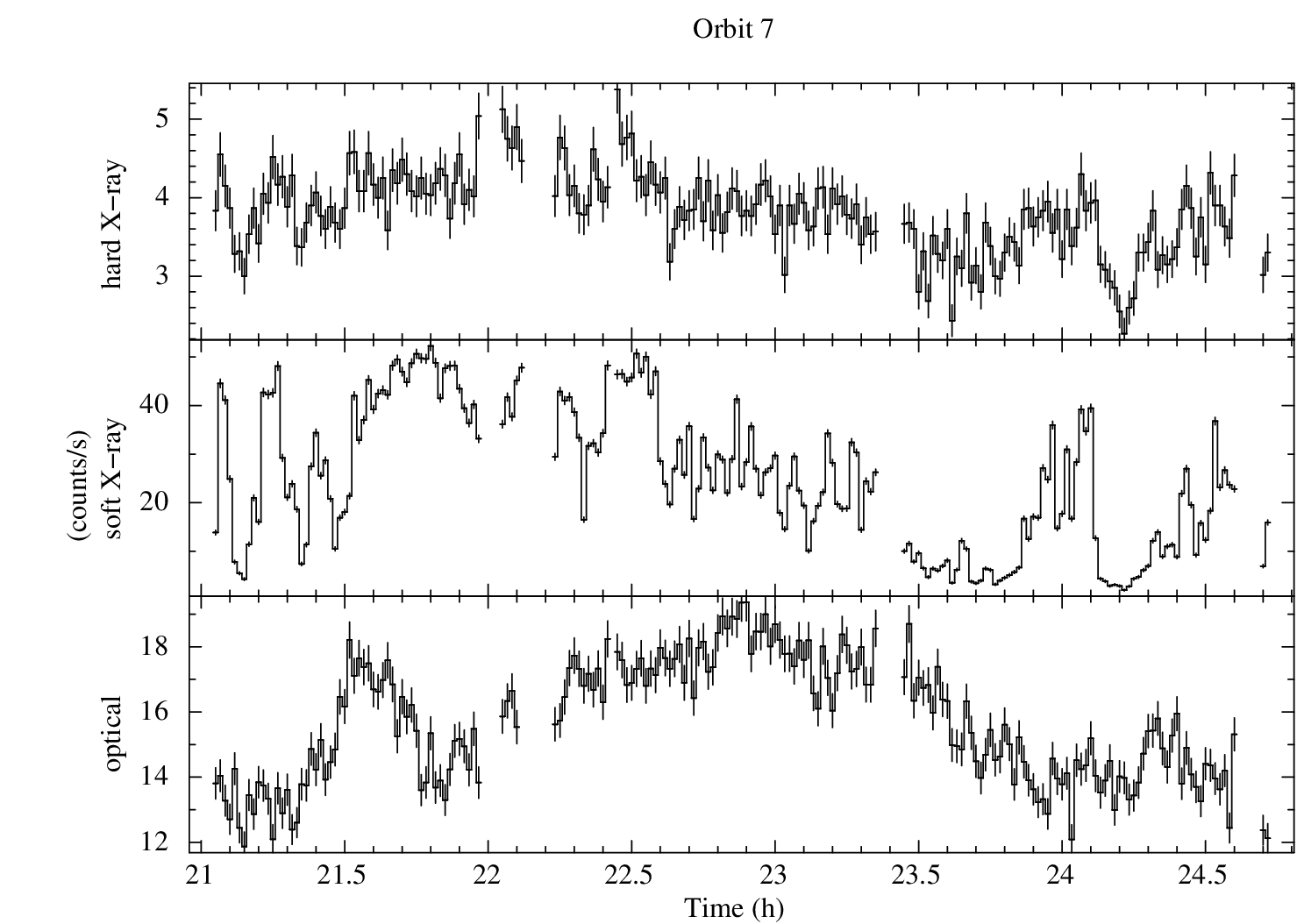}
 
 \end{minipage}
\caption{\obsid 0160760201}
 \end{figure*}

 \begin{figure*}
\begin{minipage}{1.0\linewidth}  
 \centering
 \includegraphics[scale=0.23, angle=-90]{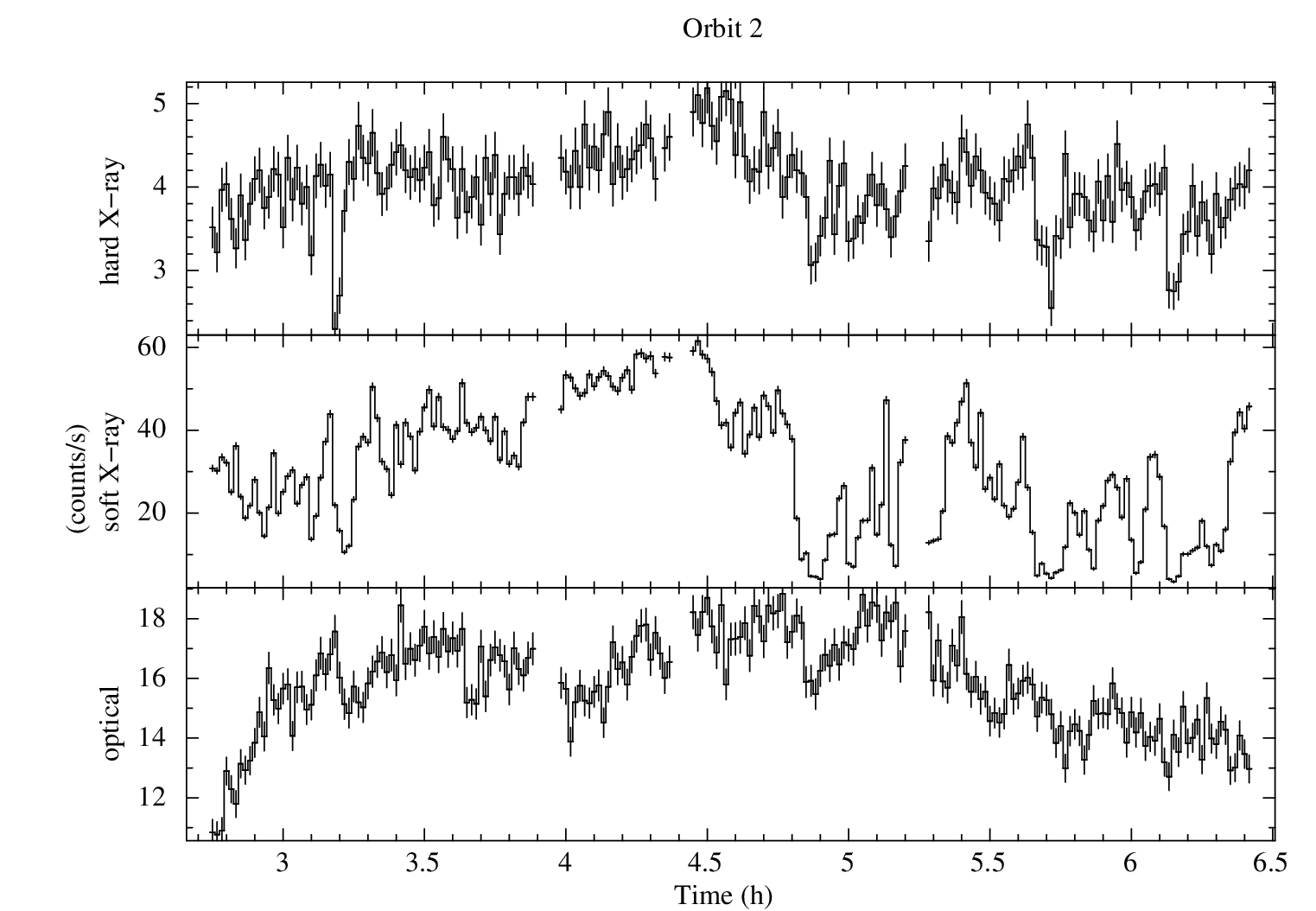}
 \includegraphics[scale=0.23, angle=-90]{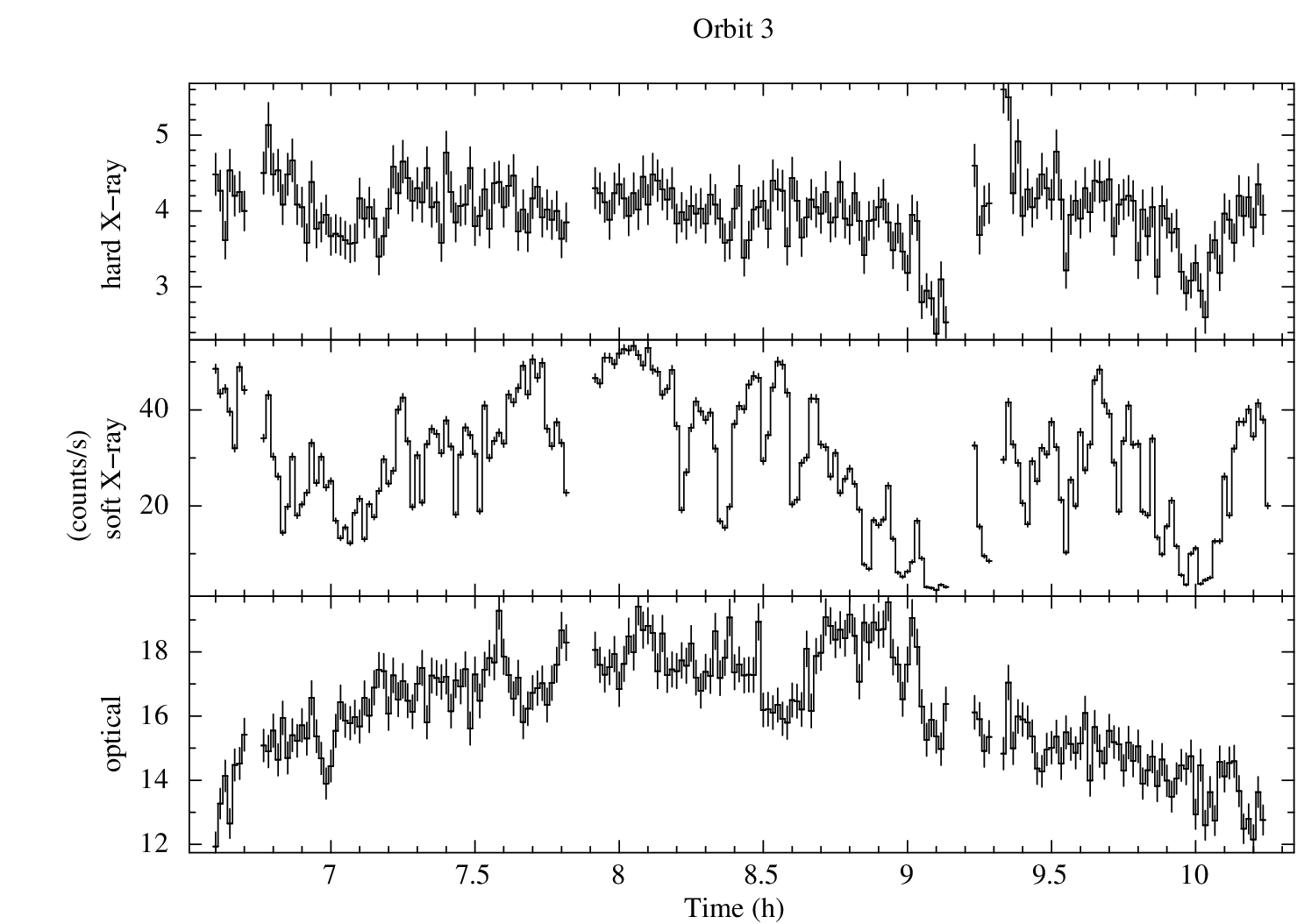}
 
 \end{minipage}
  \hspace{0.9cm}    
\begin{minipage}{1.0\linewidth} 
\centering
 \includegraphics[scale=0.23, angle=-90]{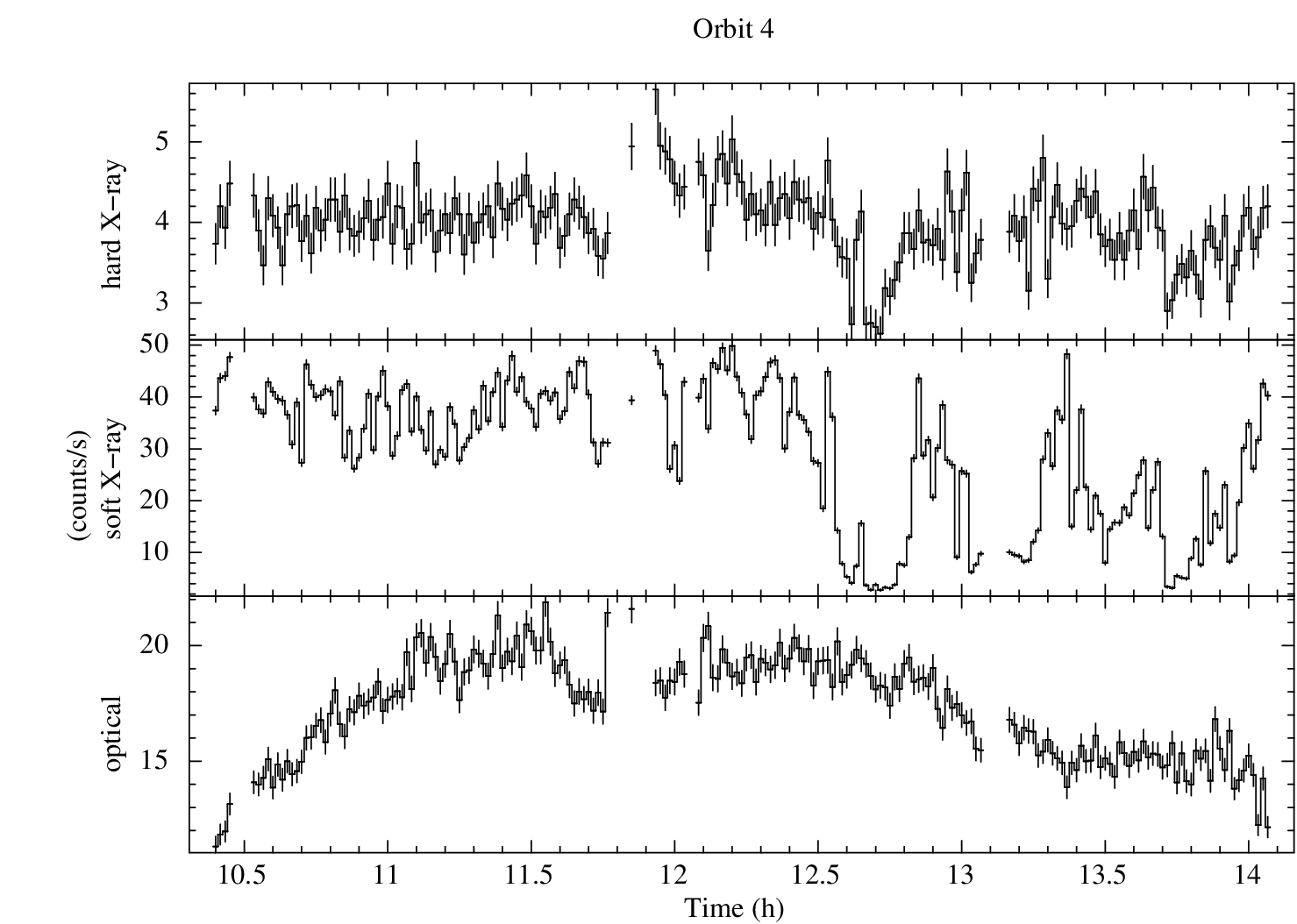}
 \includegraphics[scale=0.23, angle=-90]{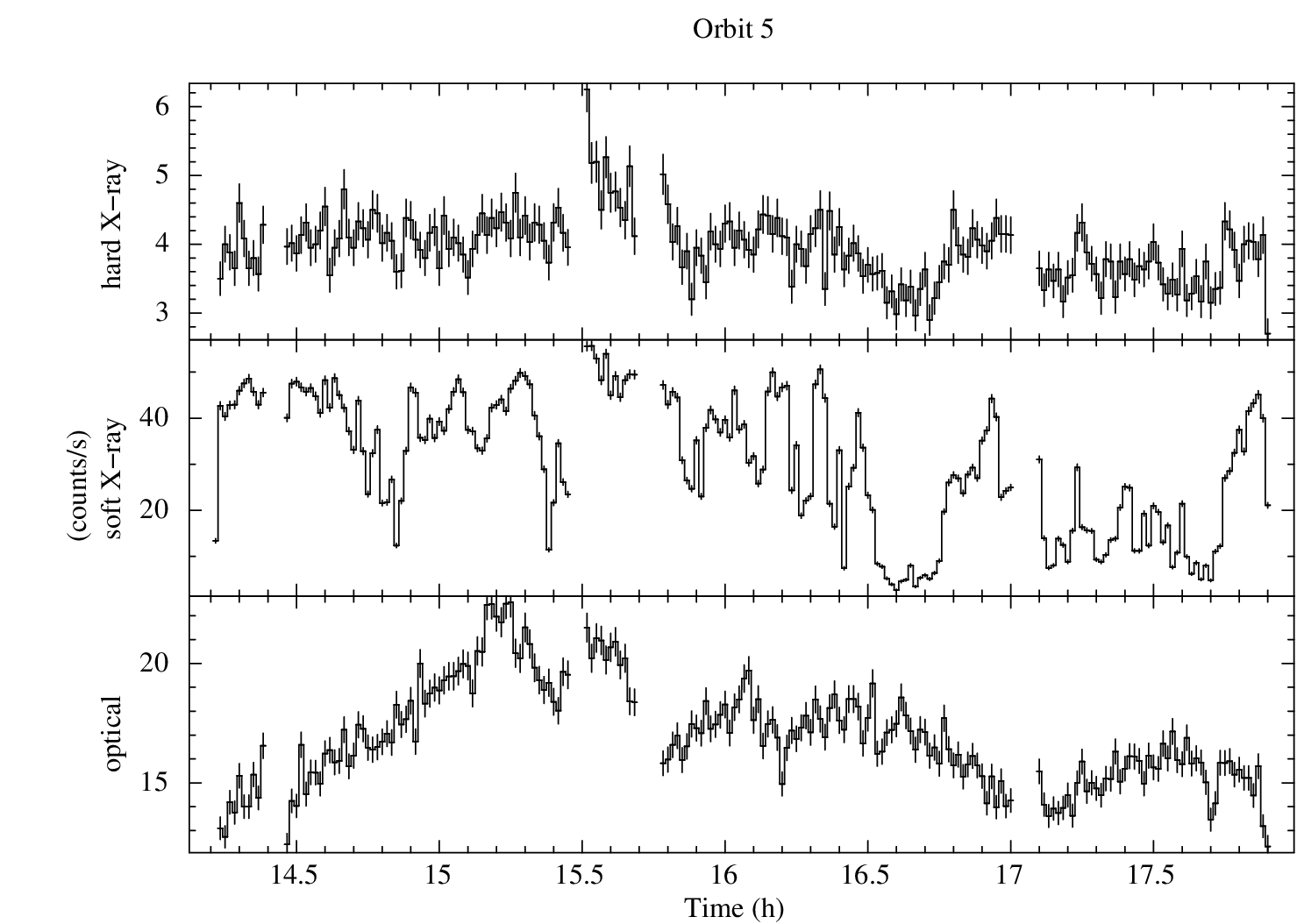} 
 
 \end{minipage}
  \hspace{0.9cm}    
\begin{minipage}{1.0\linewidth}  
  \centering
 \includegraphics[scale=0.23, angle=-90]{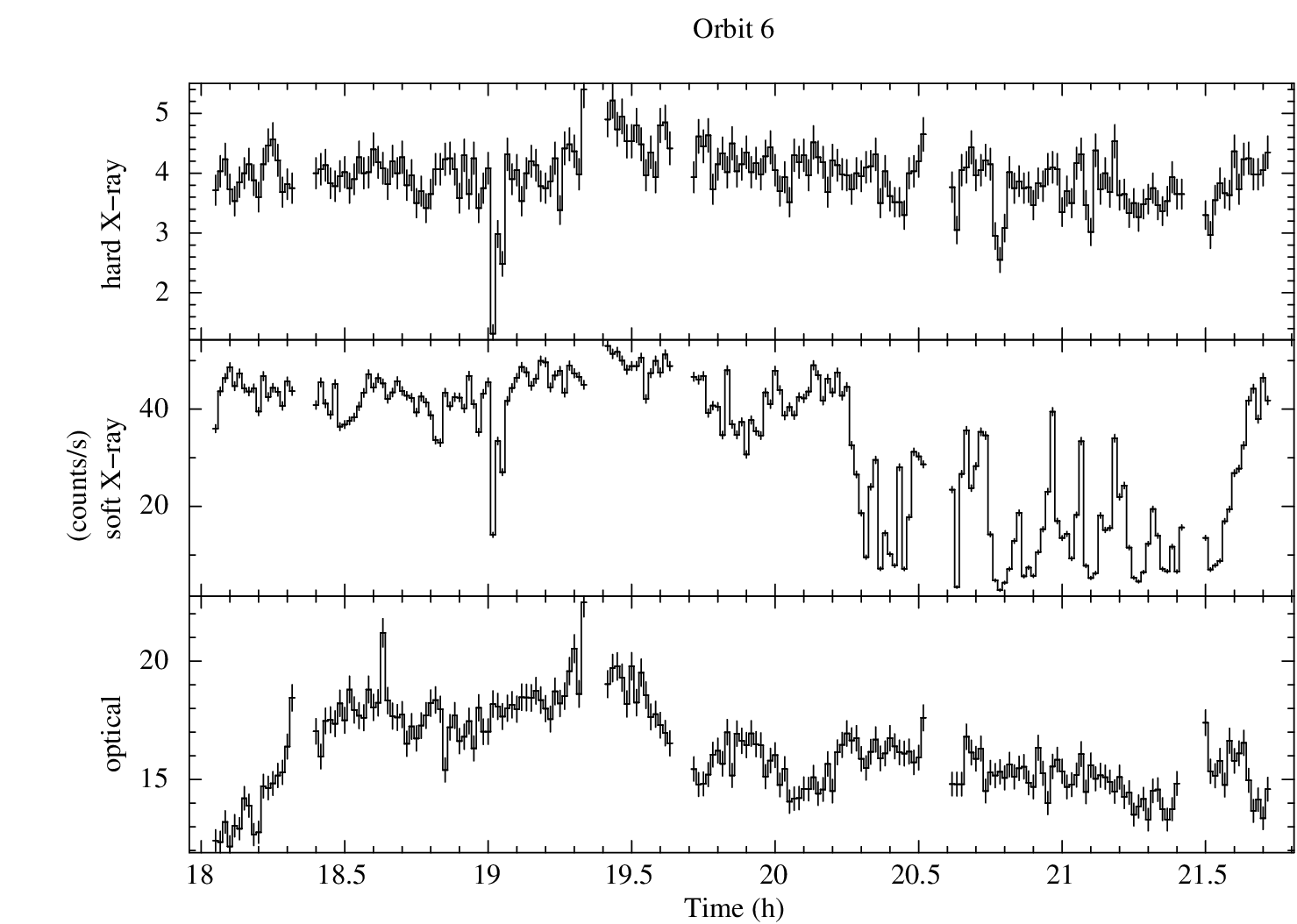}
 \includegraphics[scale=0.23, angle=-90]{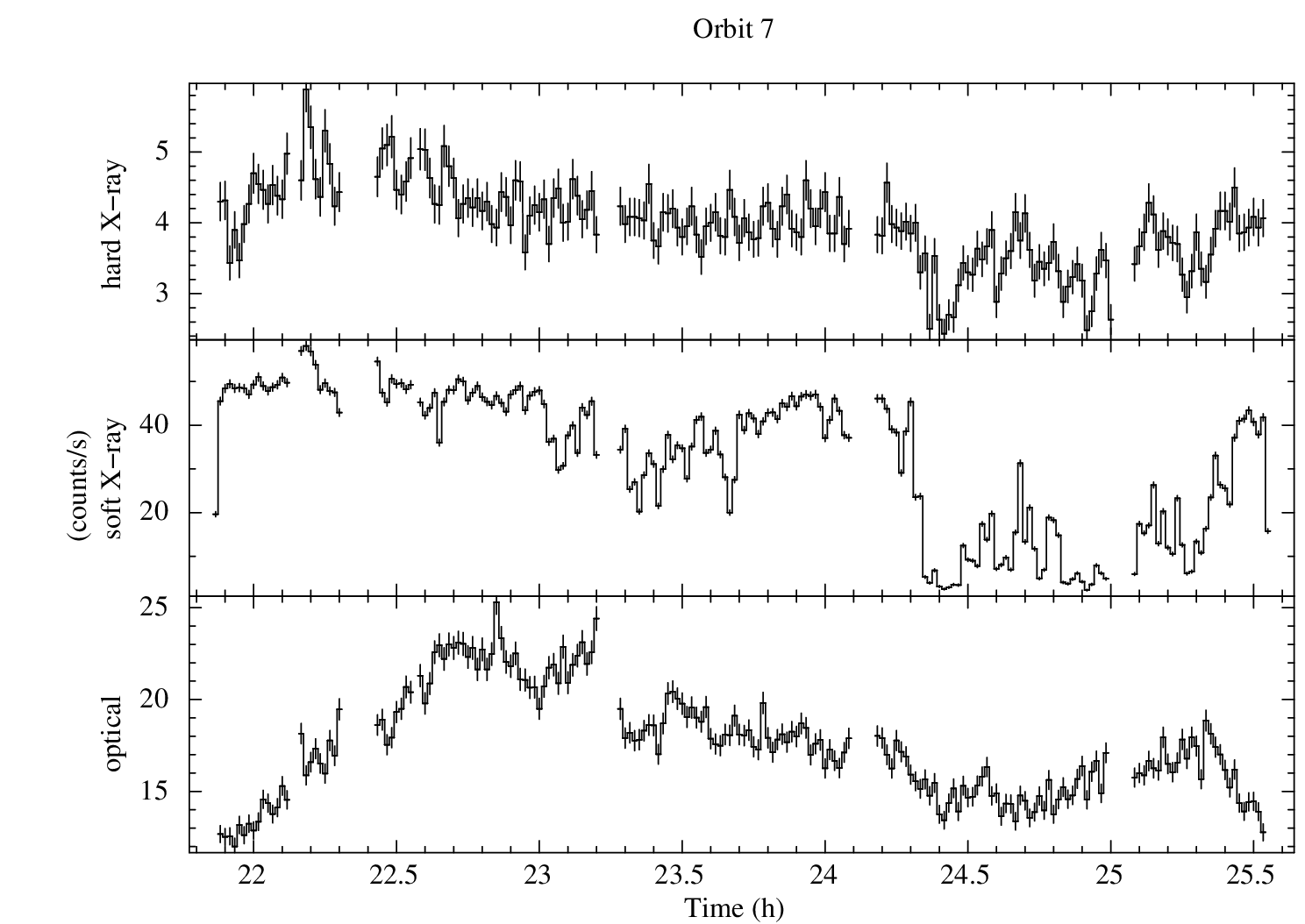}
 
 \end{minipage}
 
 \begin{minipage}{1.0\linewidth}
 \centering
  \includegraphics[scale=0.23, angle=-90]{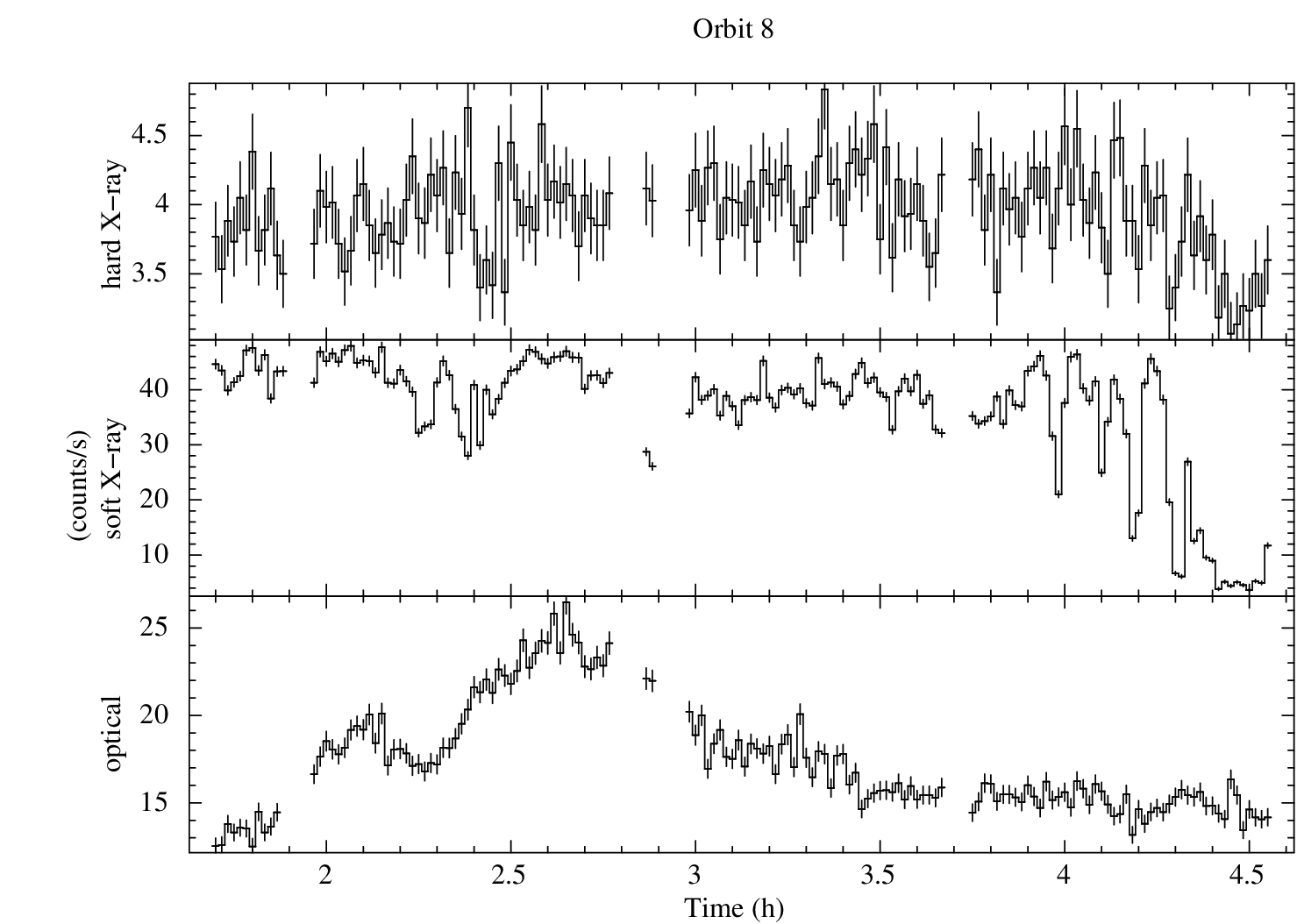} 
 
\end{minipage}
\caption{\obsid 0160760301 }
\end{figure*}

 \begin{figure*}
\begin{minipage}{1.0\linewidth}  
\centering
 \includegraphics[scale=0.23, angle=-90]{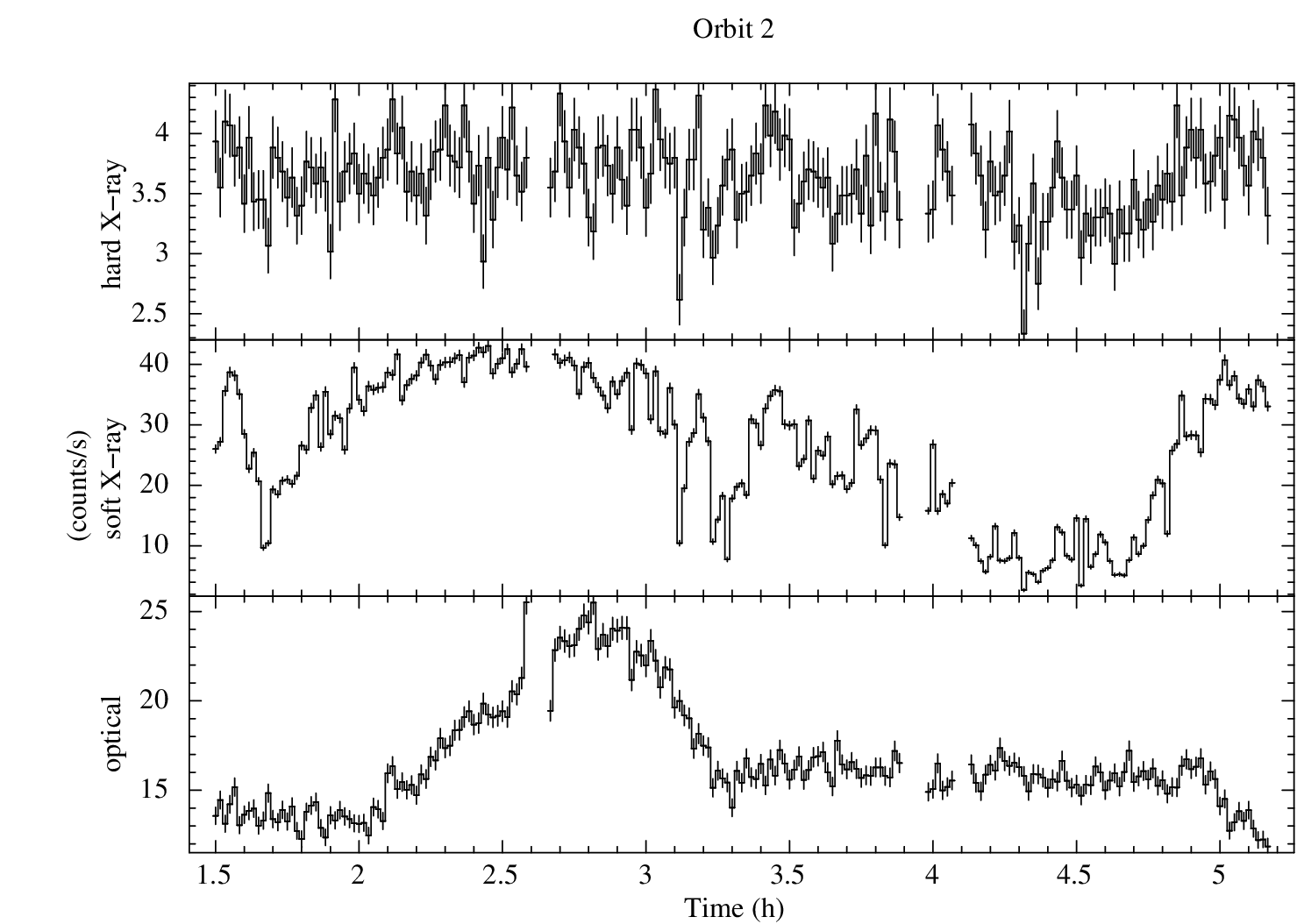}
 \includegraphics[scale=0.23, angle=-90]{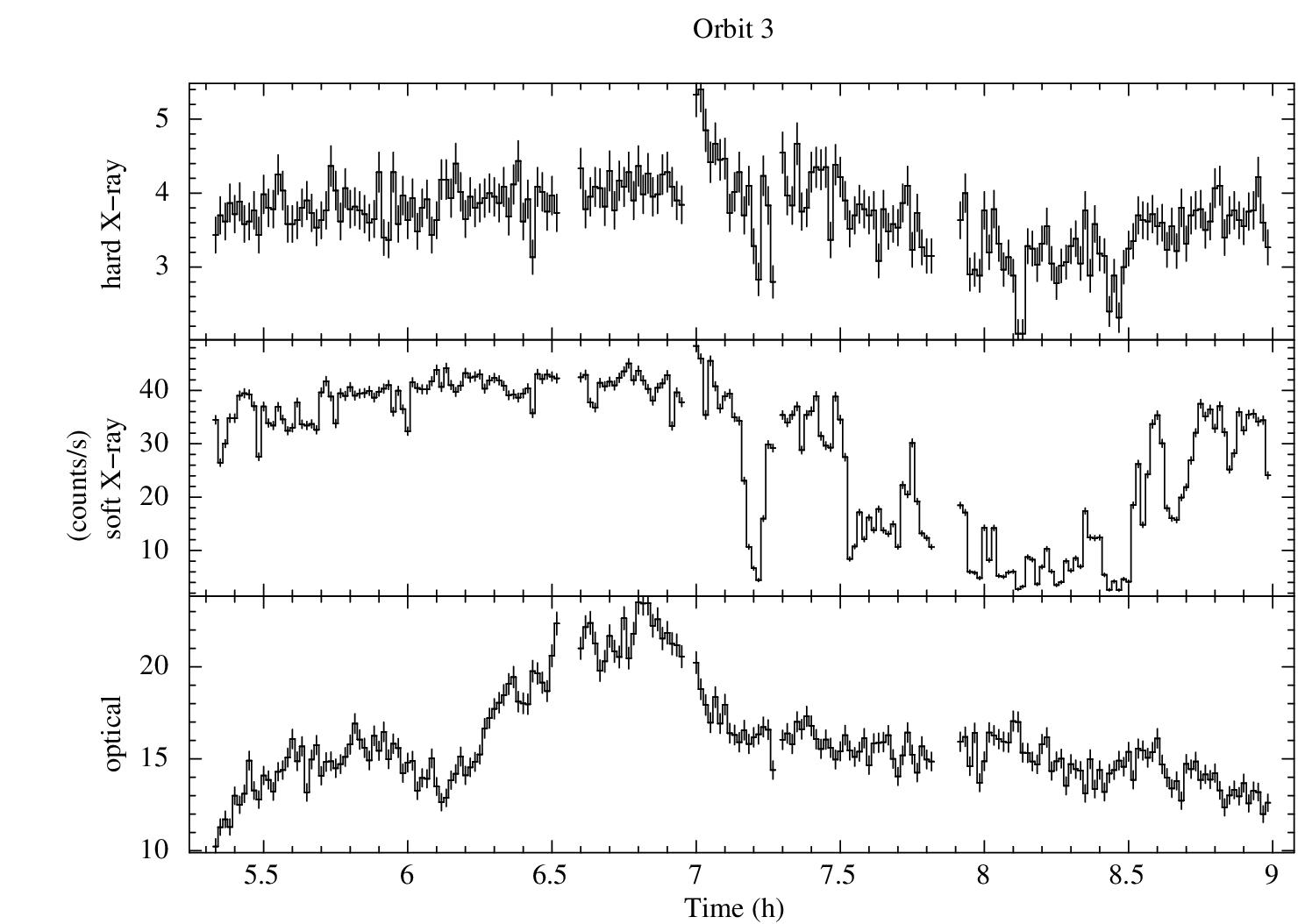}
 
 \end{minipage}
\hspace{0.8cm}    
\begin{minipage}{1.0\linewidth} 
\centering
 \includegraphics[scale=0.23, angle=-90]{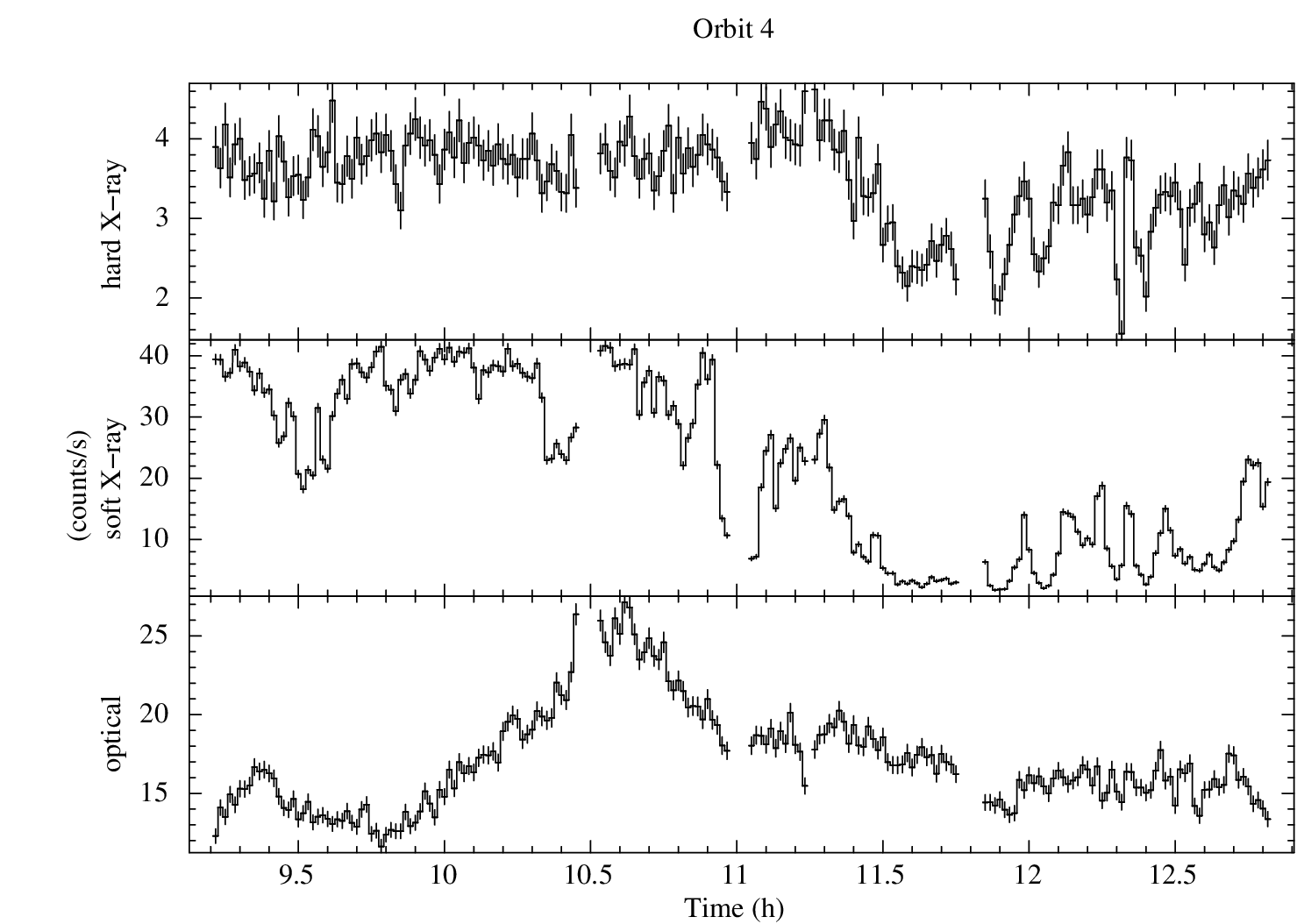}
 \includegraphics[scale=0.23, angle=-90]{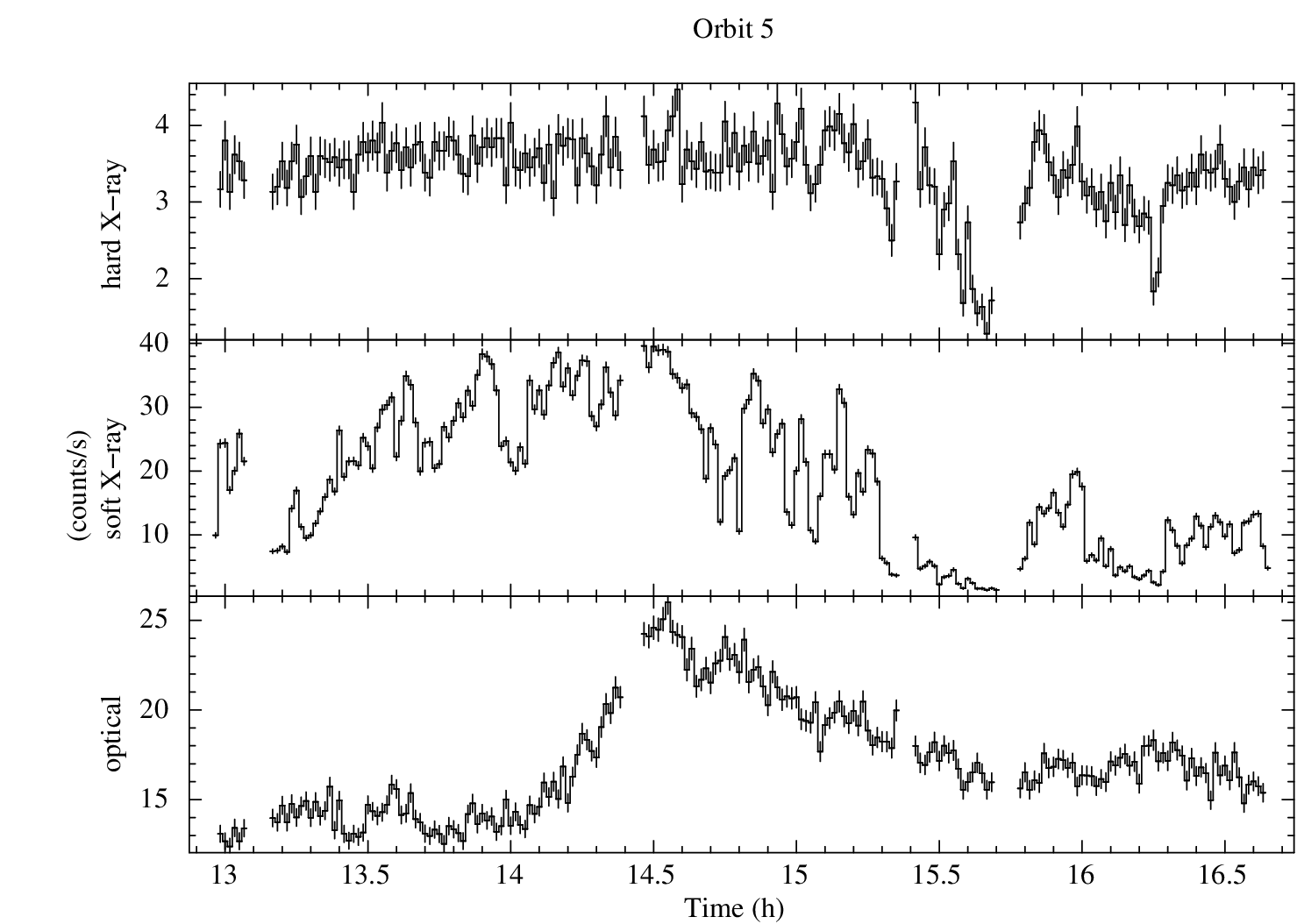}
 
 \end{minipage}
 \hspace{0.8cm}    
\begin{minipage}{1.2\linewidth} 
\centering
 \includegraphics[scale=0.23, angle=-90]{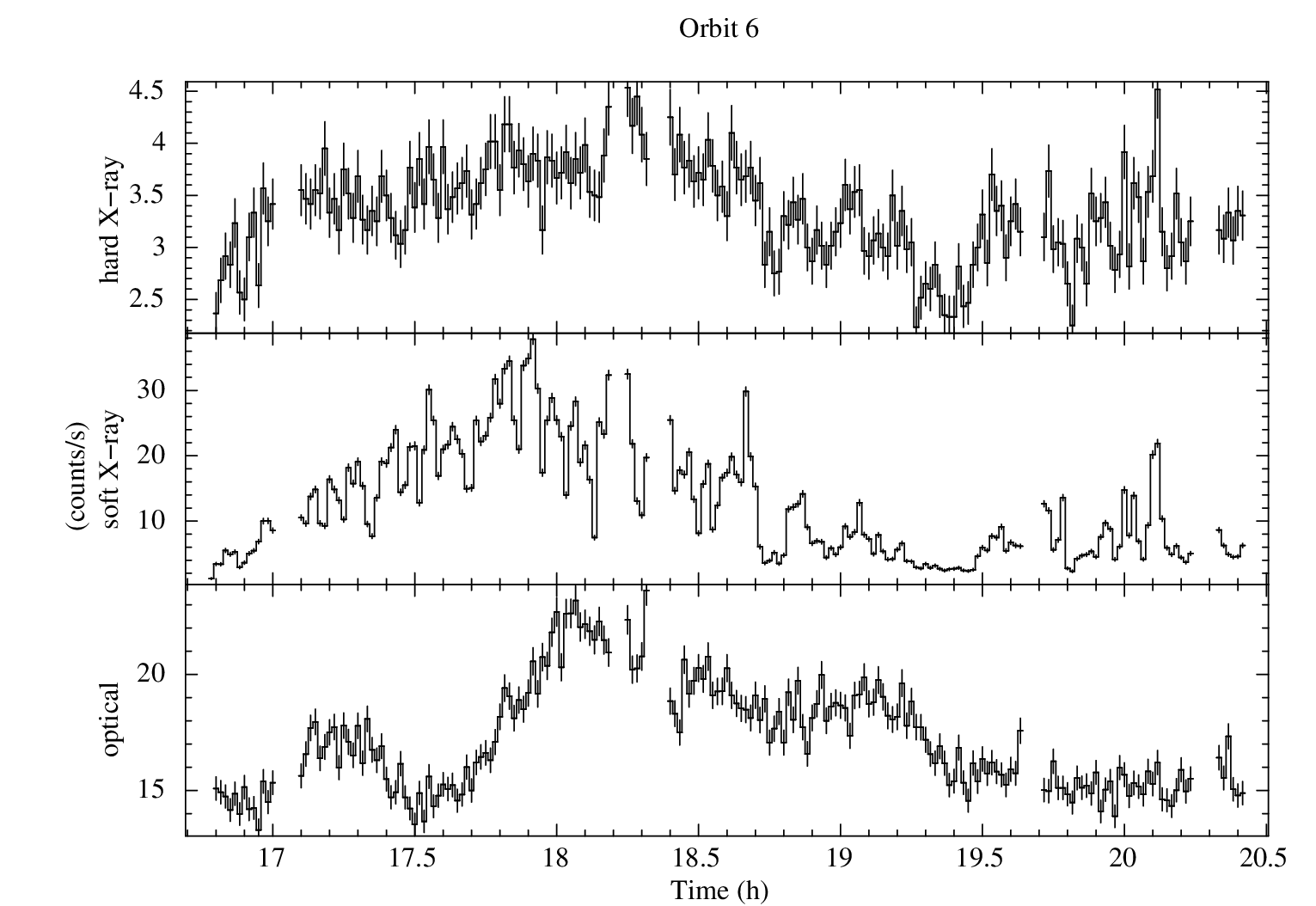}
 
 \end{minipage}
 \caption{ \obsid 0160760401}
   \end{figure*}
   
\begin{figure*}
\centering
 \includegraphics[scale=0.23, angle=-90]{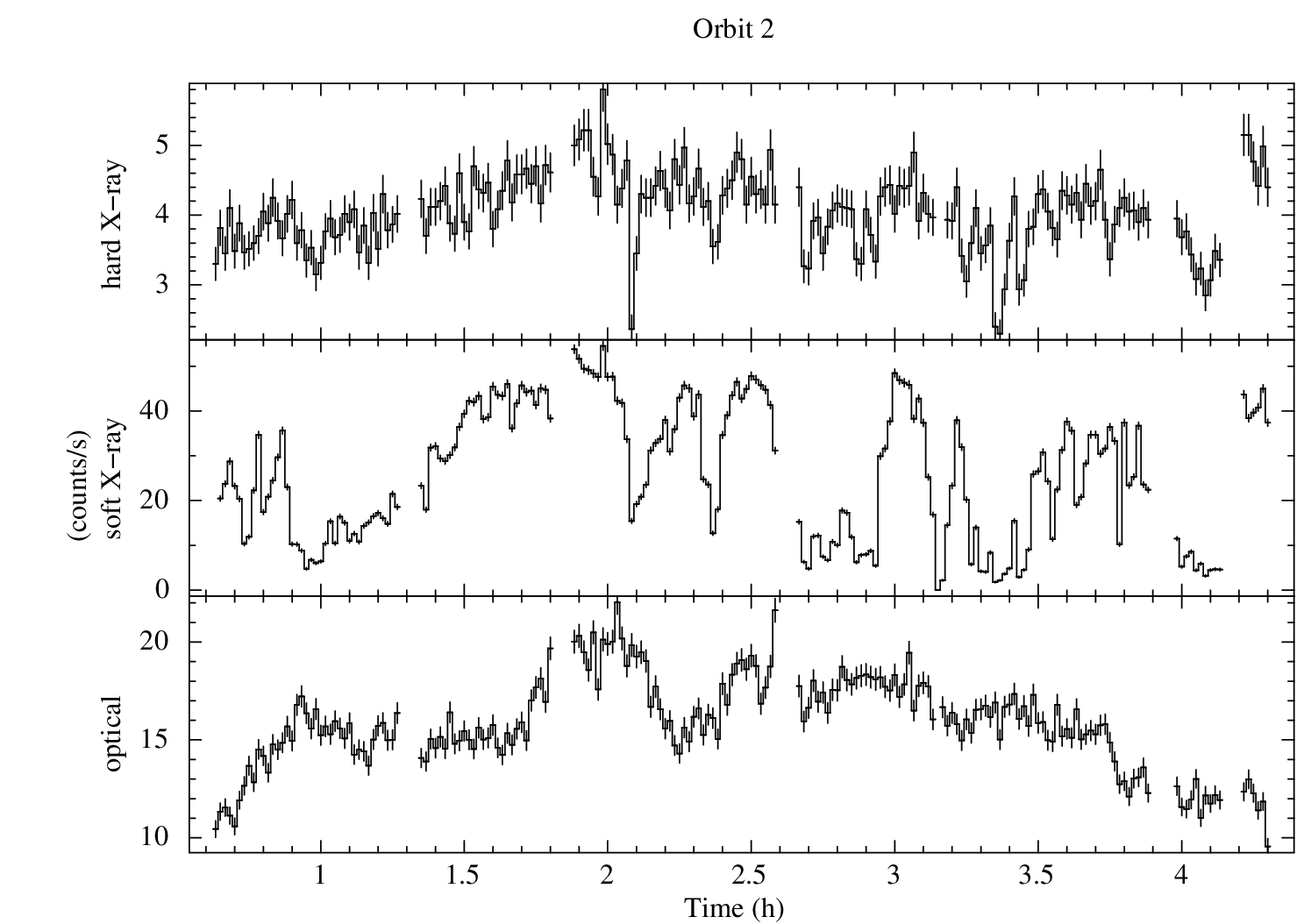}
 \includegraphics[scale=0.23, angle=-90]{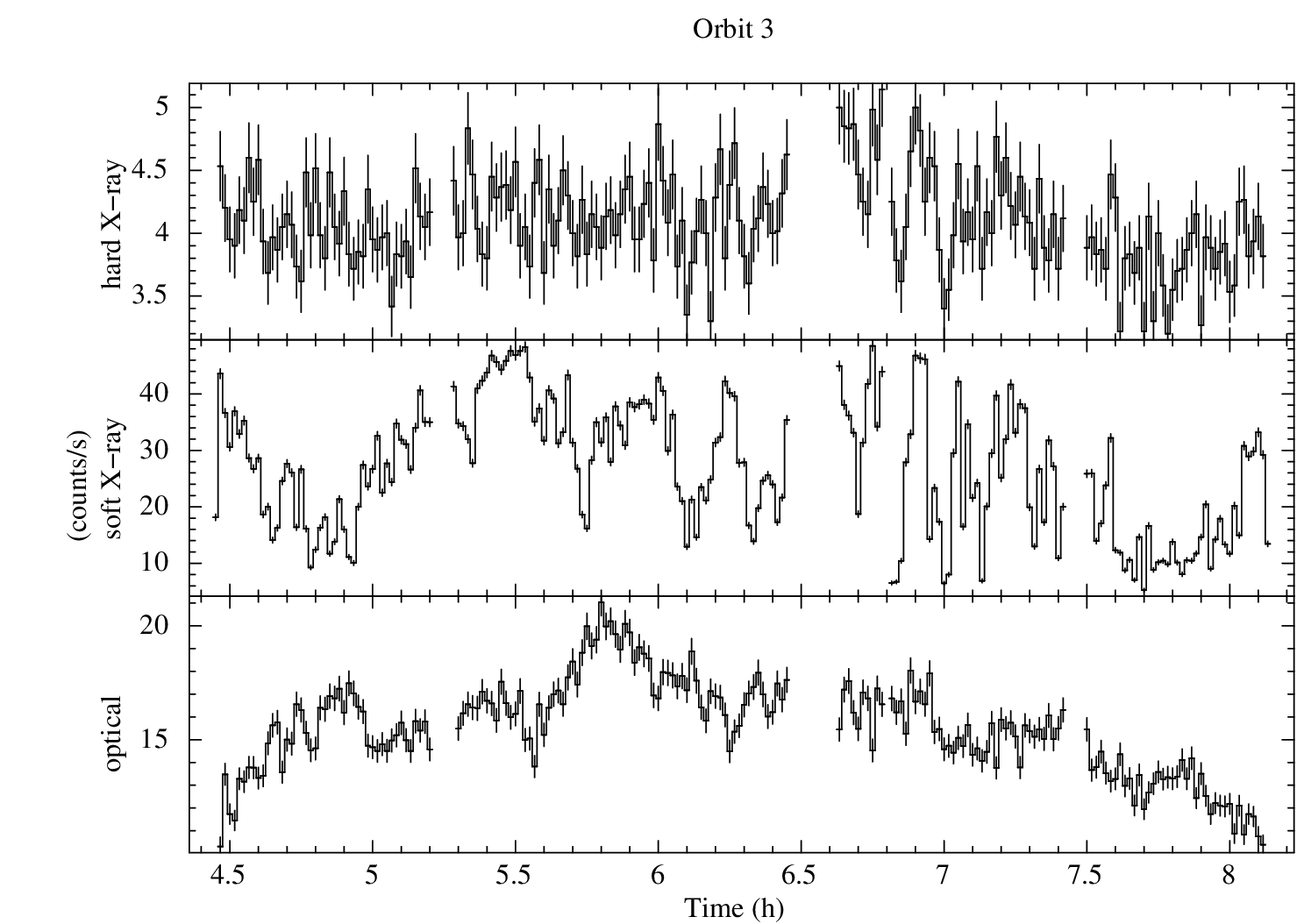}
 \includegraphics[scale=0.23, angle=-90]{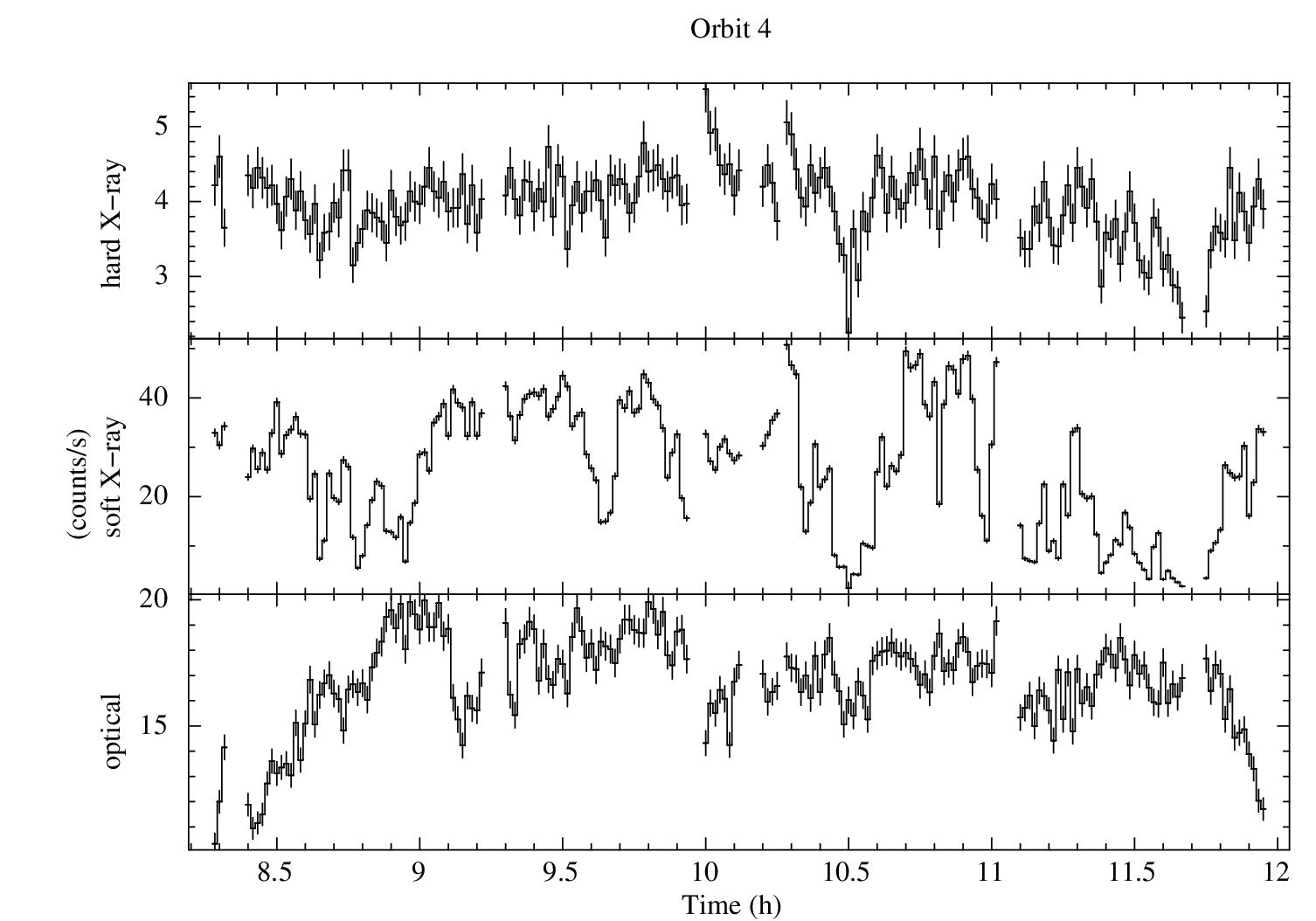}
\caption{\obsid 0160760601 }
 \end{figure*}
 
\begin{figure*}
\centering
 \includegraphics[scale=0.23, angle=-90]{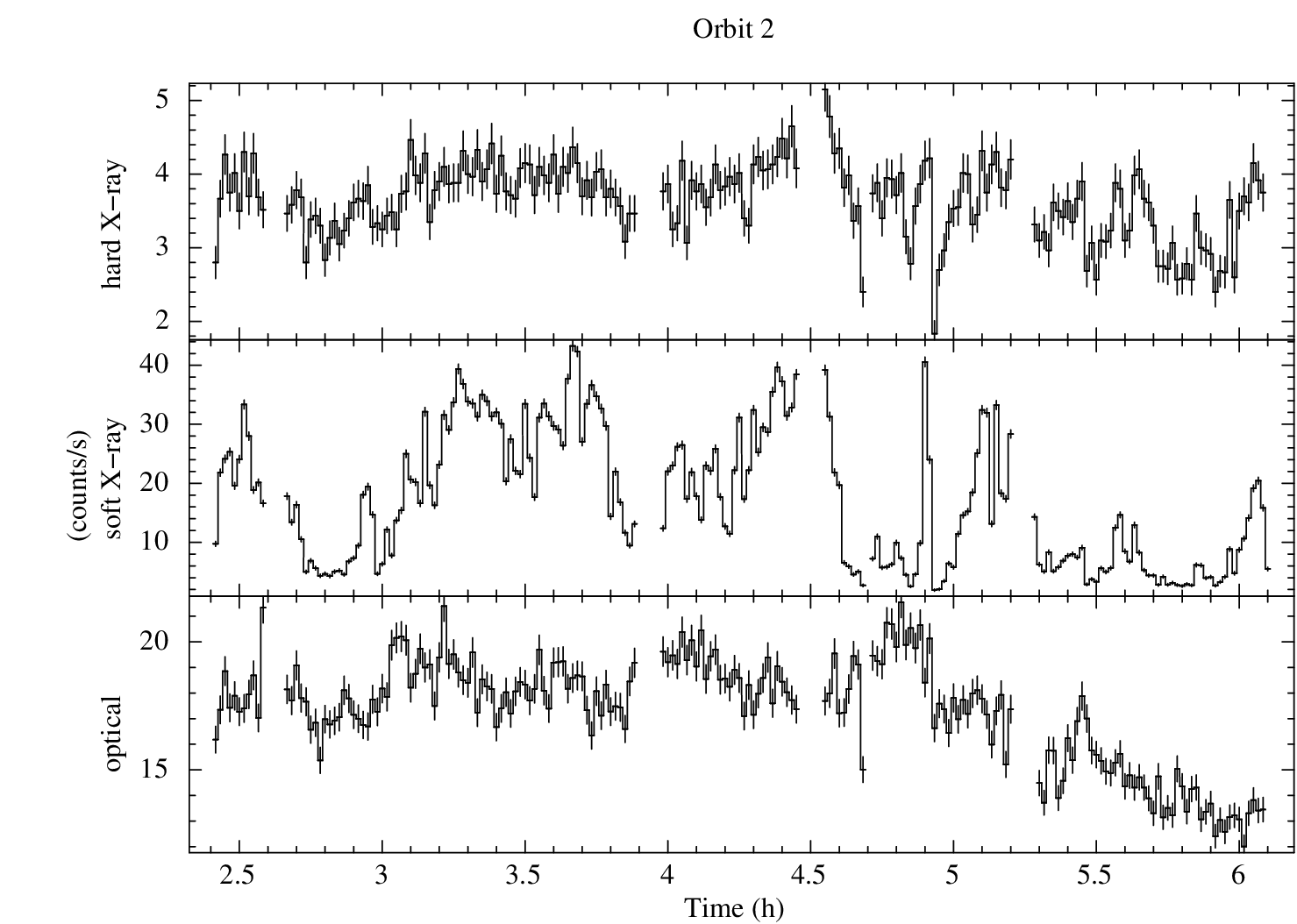}
 \includegraphics[scale=0.23, angle=-90]{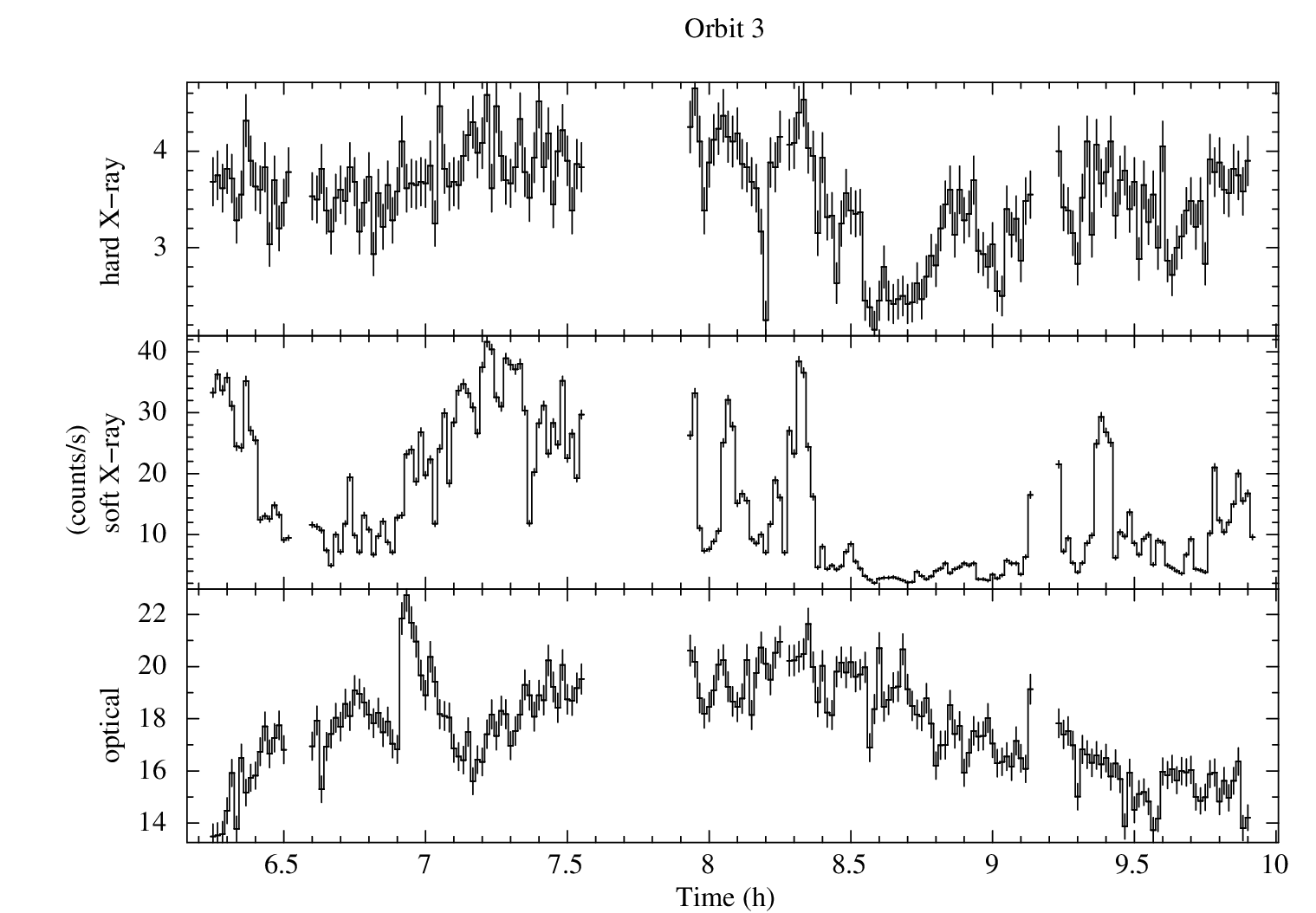}
 \includegraphics[scale=0.23, angle=-90]{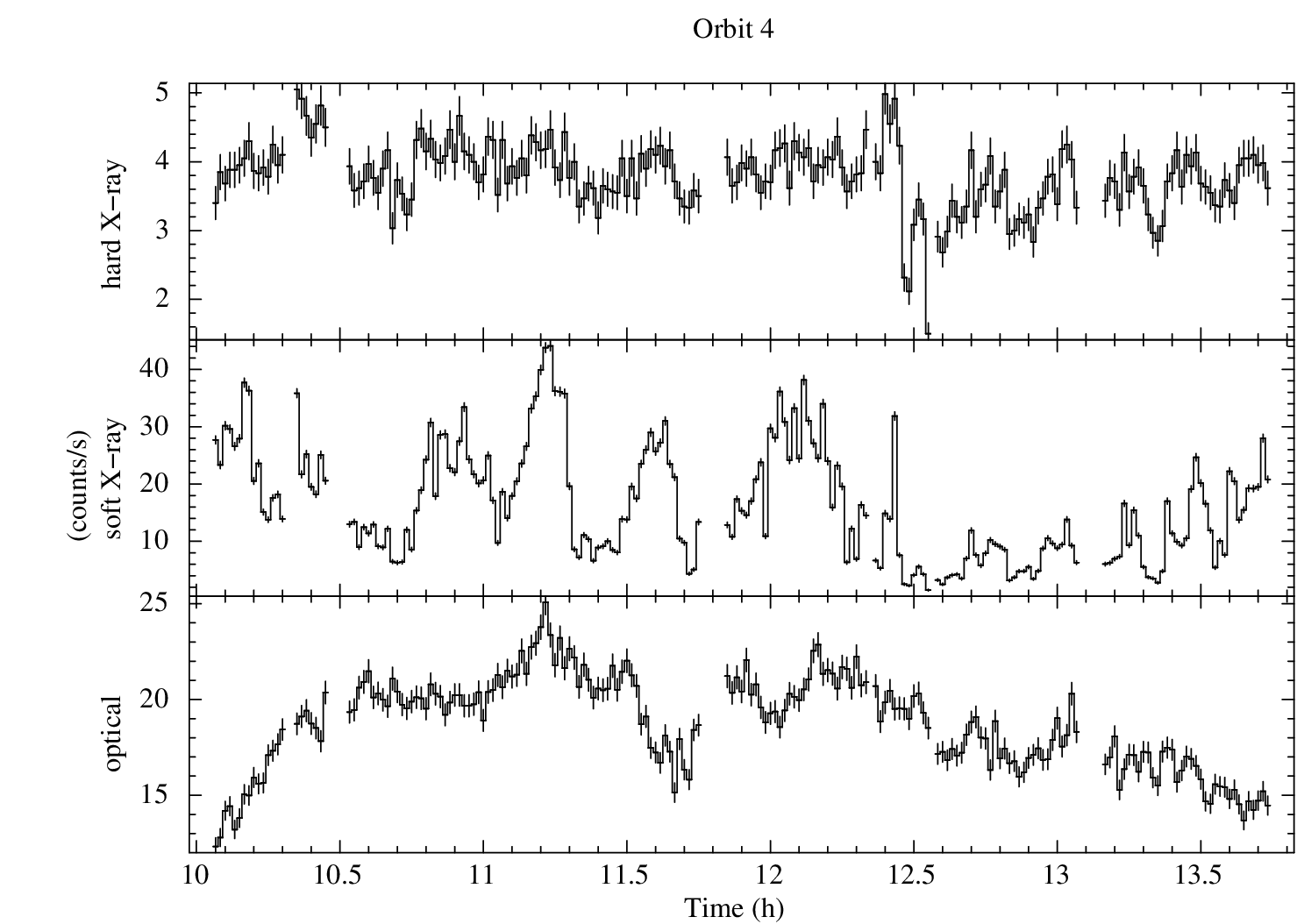}
\caption{Obs Id 0160760801} 
    \end{figure*}

 \begin{figure*}
 \begin{minipage}{1.0\linewidth}  
 \centering
 \includegraphics[scale=0.23, angle=-90]{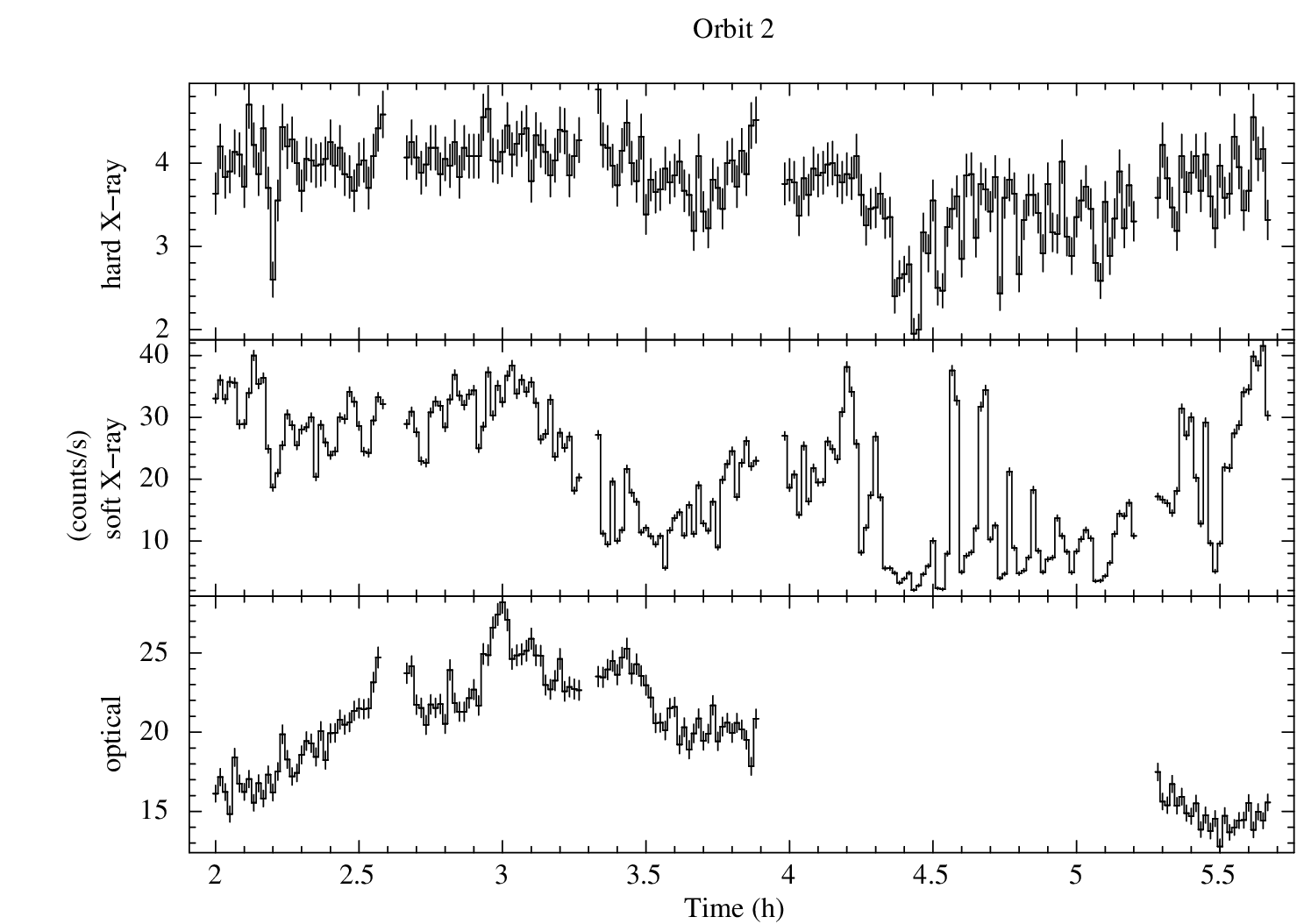}
 \includegraphics[scale=0.23, angle=-90]{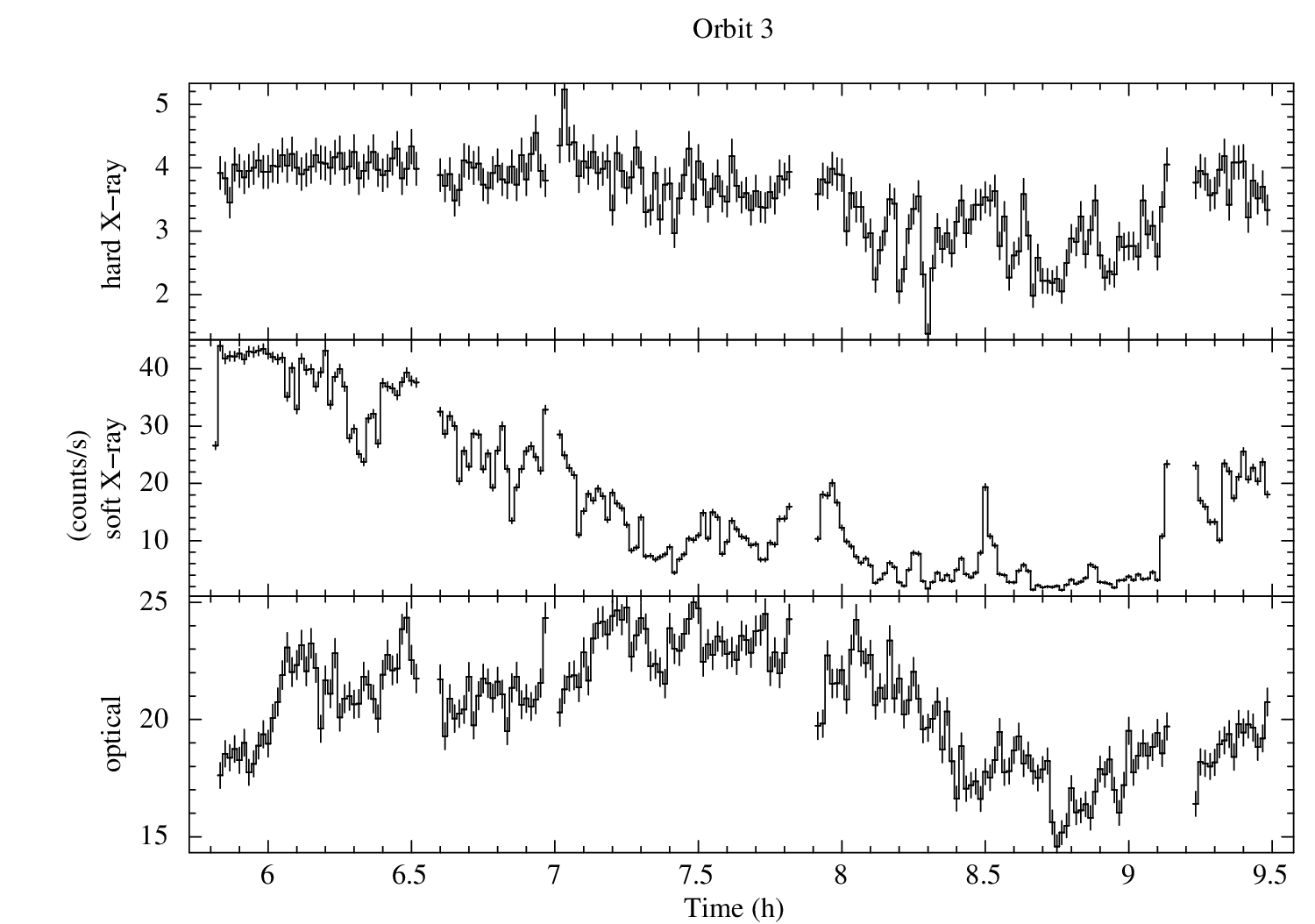}
 
  \end{minipage}
\hspace{0.9cm}
 \begin{minipage}{1.0\linewidth}  
 \centering
 \includegraphics[scale=0.23, angle=-90]{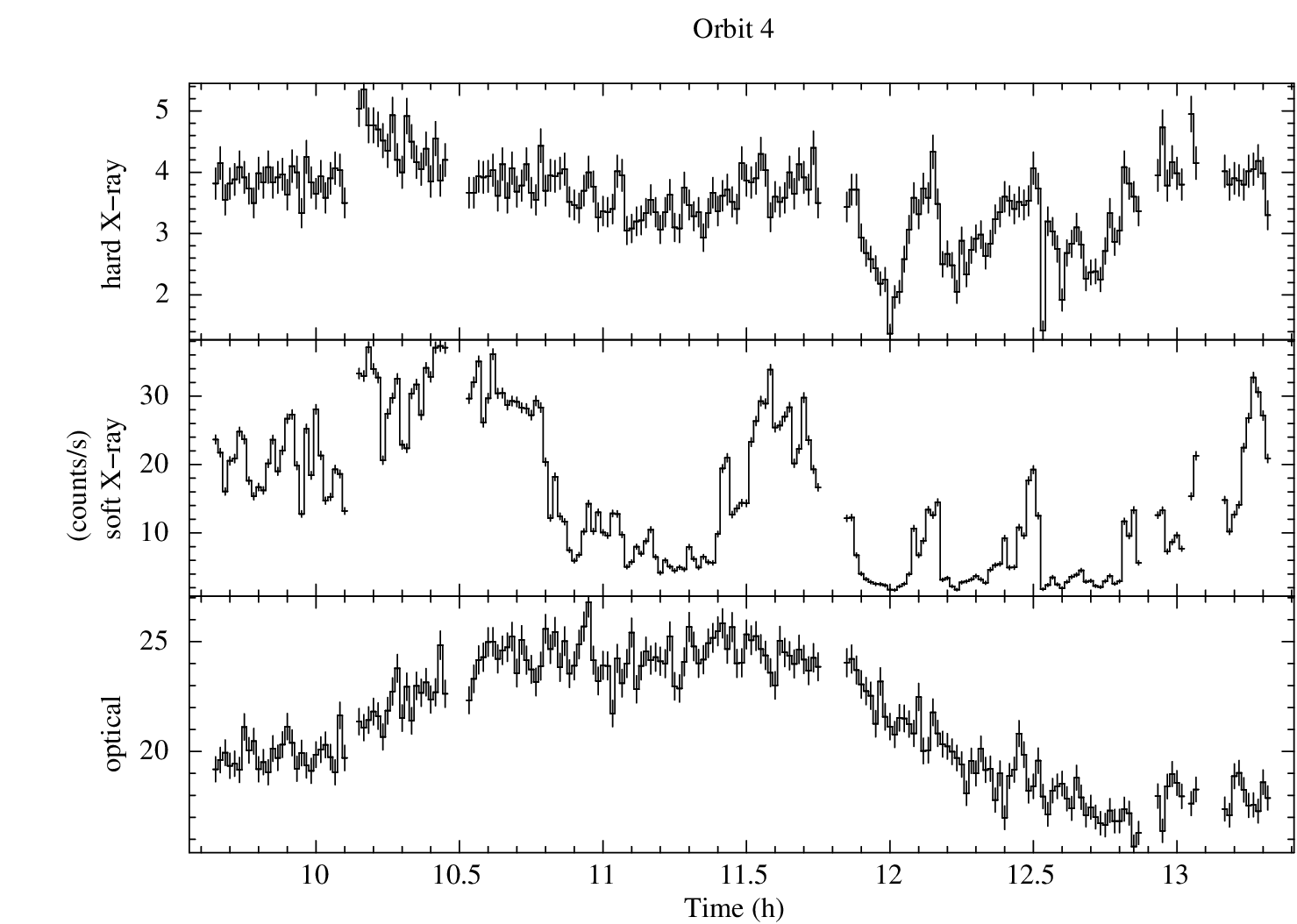}
 \includegraphics[scale=0.23, angle=-90]{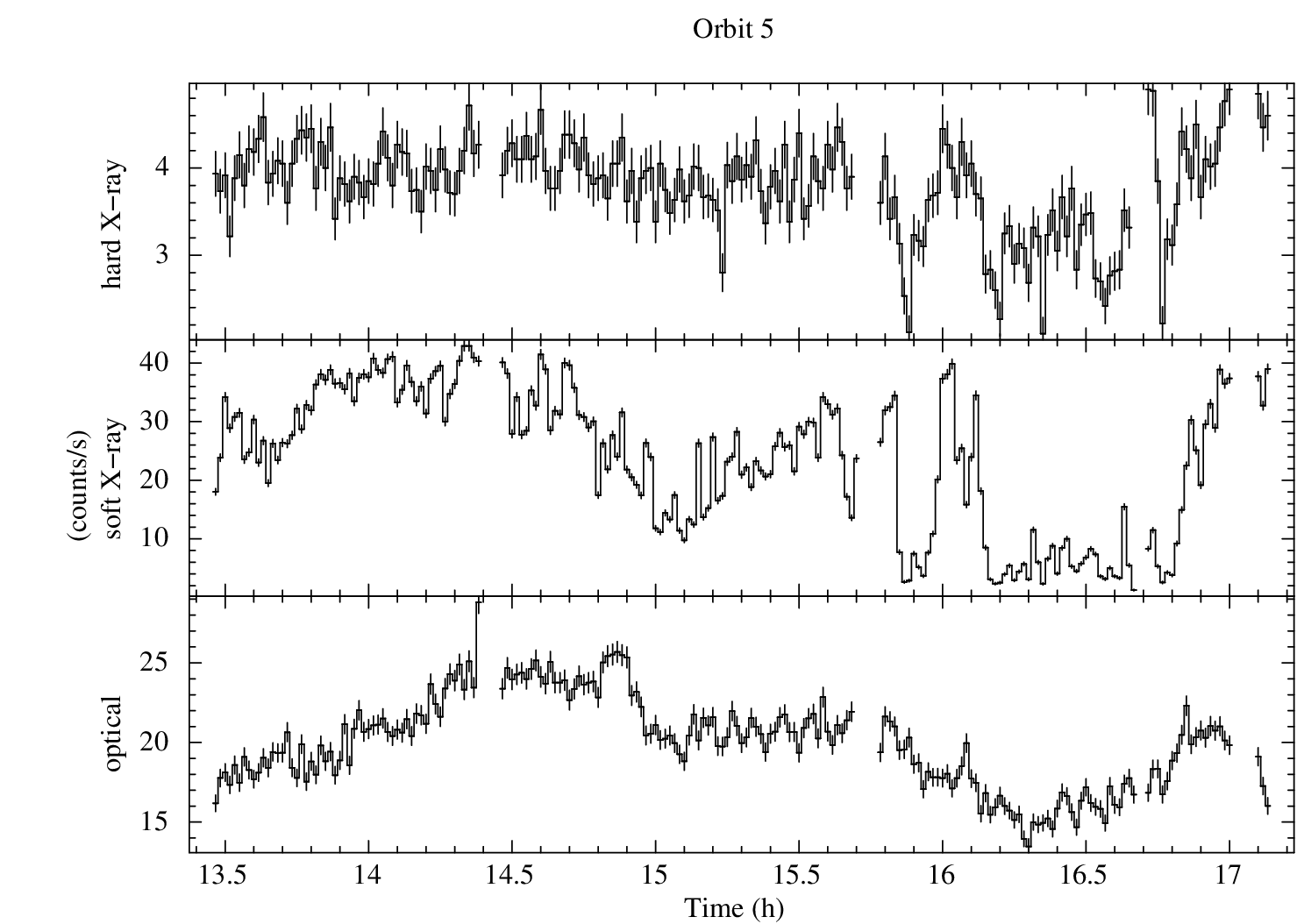}
 
  \end{minipage}
\hspace{0.9cm}
 \begin{minipage}{1.0\linewidth}  
 \centering
  \includegraphics[scale=0.23, angle=-90]{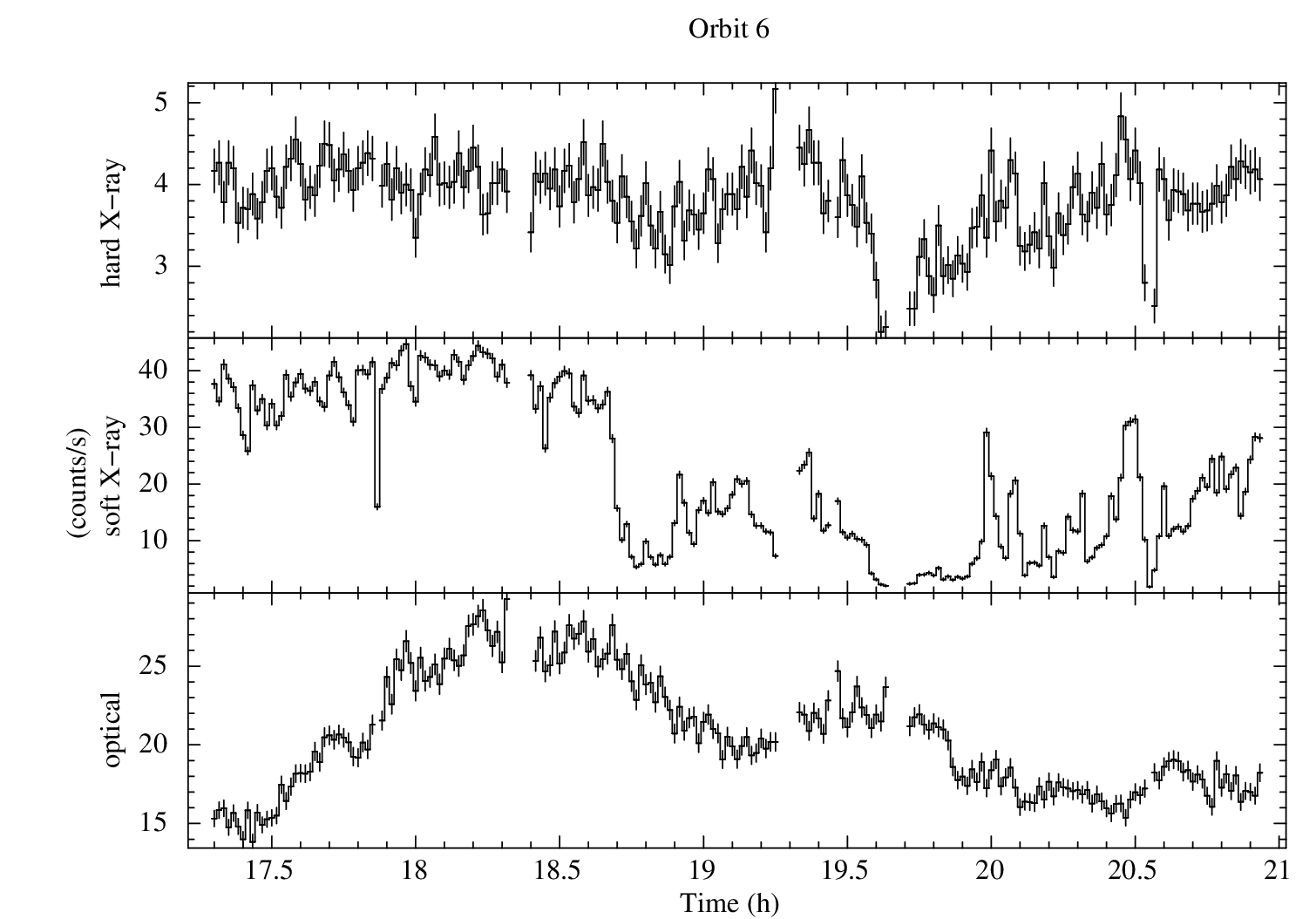}
 \includegraphics[scale=0.23, angle=-90]{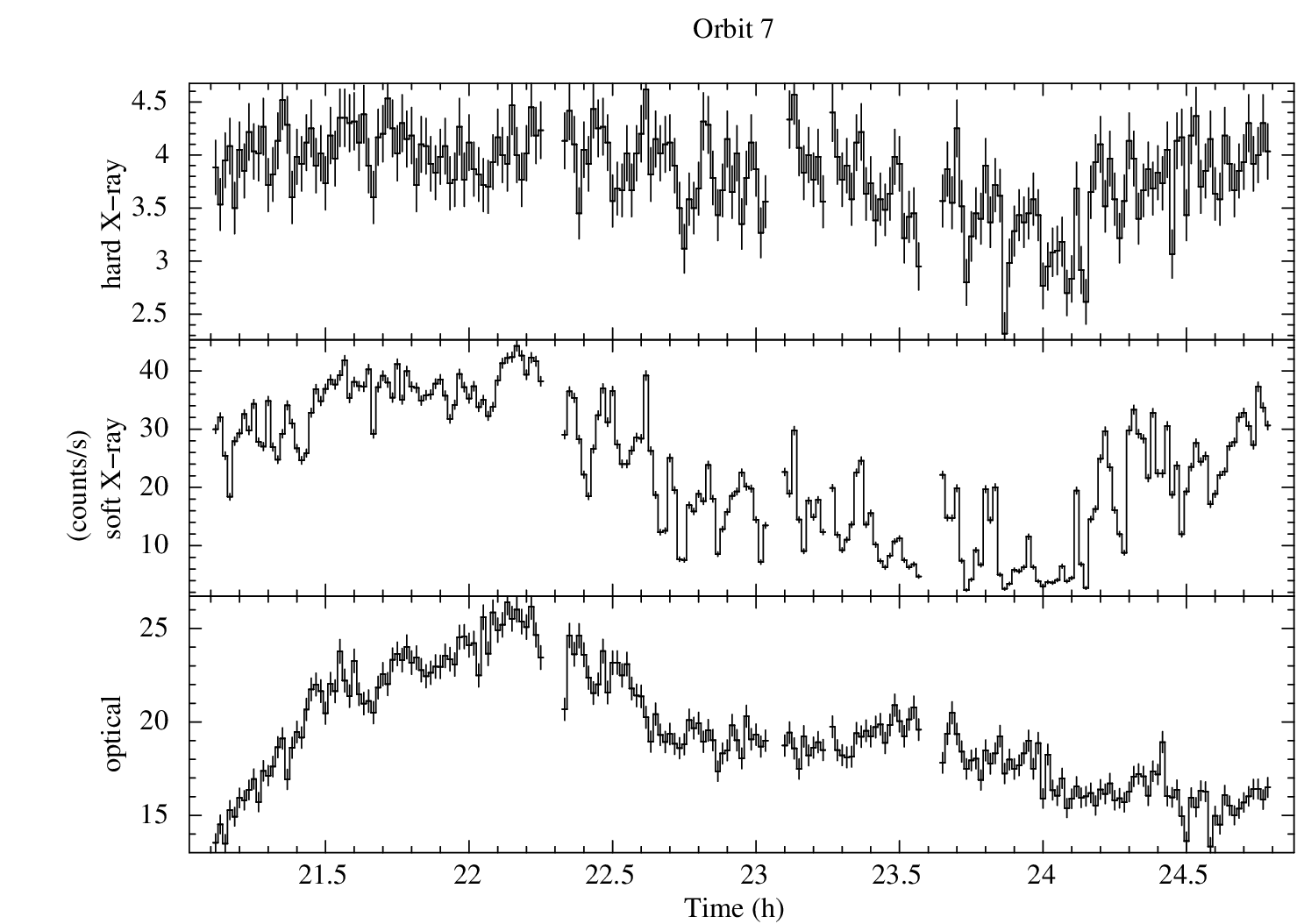}
 \end{minipage}
 \caption{\obsid 0160761301}
 \end{figure*}

 \end{appendix}

 \end{document}